\documentclass[12pt,doublespacing]{article}

\usepackage{fullpage}
\usepackage{setspace}
\usepackage{authblk}

\usepackage{cite}
\usepackage{hyperref} 

\usepackage{graphicx}

\usepackage{amsmath}
\usepackage{amssymb}

\usepackage{float}
\usepackage{subfig}
\usepackage{makecell}
\usepackage{rotating}

\usepackage{enumitem}

\begin{document}

\title{Enhanced Absorption in Thin-Film Silicon Solar Cells Using a Broadband Plasmonic Nanostructure}

\author[1]{Partha Mondal}
\author[2]{Omar Alkhazragi}
\author[1]{Hakan Bagci\vspace{0.8cm}}

\affil[1]{Electrical and Computer Engineering (ECE) Program
\authorcr Electrical and Mathematical Science and Engineering (CEMSE) Division
\authorcr King Abdullah University of Science and Technology (KAUST)
\authorcr Thuwal 23955, Saudi Arabia
\authorcr e-mail: partha.mondal@kaust.edu.sa\vspace{0.8cm}}

\affil[2]{Photonics Laboratory, Computer
\authorcr Electrical and Mathematical Science and Engineering (CEMSE) Division
\authorcr King Abdullah University of Science and Technology (KAUST)
\authorcr Thuwal 23955, Saudi Arabia}

\date{}
\maketitle
\newpage

\begin{abstract} 
 The design and fabrication of a metal-dielectric-metal absorber that achieves strong absorption from the ultraviolet (UV) to the near-infrared (near-IR) spectrum are presented. The proposed nanostructure consists of a periodic titanium ($\mathrm{Ti}$) array as the top layer, a thin silicon dioxide ($\mathrm{SiO_{2}}$) spacer, and a continuous aluminum ($\mathrm{Al}$) layer serving as the back reflector. Comprehensive optimization of structural parameters results in an average absorptance of $96\%$ in the $280$-$1000\,\mathrm{nm}$ wavelength range. The proposed design exhibits polarization insensitivity and maintains high absorption efficiency under oblique incidence. Fabrication is carried out using electron beam lithography followed by a lift-off process, ensuring both high performance and manufacturing simplicity. Experimental measurements show strong agreement with numerical simulations, validating the effectiveness of the design. Furthermore, integration of the absorber into a thin-film silicon ($\mathrm{Si}$) solar cell is analyzed, revealing significant enhancement in light absorption within the active layer. Owing to its broadband response, angular robustness, and structural simplicity, the proposed absorber shows strong potential for applications in solar energy harvesting, thermal emission, and advanced photovoltaic technologies.
\end{abstract}

\section{Introduction}
Since the pioneering work of Landy \emph{et al.} on metamaterial absorbers~\cite{landy2008perfect}, extensive research has focused on the development of high-performance absorbers operating across the microwave, terahertz, and optical frequency ranges, featuring single-band, multi-band, and broadband absorption spectra~\cite{yu2019broadband,feng2020photonic}. Absorption in plasmonic metamaterial nanostructures can be engineered through the excitation of surface plasmons (SPs), localized surface plasmons (LSPs), and cavity modes. These electromagnetic resonances enable either narrowband or broadband absorption, depending on the design. Narrowband absorbers are essential for the development of sensors~\cite{liu2016all, liao2020ultra}, optical filters, thermal emitters~\cite{howes2020near}, and spectroscopic devices~\cite{kang2019ultra, korkmaz2020mid}. In contrast, broadband absorbers are widely employed in photodetectors~\cite{zou2021polarization}, solar cells~\cite{wang2023review,bagmanci2019solar,karim2024metasurface}, thermal emitters~\cite{armghan2024high}, imaging systems~\cite{mehrabi2021high, liu2013broadband}, and photovoltaic devices~\cite{shafique2023highly, agarwal2017analysis}. With growing emphasis on renewable energy, metamaterial nanostructures are being actively explored to improve solar energy harvesting through optimized broadband absorption efficiency.

Various approaches have been developed to achieve broadband absorption in plasmonic nanostructures across the ultraviolet (UV) to near-infrared (near-IR) frequency range~\cite{hajian2019active}. One effective strategy involves using geometrically optimized metamaterial unit cells to generate multiple resonances within a broad spectral range. A wide variety of structural designs, including cylindrical arrays~\cite{zheng2022high,chaudhuri2018highly}, rectangular arrays~\cite{feng2014parallel,mehrabi2022ultra,liu2020numerical}, rings~\cite{qin2020ultra,huang2018refractory}, hole arrays~\cite{qi2019ultra,baqir2019wide}, and more complex geometries~\cite{gao2019ultraviolet,yu2020ultra}, have been explored to achieve broadband absorption. These structures offer spectral tunability through geometric modification but often require intricate fabrication techniques or multilayer architectures, which can increase manufacturing complexity and cost.

Another approach to designing broadband absorbers involves the periodic stacking of dielectric and metal layers~\cite{li2015omnidirectional}. Although such designs enable lithography-free fabrication, increasing the number of layers complicates the manufacturing process and raises costs. For practical applications, there is a pressing need to develop simple and cost-effective metamaterial nanostructures that maintain high absorption efficiency. One promising approach is the metal-dielectric-metal (MDM) configuration, in which the top metal layer is nanostructured to achieve effective impedance matching, thereby enabling optimal absorption. 

Material selection plays a crucial role in absorber performance across the UV, visible, and IR ranges~\cite{ng2019plasmonic}. Plasmonic metals such as gold ($\mathrm{Au}$)~\cite{azad2016metasurface}, nickel ($\mathrm{Ni}$)~\cite{iravani2022mxenes,wang2022design}, chromium ($\mathrm{Cr}$)~\cite{lee2018polarization}, titanium ($\mathrm{Ti}$)~\cite{zhou2021ultra,li2022polarization,qin2021broadband,lei2018ultra,wen2025high}, silver ($\mathrm{Ag}$)~\cite{tuan2019numerical}, tungsten ($\mathrm{W}$)~\cite{rana2018tungsten}, and magnesium ($\mathrm{Mn}$)~\cite{sayed2023design} have been effectively employed to realize broadband absorption~\cite{wang2023review}. Recent studies have demonstrated high absorption efficiencies supported by $\mathrm{Ni}$-based square nano-ring structures, achieving broadband performance across the $400\,\mathrm{nm}$ to $3000\,\mathrm{nm}$ wavelength range~\cite{bilal2023nanoengineered}. Similarly, $\mathrm{W}$-based nano-holed structures have been shown to support broadband absorptivity~\cite{baqir2019wide}. In addition to noble and transition metals, metallic nitrides such as aluminum nitride ($\mathrm{AlN}$)~\cite{hasan2017novel}, zirconium nitride ($\mathrm{ZrN}$)~\cite{ijaz2021exploiting}, and titanium nitride ($\mathrm{TiN}$)~\cite{gao2019ultraviolet,cao2021high,haque2024analysis,khichar2024solar} have been used in the design of near-perfect absorbers. Importantly, thermal stability is a critical factor in absorber design, as electromagnetic absorption inherently leads to heat generation. In this context, $\mathrm{Ti}$ stands out as one of the most promising materials due to its high thermal stability, with a melting point of $1668\,^{\circ}\mathrm{C}$, and low mass density. These properties make $\mathrm{Ti}$ an excellent candidate for broadband absorbers operating from the UV to long-wavelength IR regions of the spectrum~\cite{zhou2021ultra,zhu2020ultra}.

For practical applications, key design requirements for a high-performance metamaterial absorber include high absorption, broadband operation, insensitivity to polarization and incidence angle, and a simplified geometry for cost-effective fabrication. These criteria are particularly challenging to meet in the UV and visible spectral regions, where nanoscale precision is critical and fabrication tolerances are limited. Despite extensive progress, achieving broadband, high-efficiency absorption with simple, manufacturable designs remains a major challenge.

This work presents a simple $\mathrm{Ti}$ nanostructure design that enables high-efficiency fabrication through single-step lithography. The geometric configuration ensures effective light absorption, polarization insensitivity, and broad angular tolerance, while the use of $\mathrm{Ti}$ provides high thermal stability. The proposed structure consists of a periodic nanostructured $\mathrm{Ti}$ top layer, a continuous $\mathrm{Al}$ back reflector, and a thin intermediate dielectric layer. With a total thickness of $178\,\mathrm{nm}$, the device achieves an average absorptance exceeding $95\%$ over a $720\,\mathrm{nm}$ bandwidth (from $280\,\mathrm{nm}$ to $1000\,\mathrm{nm}$) under normal incidence. The symmetrical geometry ensures polarization insensitivity and stable absorption across a wide range of incident angles. The dependence of the absorption performance on geometric parameters is systematically investigated. Experimental measurements of the optimized design closely match the numerical simulations. Additionally, a theoretical analysis demonstrates that integrating the proposed absorber with a thin-film $\mathrm{Si}$ solar cell enhances light absorption in the active layer. The simplicity, efficiency, and scalability of the proposed design make it a promising candidate for energy harvesting applications, including solar cells, thermal emitters, and photovoltaic systems.

\section{Metasurface design}
\begin{figure}[h!]
\centering
\subfloat[]{\includegraphics[width=0.4\columnwidth]{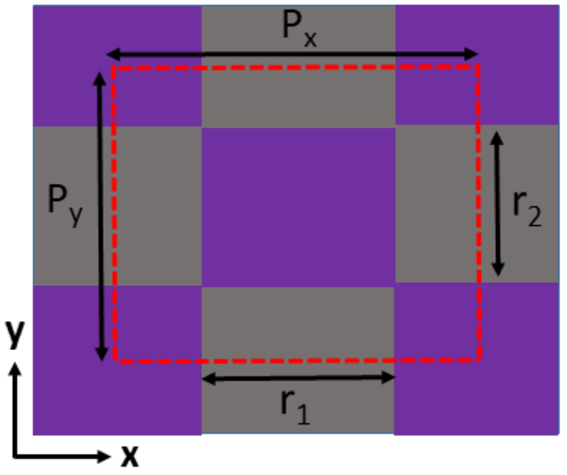}}\hspace{0.5cm}
\subfloat[]{\includegraphics[width=0.42\columnwidth]{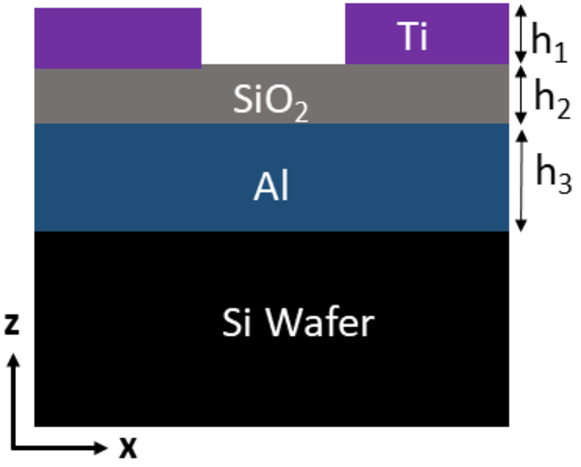}}\\
\subfloat[]{\includegraphics[width=0.55\columnwidth]{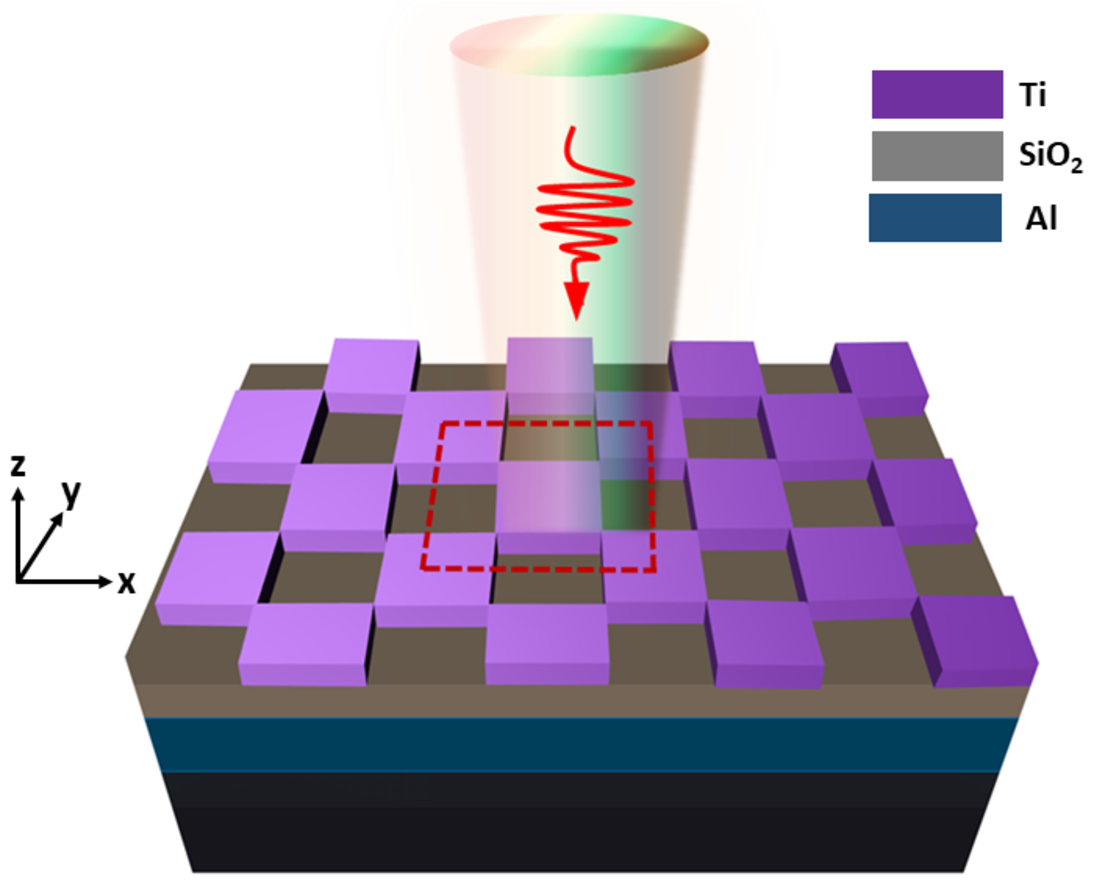}}
\caption{Schematic of the proposed MDM-based absorber. (a) Top view of the unit cell, (b) side view showing the layer configuration, and (c) 3D isometric view of the complete structure.}
\label{design_structure}
\end{figure}

The schematic of the proposed metamaterial absorber is illustrated in Fig.~\ref{design_structure}. Figure~\ref{design_structure}(a) shows the top view of the absorber, where a red dashed box highlights the unit cell consisting of symmetrical rectangular blocks made of $\mathrm{Ti}$. The side view of the structure is shown in Fig.~\ref{design_structure}(b). From top to bottom, the design consists of multiple material layers with varying thicknesses: a $\mathrm{Ti}$ layer of periodic blocks with a thickness of $h_1=38\,\mathrm{nm}$, a $\mathrm{SiO_{2}}$ spacer layer with a thickness of $h_2=40\,\mathrm{nm}$, and a continuous $\mathrm{Al}$ back reflector layer with a thickness of $h_3=100\,\mathrm{nm}$. The dimensions of the unit cell as shown by the red dashed box in Fig.~\ref{design_structure}(a) are $r_{1} = 120\,\mathrm{nm}$ and $r_{2} = 100\,\mathrm{nm}$, and periodicities in the $x$- and $y$-directions are $P_{x} = 220\,\mathrm{nm}$ and $P_{y} = 220\,\mathrm{nm}$, respectively. The device is fabricated on a $\mathrm{Si}$ substrate. A three-dimensional (3D) isometric view of the proposed structure is shown in Fig.~\ref{design_structure}(c). The complex dielectric constants of $\mathrm{Ti}$ and $\mathrm{Al}$ are obtained using the Drude-Lorentz  model~\cite{palik1998handbook} and the surrounding medium is assumed to be free space. 

\section{Simulation study and analysis}
Electromagnetic simulations of the plasmonic nanostructure are performed using the Finite-Difference Time-Domain (FDTD) method~\cite{refFDTD}. In the simulations, only the unit cell is considered, periodic boundary conditions are enforced on the unit cell along the $x$- and $y$-directions, while perfectly matched layers (PMLs) are used along the $z$-direction to absorb outgoing waves and eliminate boundary reflections. A plane wave is used to excite the structure, with the polarization and angle of incidence varied across different simulation scenarios, as described in the following sections. 




\subsection{Plane wave excitation at normal incidence}
\begin{figure}
  \centering
   \includegraphics[width=0.6\linewidth]{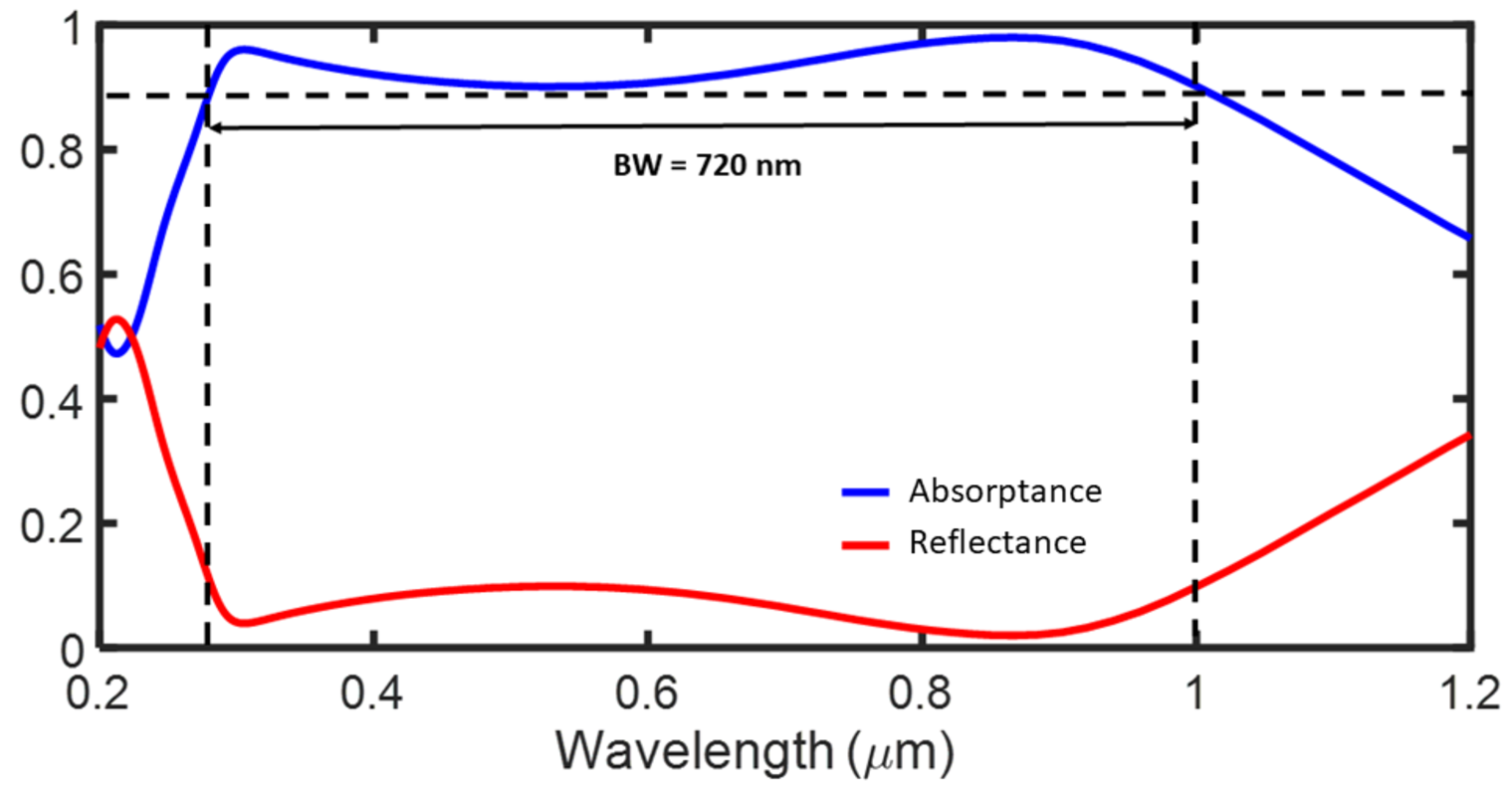}
  \caption{Absorptance and reflectance spectra of the proposed absorber under excitation by a normally incident $x$-polarized plane wave.}
    \label{simulated_absorption}
\end{figure}

In the initial simulation, the incident plane wave is polarized along the $x$-direction and is normally incident on the structure from the top. Figure~\ref{simulated_absorption} shows the simulated absorptance and reflectance spectra. The device exhibits near-unity absorptance, exceeding $90\%$, within the wavelength range from $280\,\mathrm{nm}$ to $1000\,\mathrm{nm}$, with an average absorptance of $96\%$ over a bandwidth of $720\,\mathrm{nm}$. Peak absorption occurs at $290\,\mathrm{nm}$ and $890\,\mathrm{nm}$. Note that in the frequency domain, absorptance ($A$), reflectance ($R$), and transmittance ($T$) satisfy the relation: $A = 1 - R - T$. In the proposed design, the back reflector $\mathrm{Al}$ layer is significantly thicker than the penetration depth of the incident light. As a result, transmittance through the structure is negligible ($T=0$) and the relationship above simplifies to: $A = 1 - R$, which is demonstrated by the results in Fig.~\ref{simulated_absorption}.

\begin{figure}
\centering
\subfloat[]{\includegraphics[width=0.45\columnwidth]{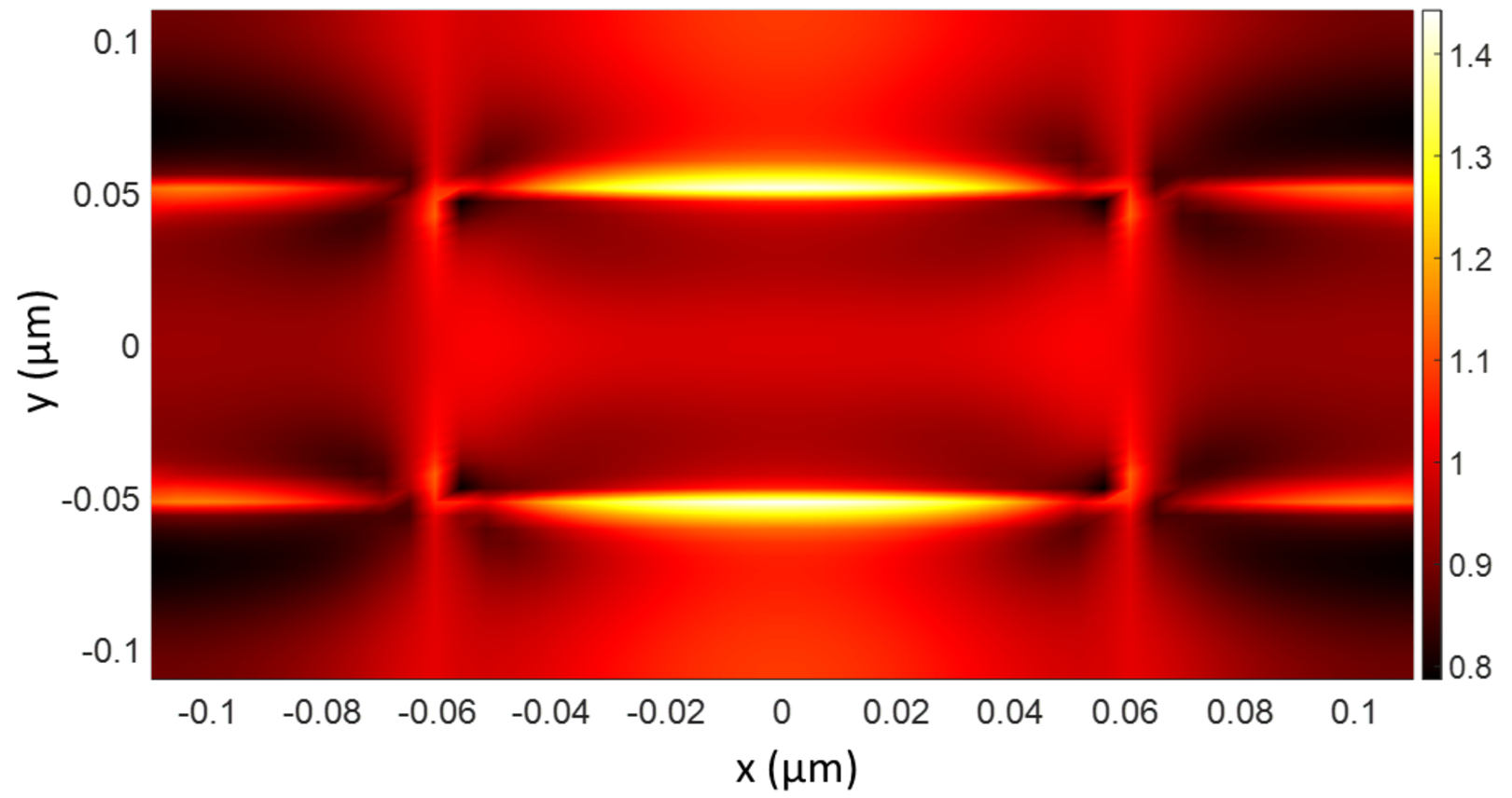}}\hspace{0.5cm}
\subfloat[]{\includegraphics[width=0.45\columnwidth]{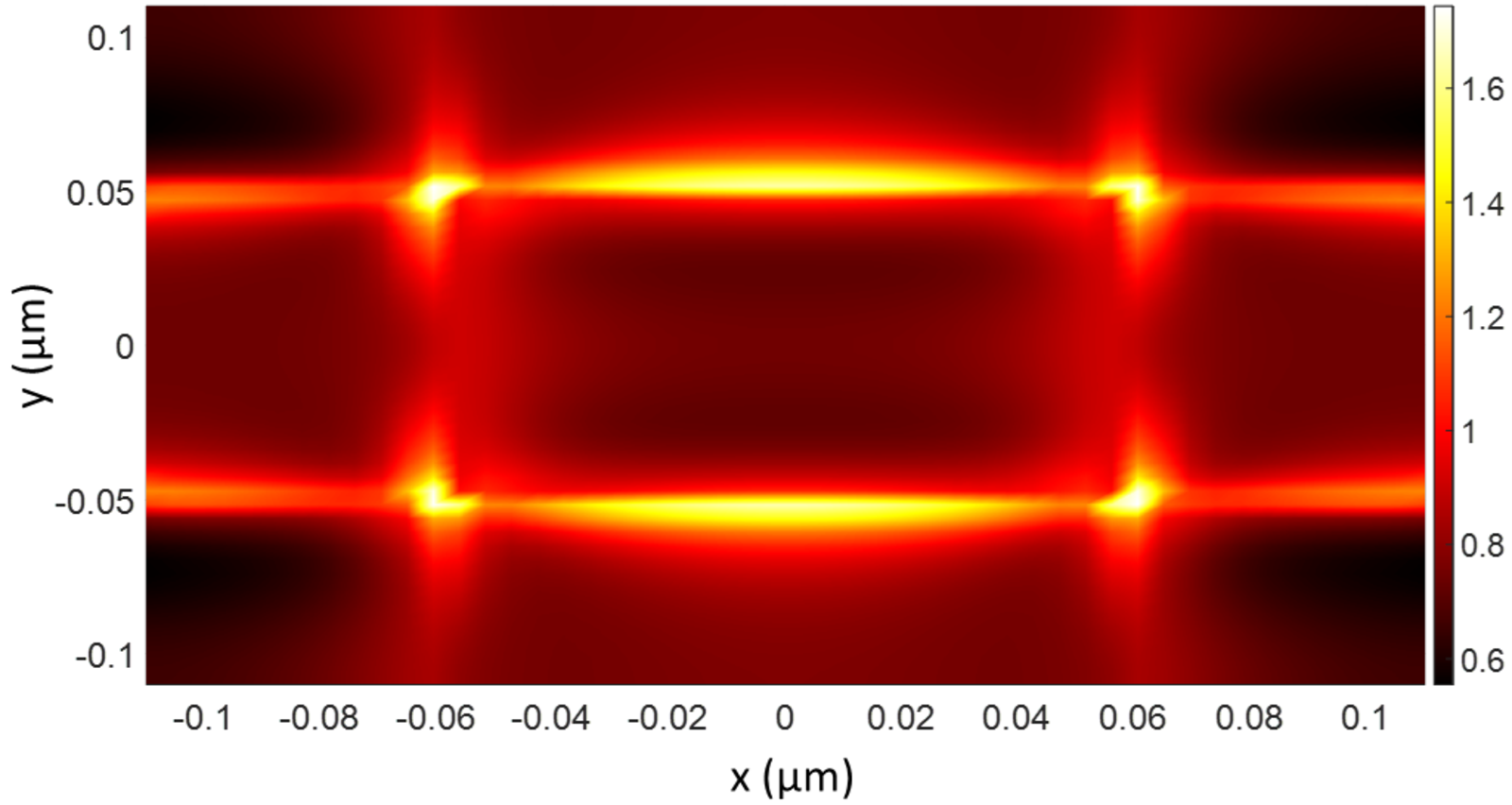}}\\
\subfloat[]{\includegraphics[width=0.45\columnwidth]{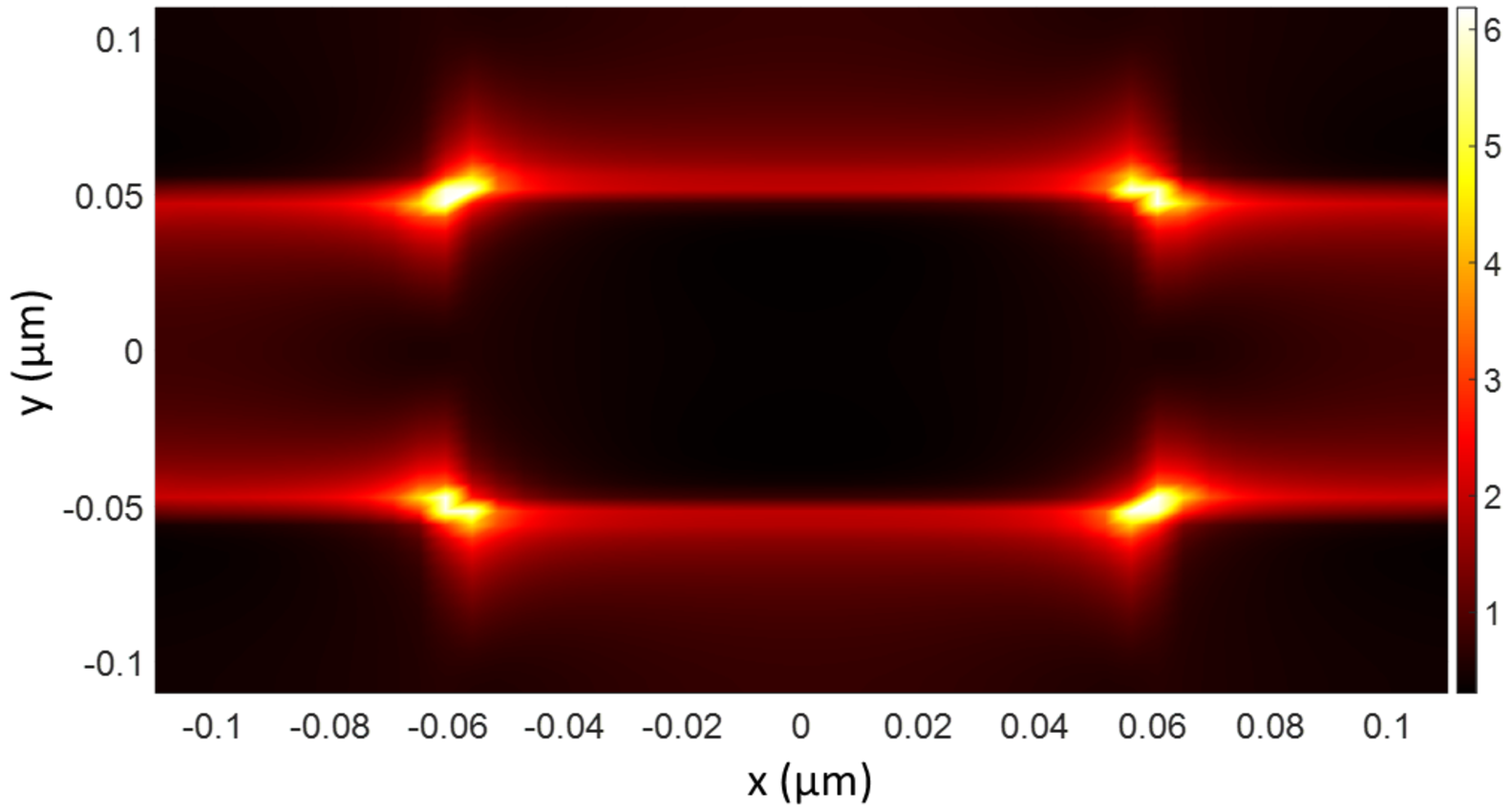}}\hspace{0.5cm}
\subfloat[]{\includegraphics[width=0.45\columnwidth]{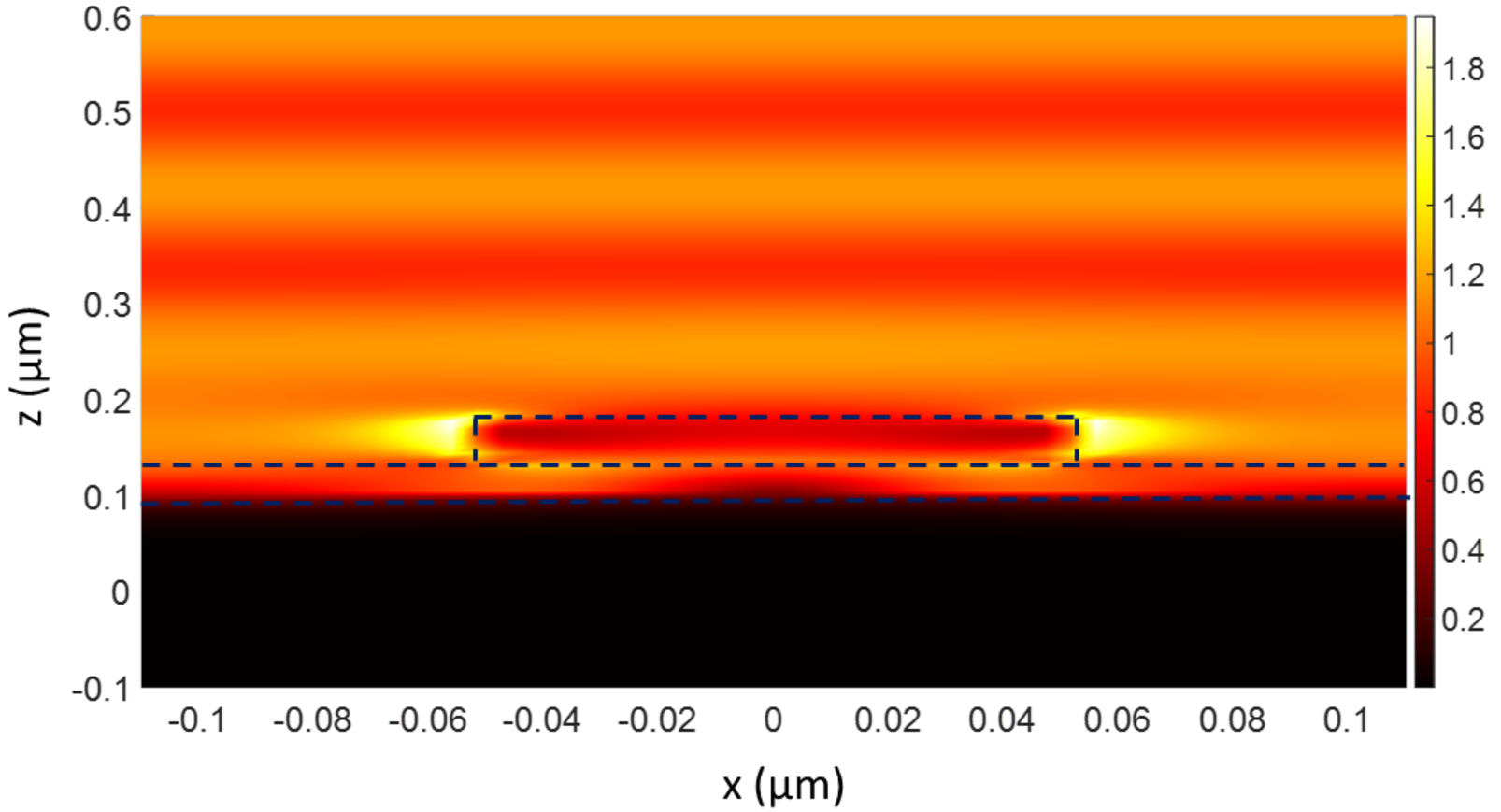}}\\
\subfloat[]{\includegraphics[width=0.45\columnwidth]{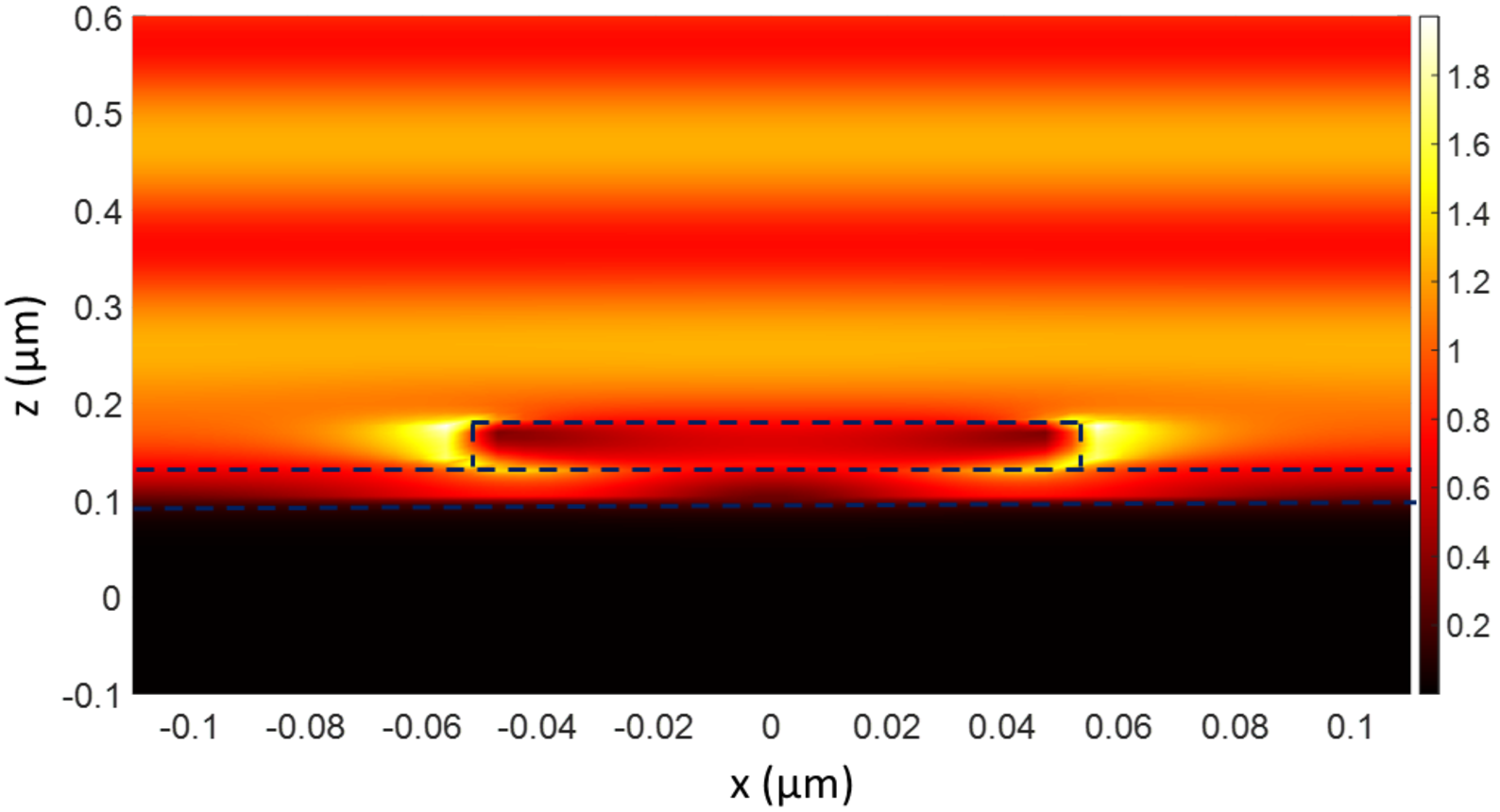}}\hspace{0.5cm}
\subfloat[]{\includegraphics[width=0.45\columnwidth]{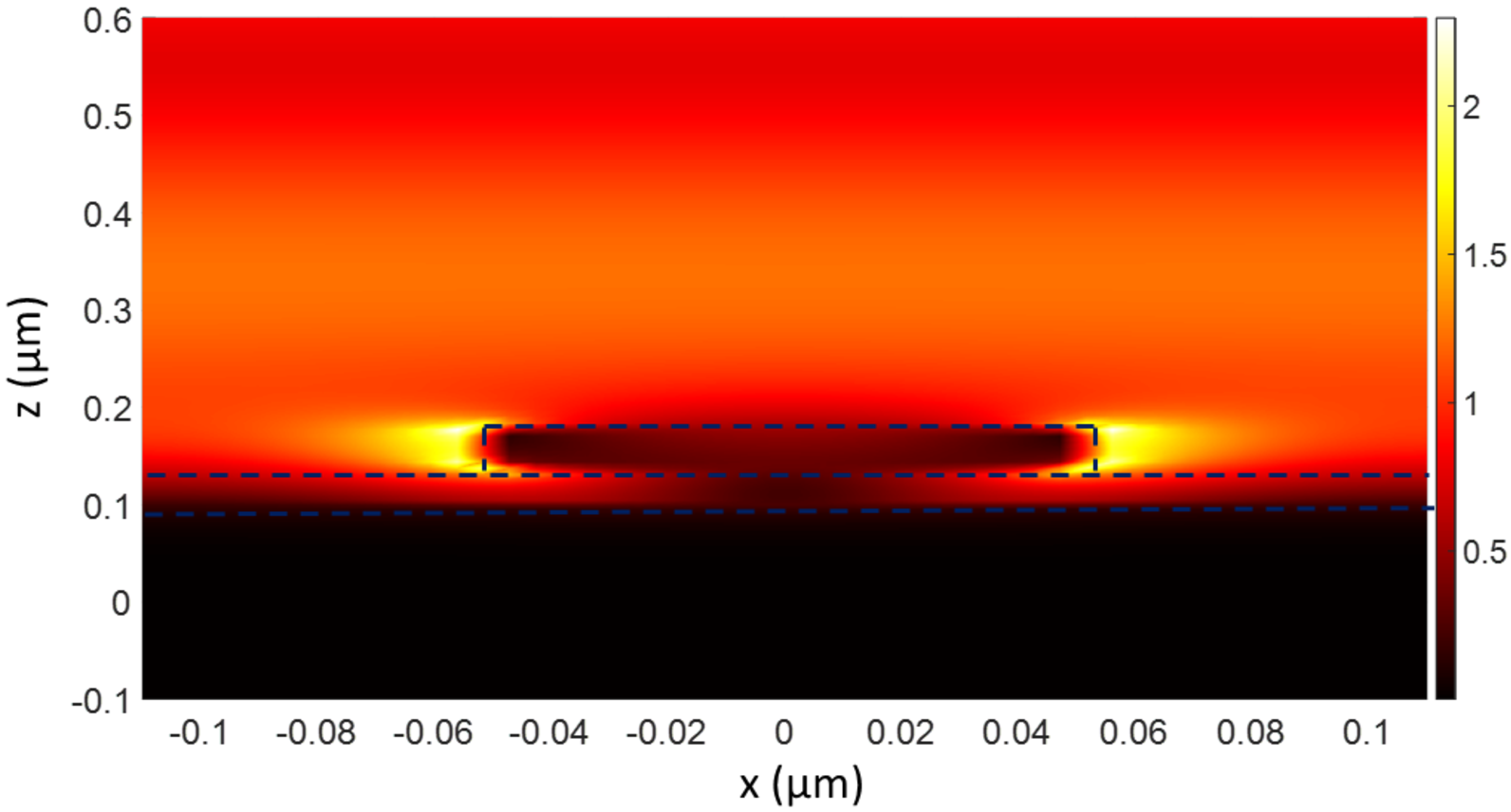}}
\caption{Magnitude of the electric field distribution in the $xy$-plane at (a) $333.5\,\mathrm{nm}$, (b) $420\,\mathrm{nm}$, and (c) $880.5\, \mathrm{nm}$, and in the $xz$-plane at (d) $333.5\,\mathrm{nm}$, (e) $420\,\mathrm{nm}$, and (f) $880.5\, \mathrm{nm}$.}
\label{electric_field_combined}
\end{figure}

\begin{figure}
\centering
\subfloat[]{\includegraphics[width=0.45\columnwidth]{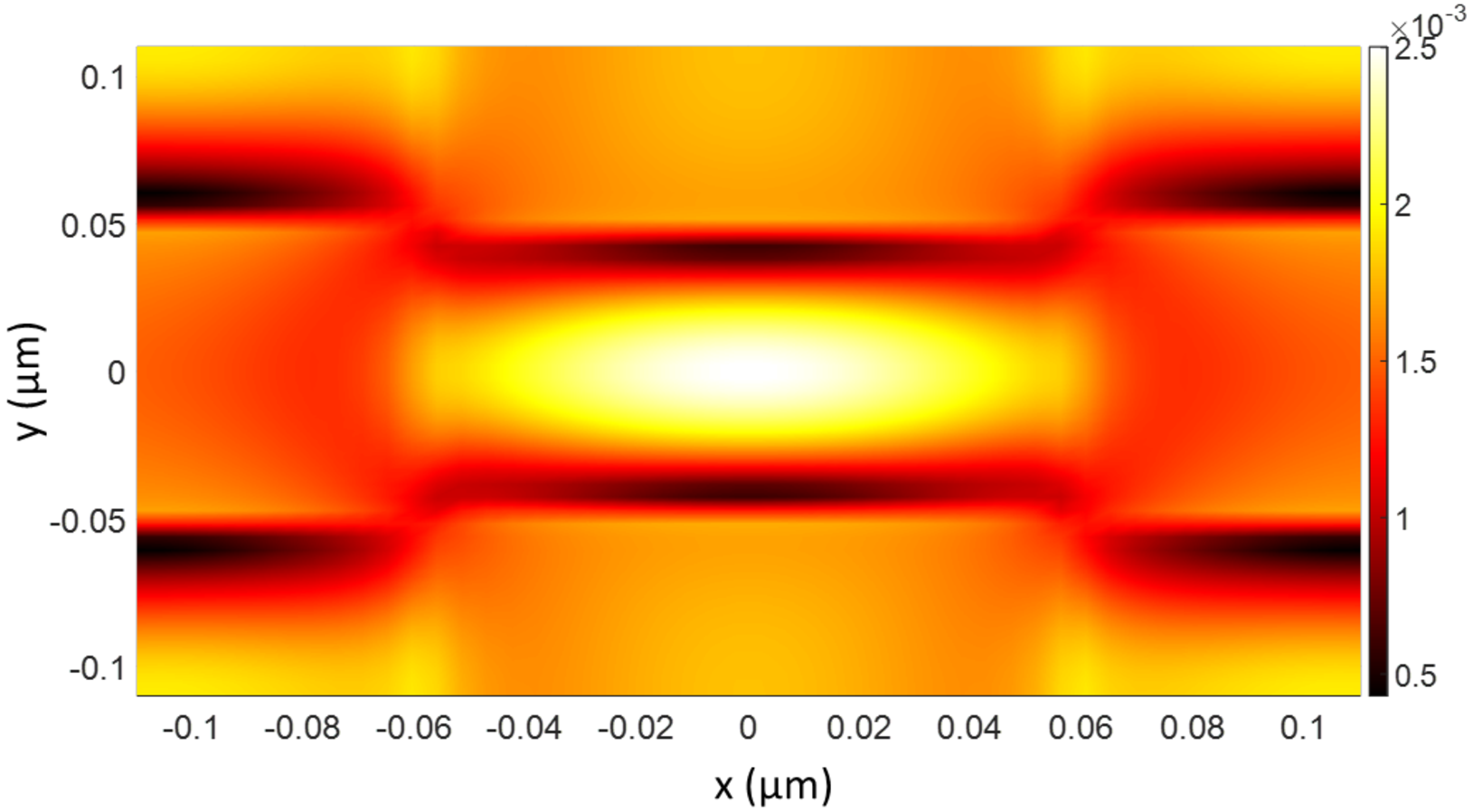}}\hspace{0.5cm}
\subfloat[]{\includegraphics[width=0.45\columnwidth]{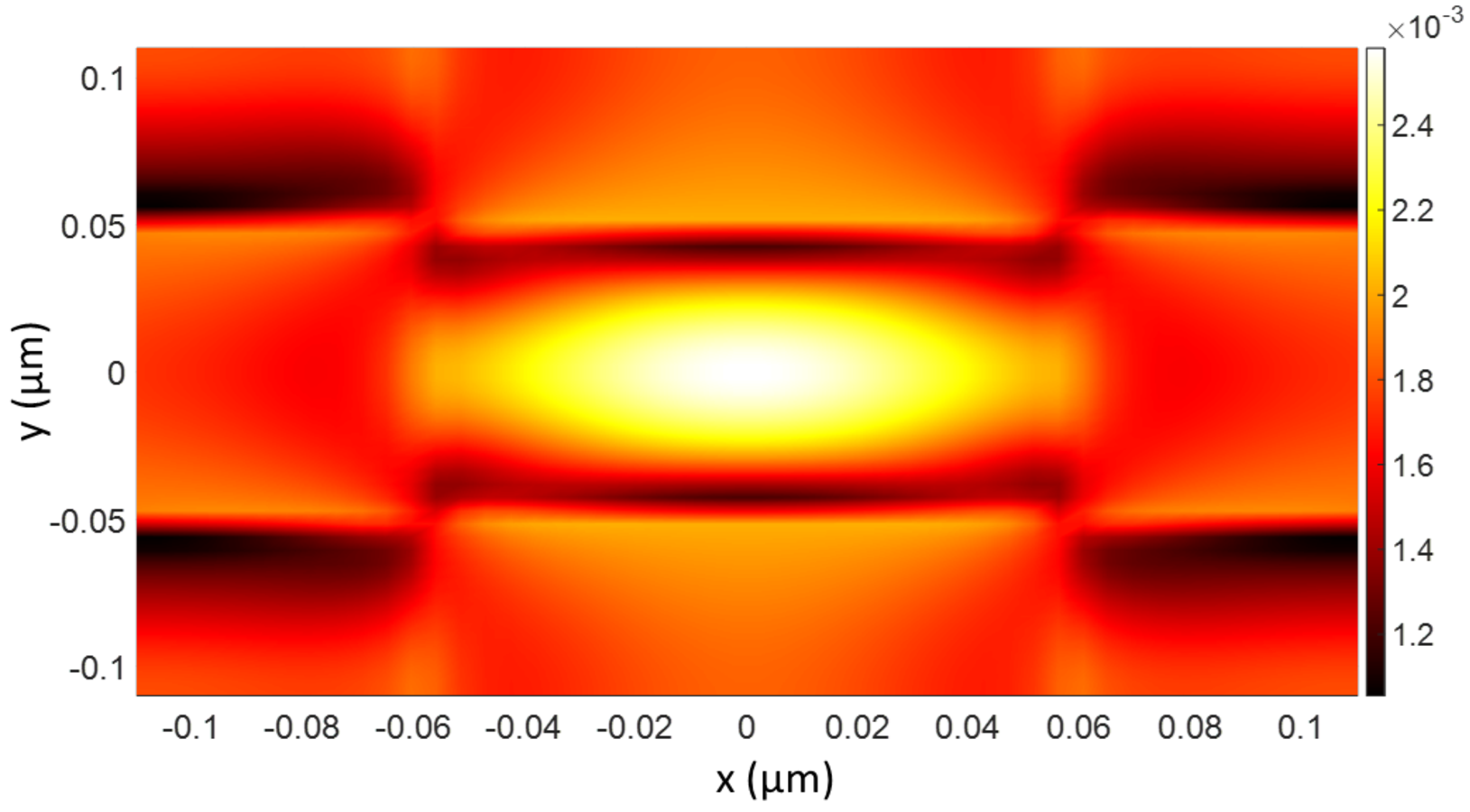}}\\
\subfloat[]{\includegraphics[width=0.45\columnwidth]{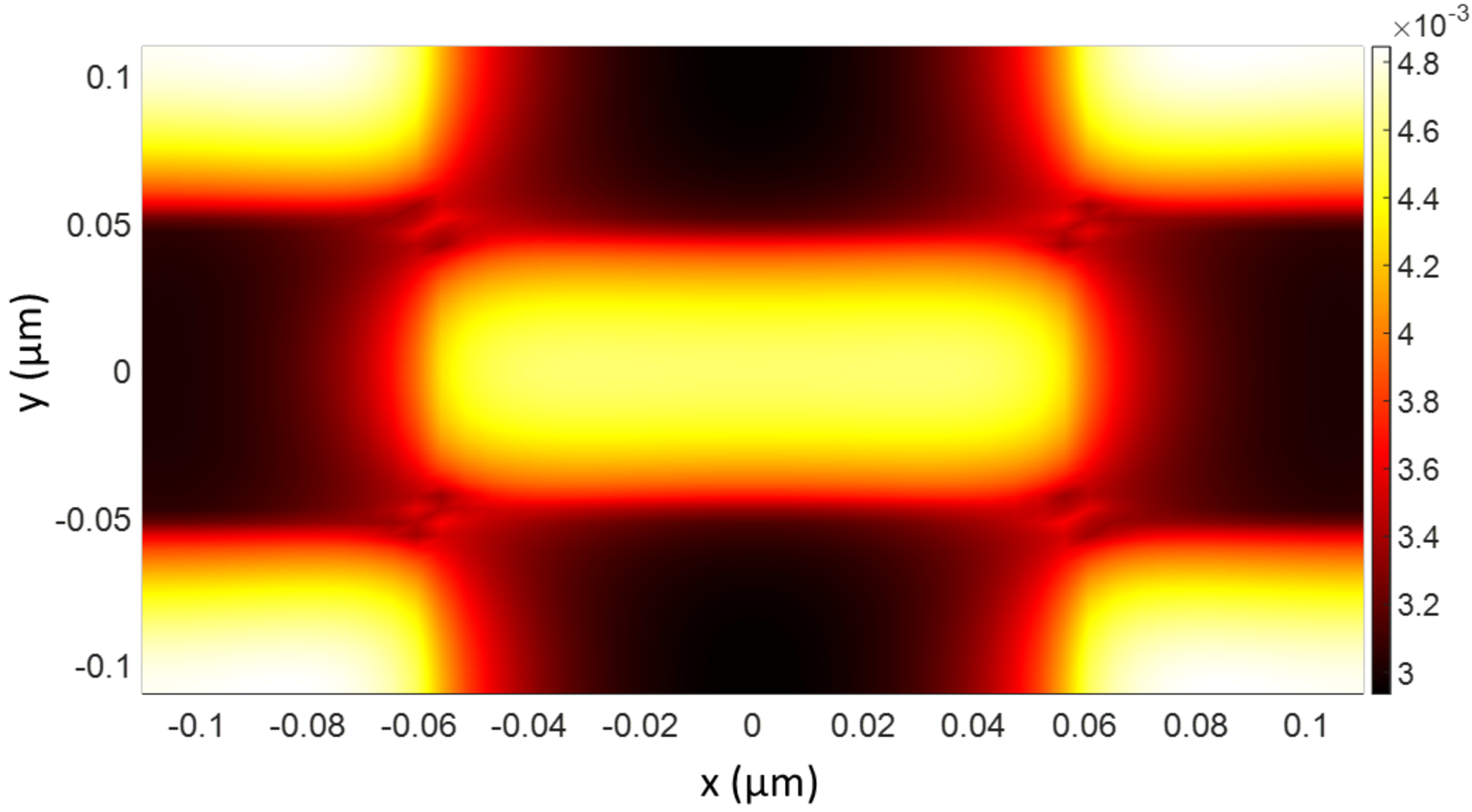}}\hspace{0.5cm}
\subfloat[]{\includegraphics[width=0.45\columnwidth]{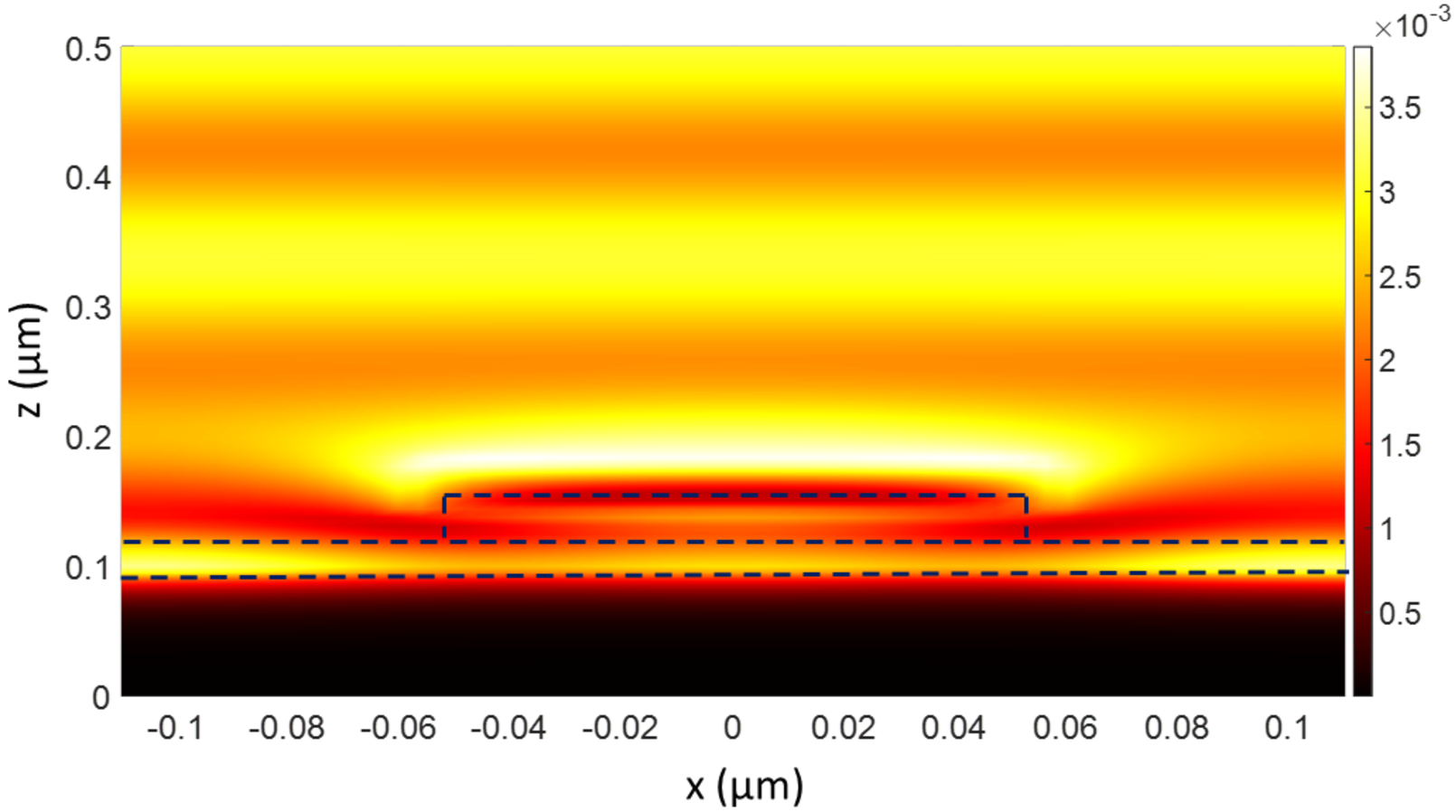}}\\
\subfloat[]{\includegraphics[width=0.45\columnwidth]{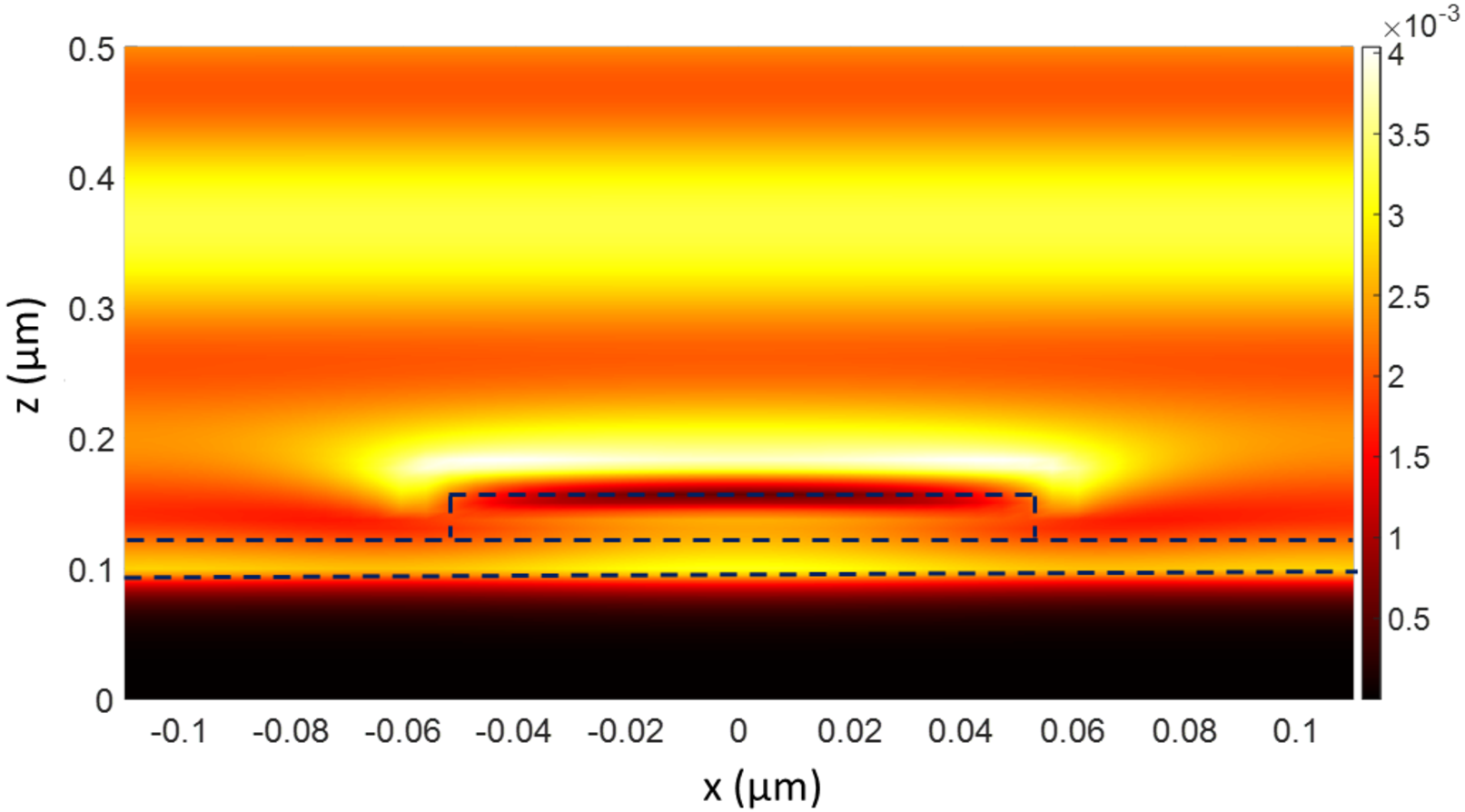}}\hspace{0.5cm}
\subfloat[]{\includegraphics[width=0.45\columnwidth]{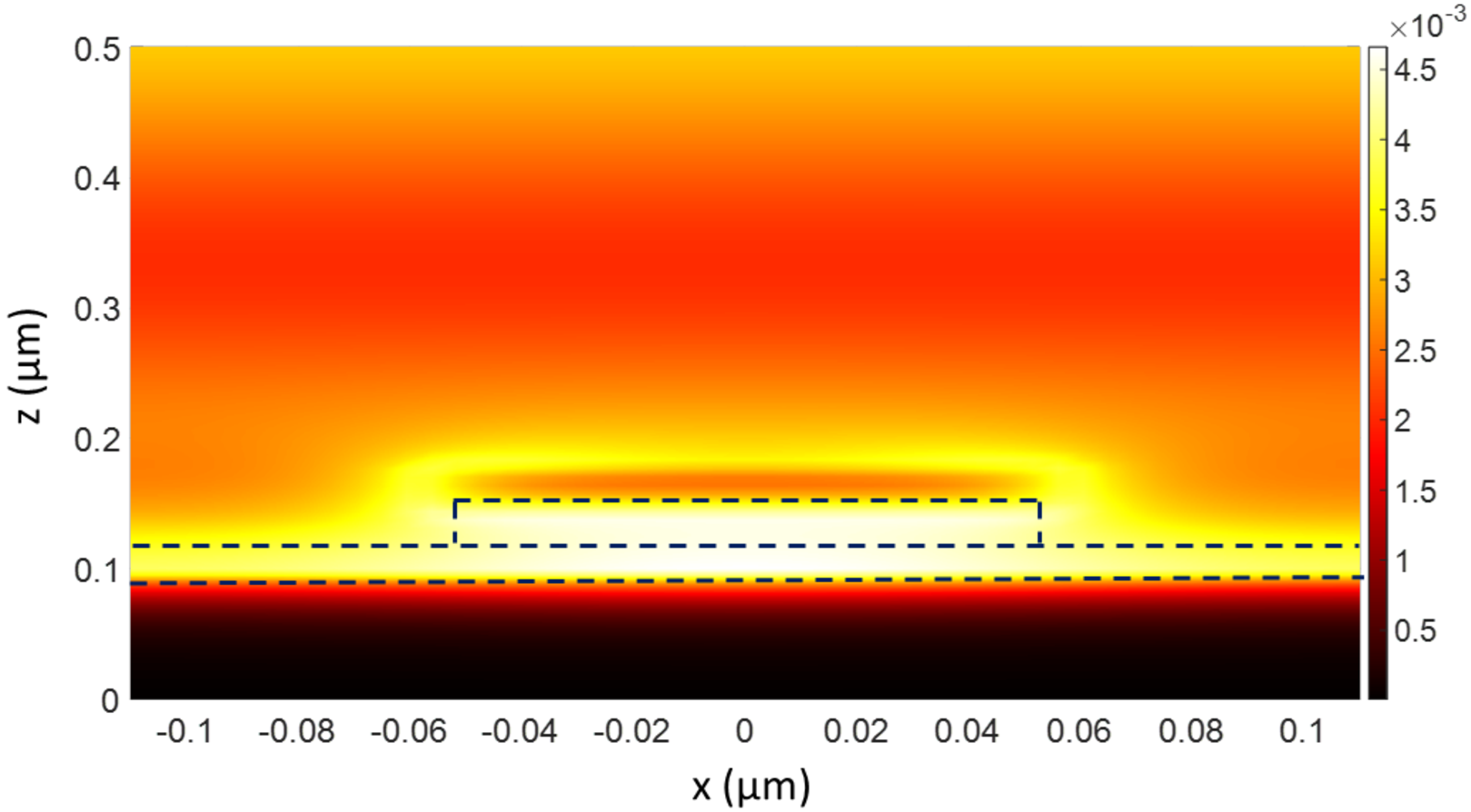}}
\caption{Magnitude of the magnetic field distribution in the $xy$-plane at (a) $333.5\,\mathrm{nm}$, (b) $420\,\mathrm{nm}$, and (c) $880.5\, \mathrm{nm}$, and in the $xz$-plane at (d) $333.5\,\mathrm{nm}$, (e) $420\,\mathrm{nm}$, and (f) $880.5\, \mathrm{nm}$.}
\label{magnetic_field_combined}
\end{figure}

To gain further insight into the physical mechanisms underlying the increased absorption, the magnitudes of the electric and magnetic field distributions at selected wavelengths are presented in Figs.~\ref{electric_field_combined} and~\ref{magnetic_field_combined}, respectively. Figures~\ref{electric_field_combined}(a)-(c) show the magnitude of the electric field distribution in the $xy$-plane at $333.5\,\mathrm{nm}$, $420\,\mathrm{nm}$, and $880.5\,\mathrm{nm}$, respectively. The electric field is observed to couple to the edges and localized at the corners of adjacent $\mathrm{Ti}$ blocks, indicating the excitation of LSPs across the absorber. Additionally, Figures~\ref{electric_field_combined}(d)-(f) show the magnitude of the electric field distribution in the $xz$-plane at the same three wavelengths, where the electric field is concentrated in the air slot and at the corners of the $\mathrm{Ti}$ blocks. 

In contrast, the magnetic field distributions exhibit distinct characteristics, as shown in Fig.~\ref{magnetic_field_combined}. Figures~\ref{magnetic_field_combined}(a)-(c) and (d)-(f) show the magnitude of the magnetic field distribution in the $xy$- and $xz$-planes, respectively, at $333.5\,\mathrm{nm}$, $420\,\mathrm{nm}$, and $880.5\,\mathrm{nm}$. At $333.5\,\mathrm{nm}$ and $420\,\mathrm{nm}$, the magnetic field is concentrated not only on top of the $\mathrm{Ti}$ blocks but also within the $\mathrm{SiO_2}$ spacer layer above the continuous $\mathrm{Al}$ back reflector layer, indicating the excitation of cavity modes at these wavelengths. 

At $880.5\,\mathrm{nm}$, the magnetic field is predominantly confined within the top $\mathrm{Ti}$ blocks and in the gap between them and the continuous $\mathrm{Al}$ back reflector layer as shown in Figs.~\ref{magnetic_field_combined}(c) and (f). As the electromagnetic fields interact with the top $\mathrm{Ti}$ layer, they are reflected by the bottom $\mathrm{Al}$ layer. The top and bottom metal layers, separated by the thin $\mathrm{SiO_2}$ dielectric spacer, form a lossy Fabry-Pérot (FP) cavity with a low $Q$-factor, resulting in broadband absorption with high efficiency. Therefore, the absorption in this band is attributed to the combined effects of LSPs and FP cavity modes.

\begin{figure}
\centering
\subfloat[]{\includegraphics[width=0.6\columnwidth]{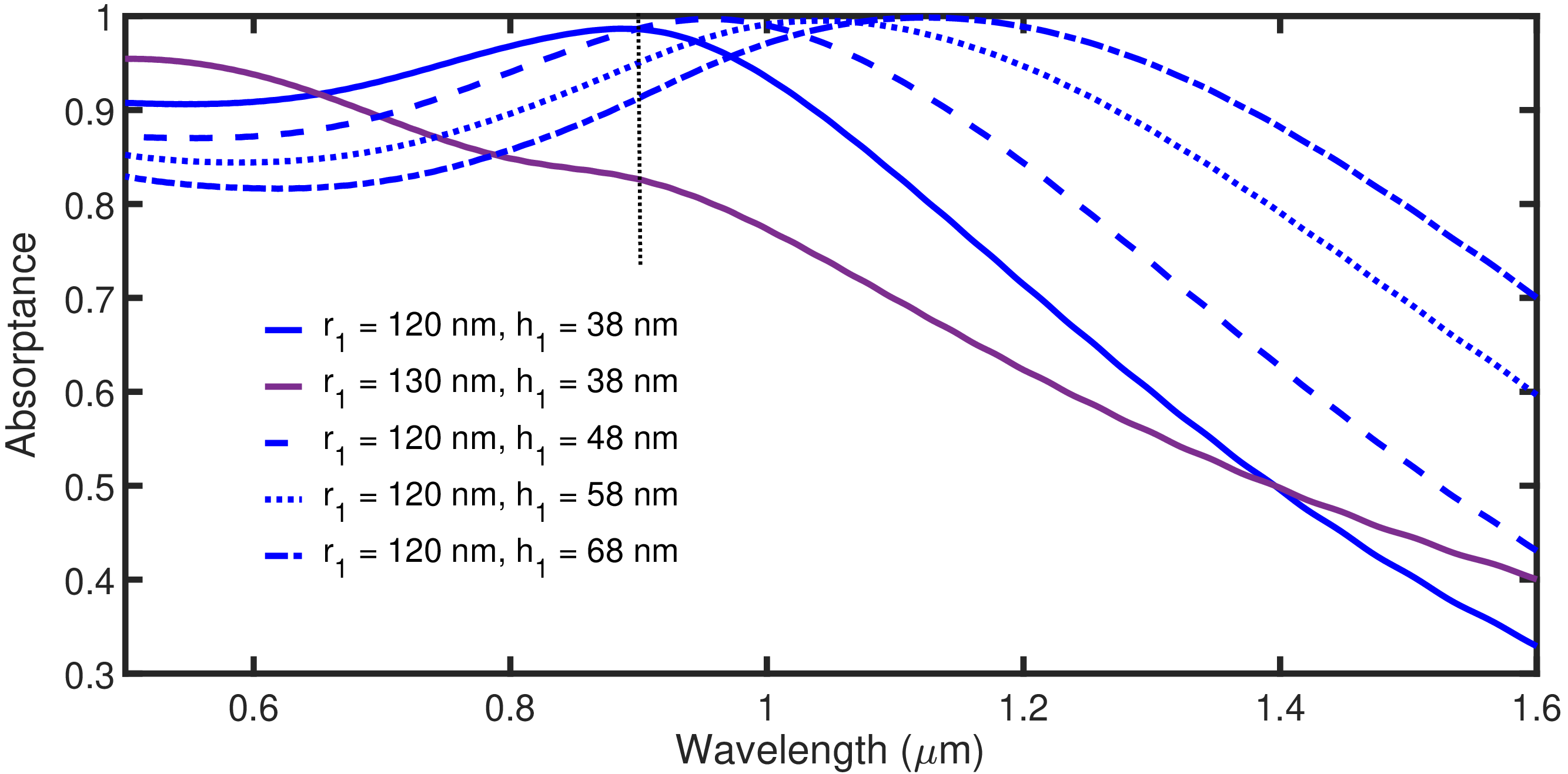}}\\
\subfloat[]{\includegraphics[width=0.45\columnwidth]{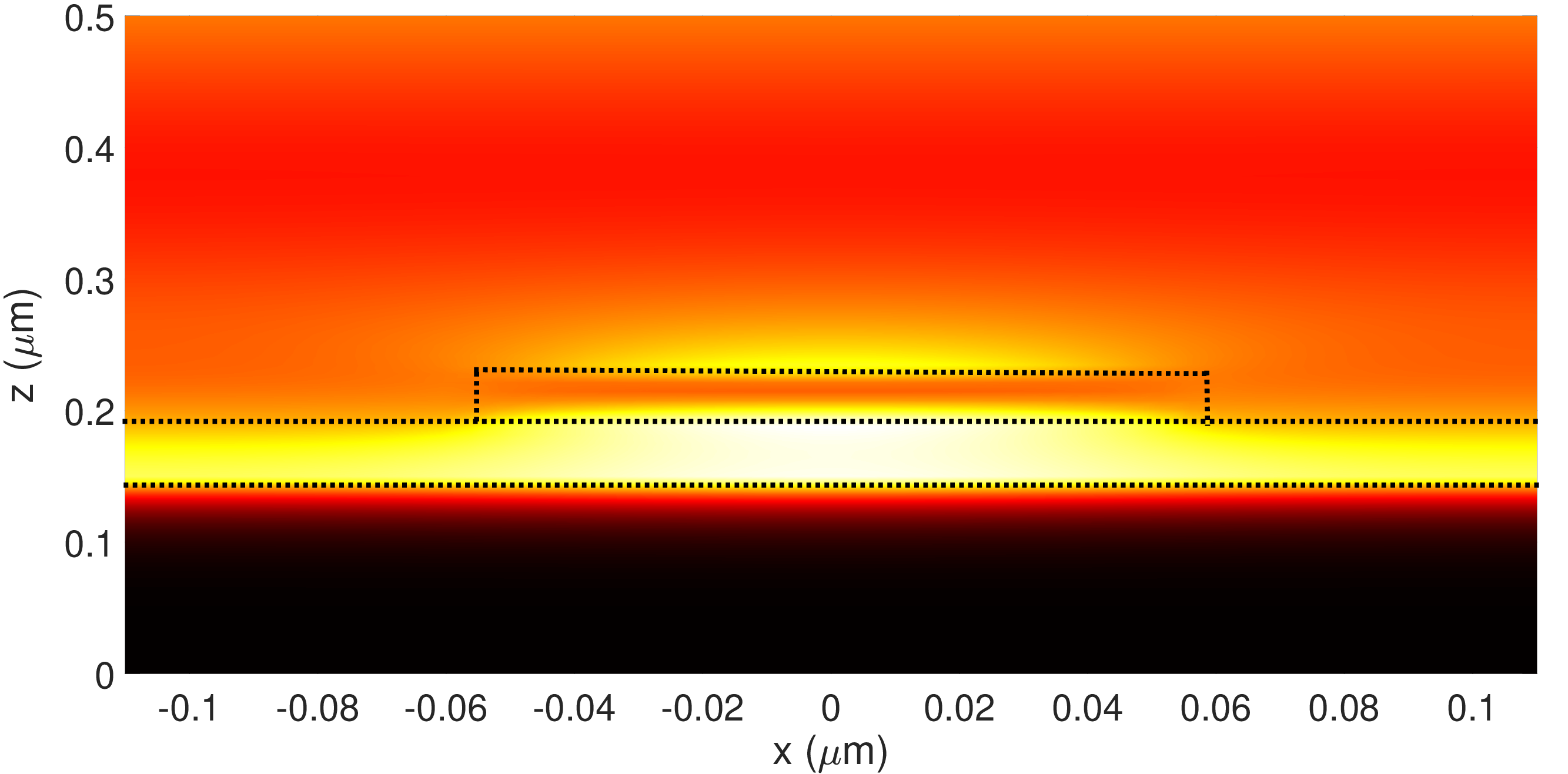}}\\
\subfloat[]{\includegraphics[width=0.46\columnwidth]{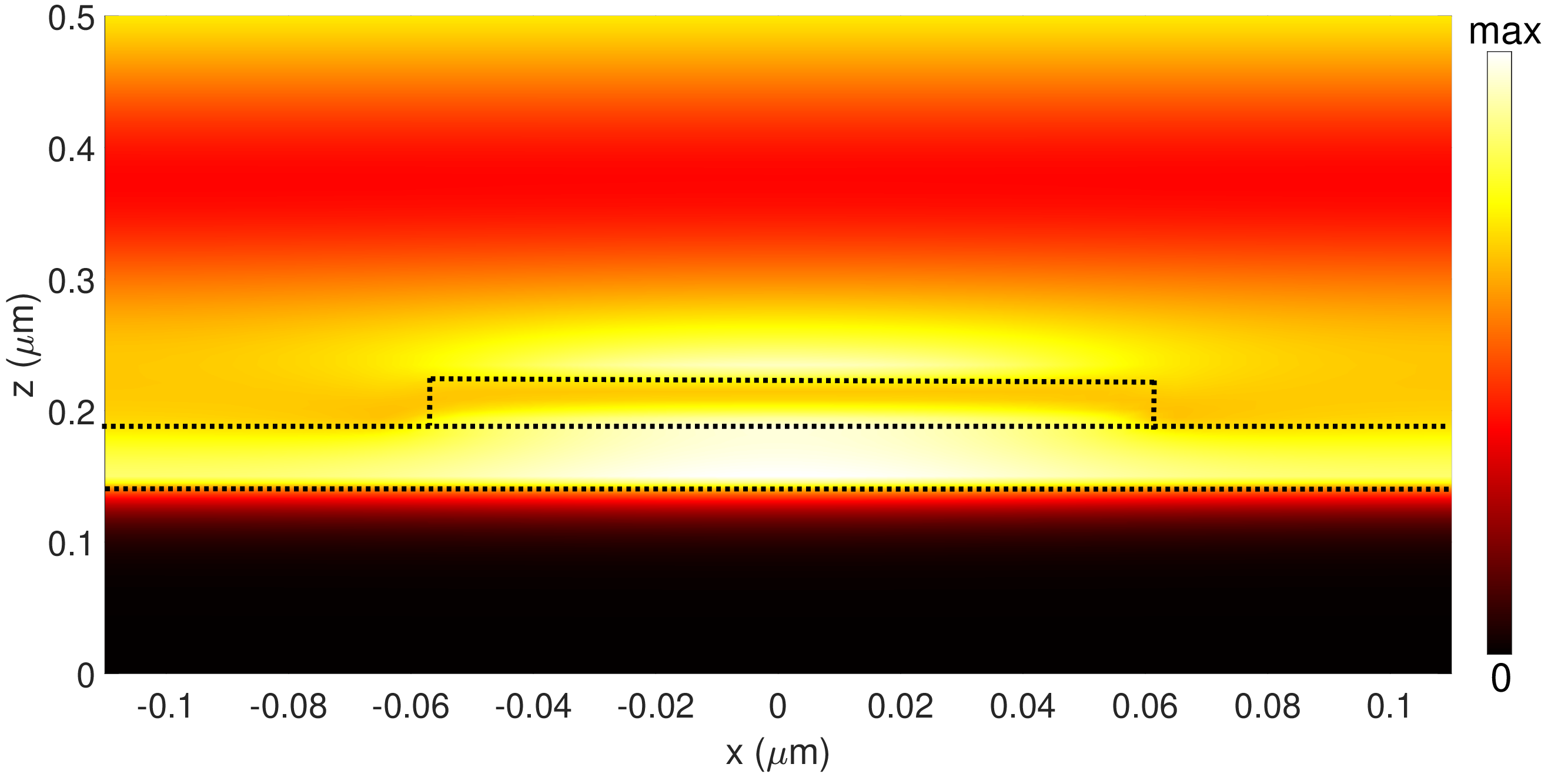}}
\caption{Investigation of the FP modes. (a) Absorptance spectra for different values of the top $\mathrm{Ti}$ layer thickness $h_1$ and the side length of the $\mathrm{Ti}$ blocks $r_1$. The magnitude of the magnetic field distributions in the $xz$-plane at $890\,\mathrm{nm}$ for $h_1=38\,\mathrm{nm}$ with (b) $r_1=120\,\mathrm{nm}$ and (c) $r_1=130\,\mathrm{nm}$.}
\label{FP_resonance}
\end{figure}

The behavior of the FP modes are fundamentally governed by the top $\mathrm{Ti}$ layer thickness $h_1$ and the side length of the $\mathrm{Ti}$ blocks $r_1$~\cite{barnard2008, liu2017}. This is demonstrated by simulations where $h_1$ and $r_1$ are varied, while all other geometric parameters are kept at their optimized values. The excitation is kept the same. Figure~\ref{FP_resonance}(a) presents the absorptance spectra for different combinations of $h_1$ and $r_1$. The solid blue line is the absorptance spectrum of the optimized device, with the FP resonance prominently appearing at $890\,\mathrm{nm}$, marked by the dotted vertical black line. As seen in the figure, variations $r_1$ strongly perturb the FP resonance, resulting in a sharp decline in absorption performance. In contrast, increasing $h_1$ leads to a redshift in the resonance wavelength, indicating its influence on the optical path length within the cavity.

To visualize the FP resonance behavior, the magnetic field distributions in the $xz$-plane at $890\,\mathrm{nm}$ for $h_1=38\,\mathrm{nm}$ with $r_1=120\,\mathrm{nm}$ and $r_1=130\,\mathrm{nm}$ are shown in Fig.~\ref{FP_resonance}(b) and (c), respectively. For $r_{1}=120\,\mathrm{nm}$, clear standing wave patterns are observed within the dielectric spacer layer, confined between the top $\mathrm{Ti}$ blocks and the reflector layer, confirming the formation of an FP cavity mode. In contrast, for $r_1=130\,\mathrm{nm}$, the magnetic field exhibits significant scattering, indicating a degradation of the resonant condition and diminished absorption efficiency.

\subsection{Effect of geometric parameters on absorption}
\begin{figure}
\centering
\subfloat[]{\includegraphics[width=0.48\columnwidth]{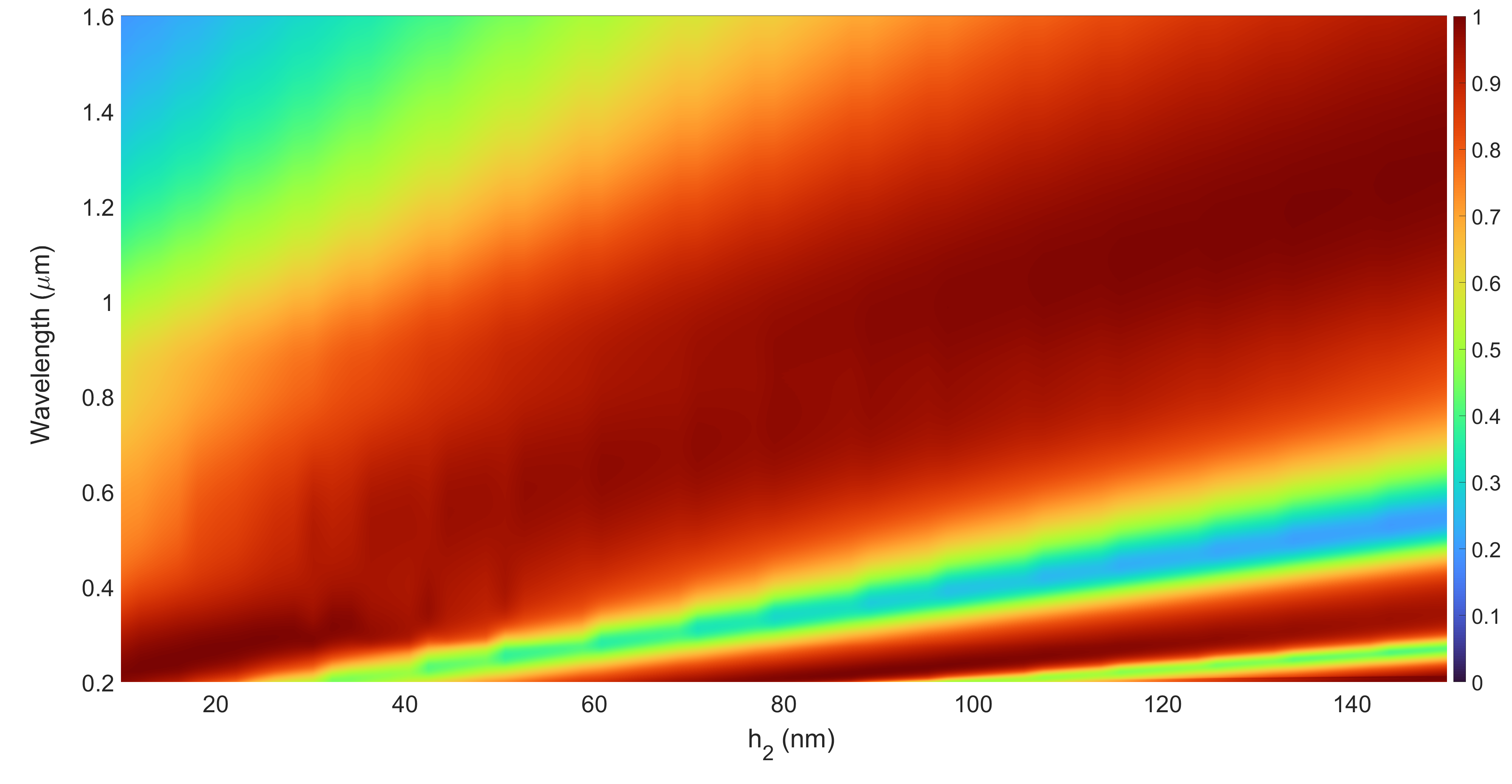}}\hspace{0.25cm}
\subfloat[]{\includegraphics[width=0.48\columnwidth]{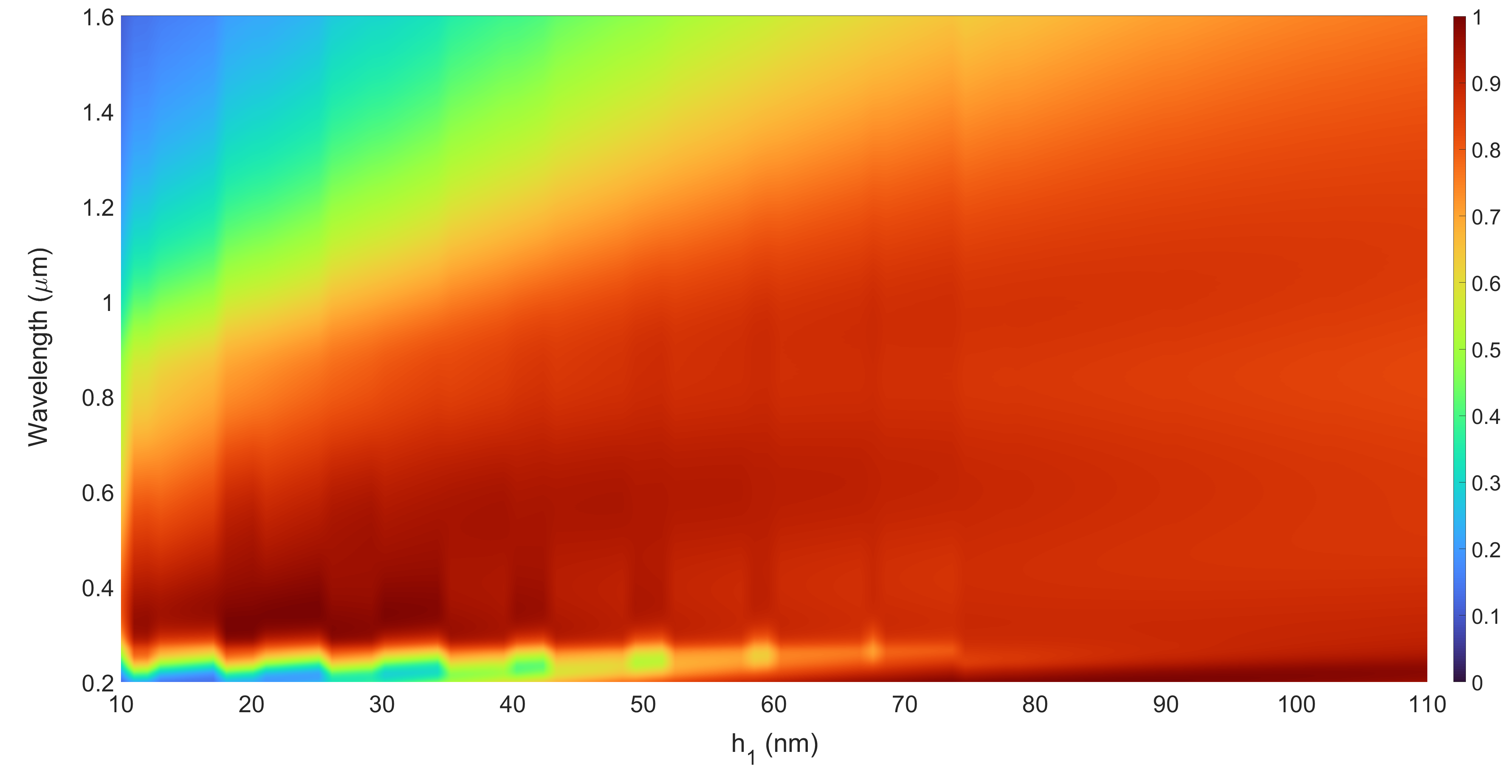}}\\
\subfloat[]{\includegraphics[width=0.48\columnwidth]{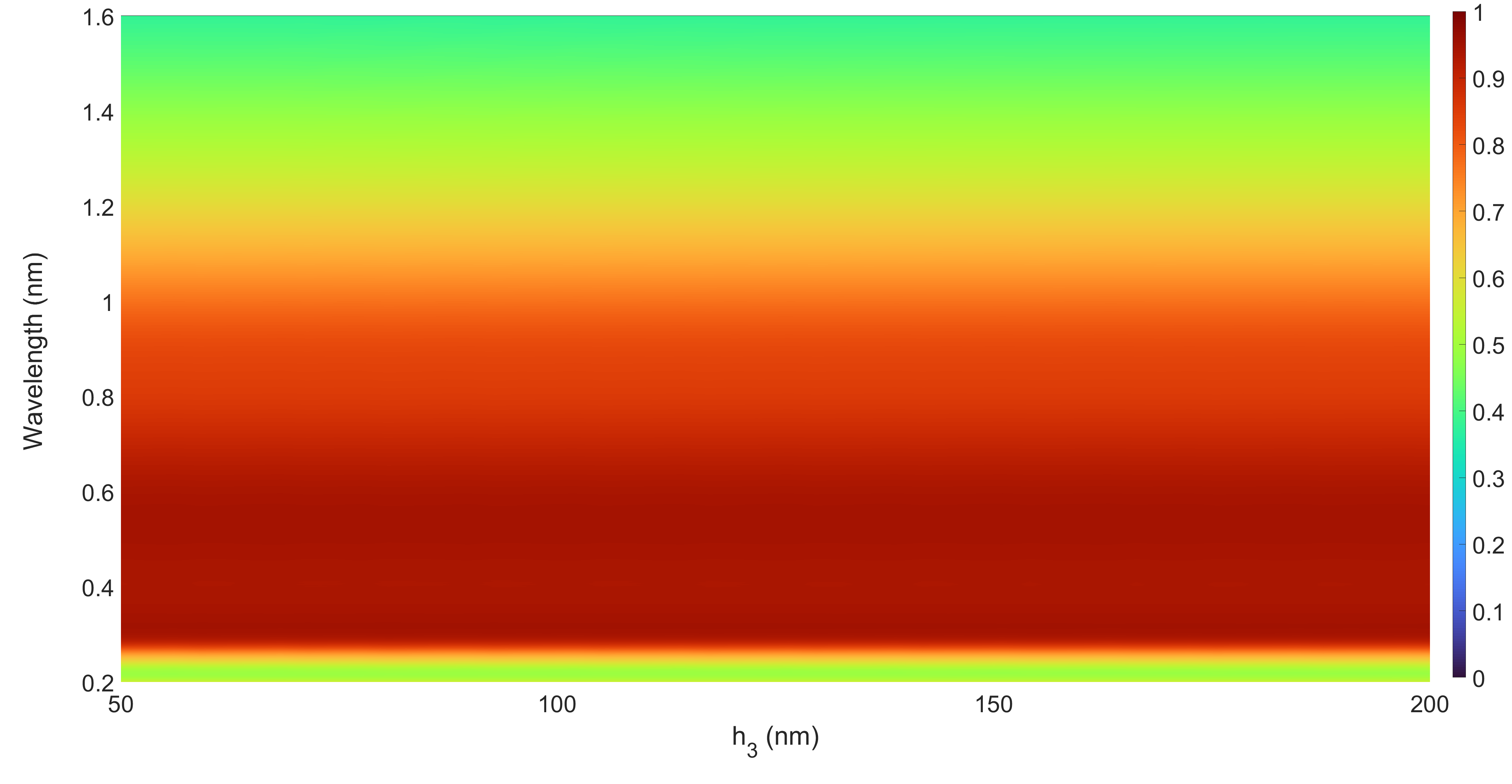}}\hspace{0.25cm}
\subfloat[]{\includegraphics[width=0.48\columnwidth]{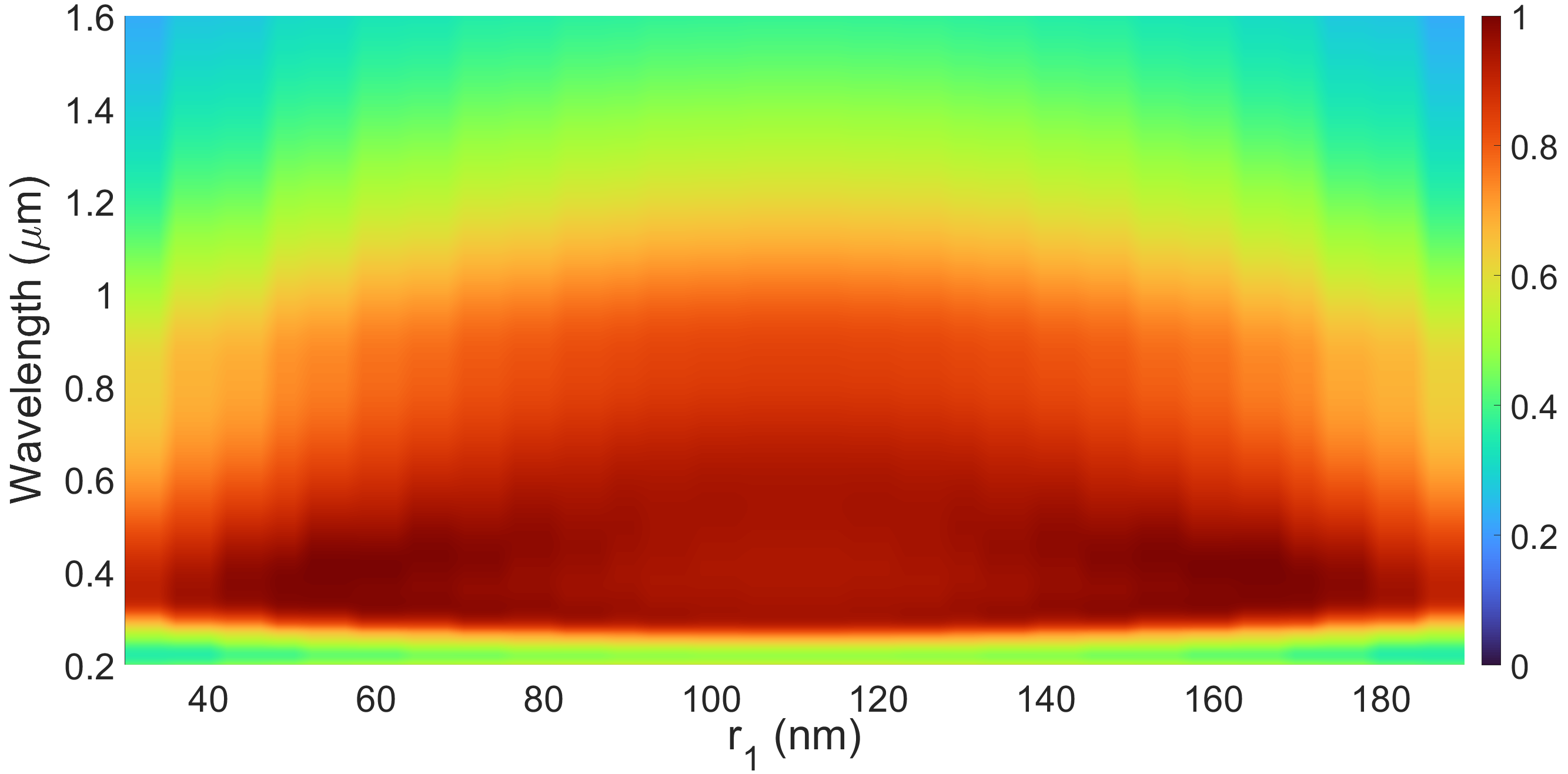}}
\caption{Influence of geometric parameters on the absorptance spectrum: (a) $\mathrm{SiO_2}$ spacer layer thickness $h_2$, (b) top $\mathrm{Ti}$ layer thickness $h_1$, (c) bottom $\mathrm{Al}$ layer thickness $h_3$, and (d) side length of the top $\mathrm{Ti}$ blocks $r_1$.}
\label{geometric_parameter}
\end{figure}

In subsequent simulations, the excitation is kept the same, while the top $\mathrm{Ti}$ layer thickness $h_1$, the $\mathrm{SiO_2}$ spacer layer thickness $h_2$, the $\mathrm{Al}$ back reflector layer thickness $h_3$, and the side length of the $\mathrm{Ti}$ blocks $r_1$ are varied to investigate their influence on the absorption performance. Figure~\ref{geometric_parameter}(a) shows the simulated absorptance spectra for different values of the $\mathrm{SiO_2}$ spacer layer thickness $h_2$. As shown, the absorption band redshifts progressively with the increasing $h_2$. This is because an increase in $h_2$ extends the cavity length, which in turn shifts the resonance frequencies to longer wavelengths. Therefore, $h_2$ serves as an effective tuning parameter for tailoring the absorption spectrum to specific application requirements. For optimal performance, $h_2$ is set to $40\,\mathrm{nm}$.  

Figure~\ref{geometric_parameter}(b) presents the simulated absorptance spectra for different values of the top $\mathrm{Ti}$ layer $h_1$. The results show that increasing $h_1$ broadens the absorption bandwidth, though it reduces the peak absorption level. For optimal performance, $h_1$ is set to $38\,\mathrm{nm}$. 

As shown in Fig.~\ref{geometric_parameter}(c), the thickness of the $\mathrm{Al}$ back reflector layer $h_3$ has negligible influence on the absorption spectra, due to its large value relative to the penetration depth of incident light. Finally, Fig.~\ref{geometric_parameter}(d) shows the simulated absorptance for different values of the $\mathrm{Ti}$ block side length $r_1$, where the total length $r_{1}+r_{2}$ is fixed to the periodicity of the unit cell. The results demonstrate that $r_1$ significantly affects broadband absorption by affecting the excitation of resonance modes. Maximum absorption efficiency is achieved for $r_1=120\,\mathrm{nm}$, which is adopted for device fabrication. 

\subsection{Effect of polarization and angle of incidence on absorption}
\begin{figure}
\centering
\subfloat[]{\includegraphics[width=0.6\columnwidth]{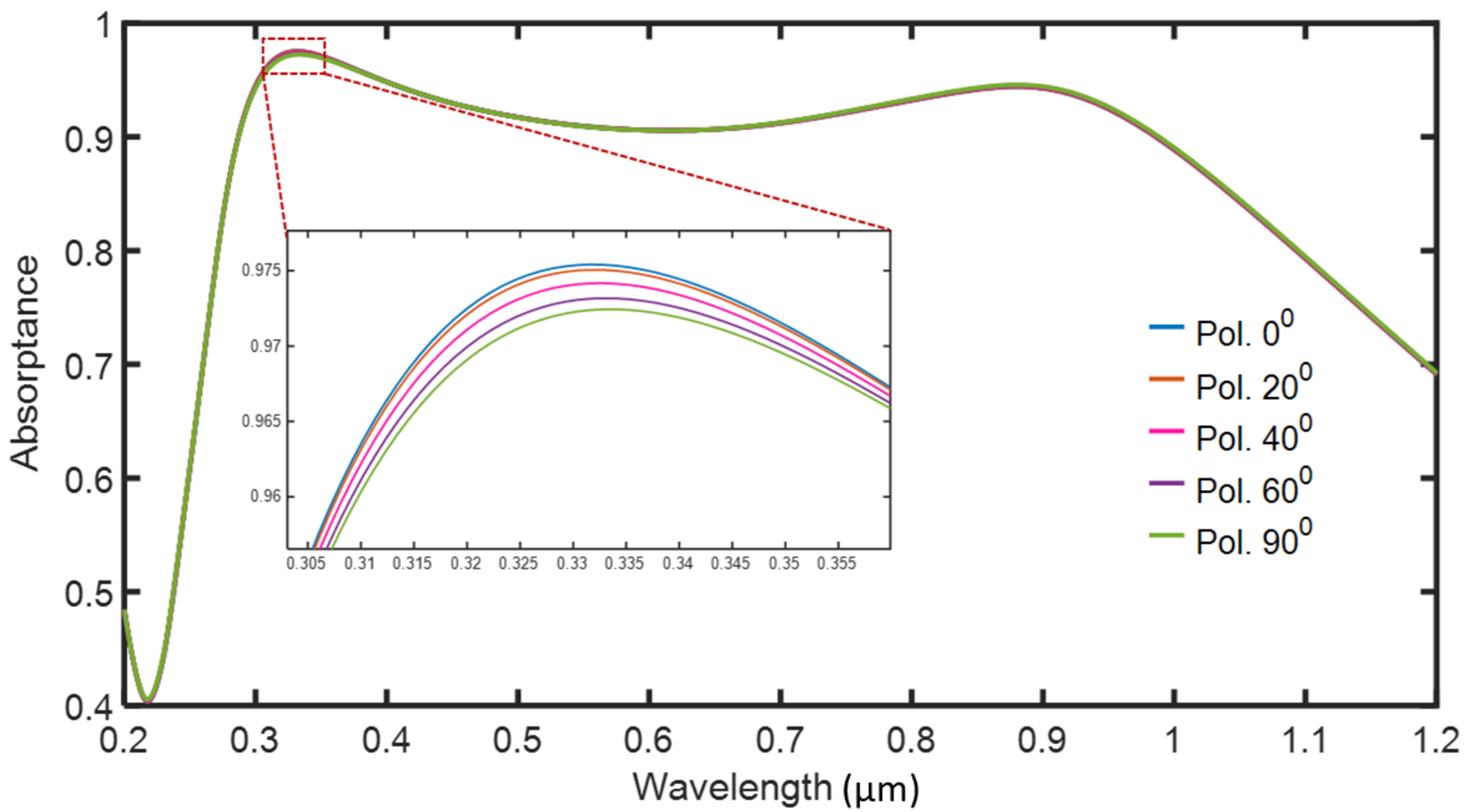}}\\
\subfloat[]{\includegraphics[width=0.6\columnwidth]{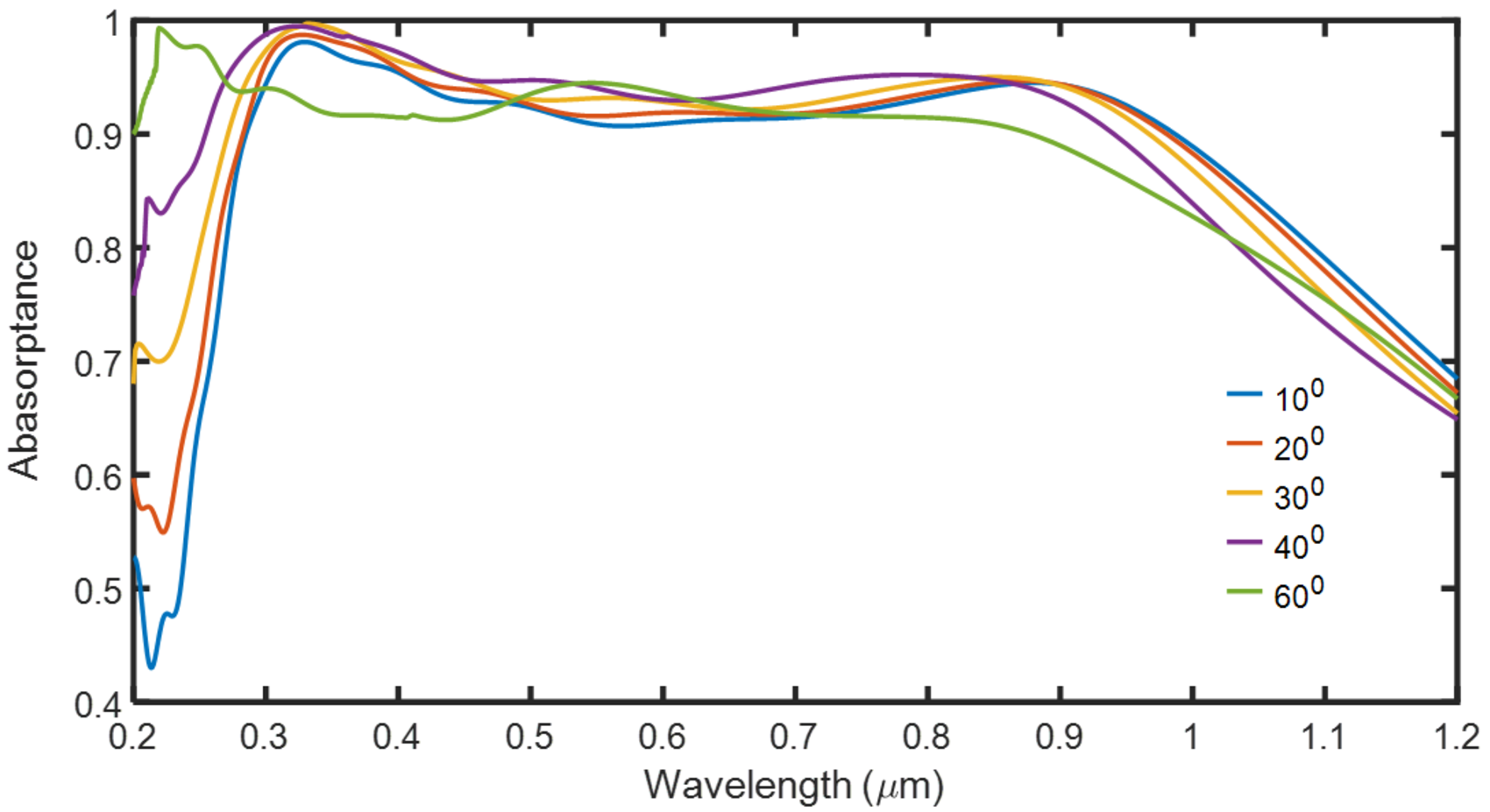}}
\caption{Absorptance spectra for (a) different polarization angles of the normally incident plane wave and (b) angles of incidence of the $x$-polarized plane wave.}
\label{Pol_angle_variance_sim}
\end{figure}

Next, the absorption performance under varying polarization and angle of incidence is investigated. In the first set of simulations, a normally incident the plane wave (angle of incidence is $0^{\circ}$) illuminates the absorber, while the polarization direction is varied from the $x$- to the $y$-axis. This variation is defined in terms of the angle between the electric field vector and the $x$-axis. As shown in Fig.~\ref{Pol_angle_variance_sim}(a), the absorptance spectra remain nearly unchanged with variations in the polarization direction of the incident plane wave. This polarization insensitivity is attributed to the structural symmetry.

Figure~\ref{Pol_angle_variance_sim}(b) shows the absorptance spectra under oblique incidence for angles up to $60^{\circ}$. The results indicate that the structure maintains strong broadband absorption, with only a minor blue shift observed at the highest angle, while the overall bandwidth is preserved. These results confirm that the proposed absorber exhibits excellent insensitivity to both polarization and angle of incidence.


\subsection{Effective impedance matching}

\begin{figure}
\centering
\subfloat[]{\includegraphics[width=0.475\columnwidth]{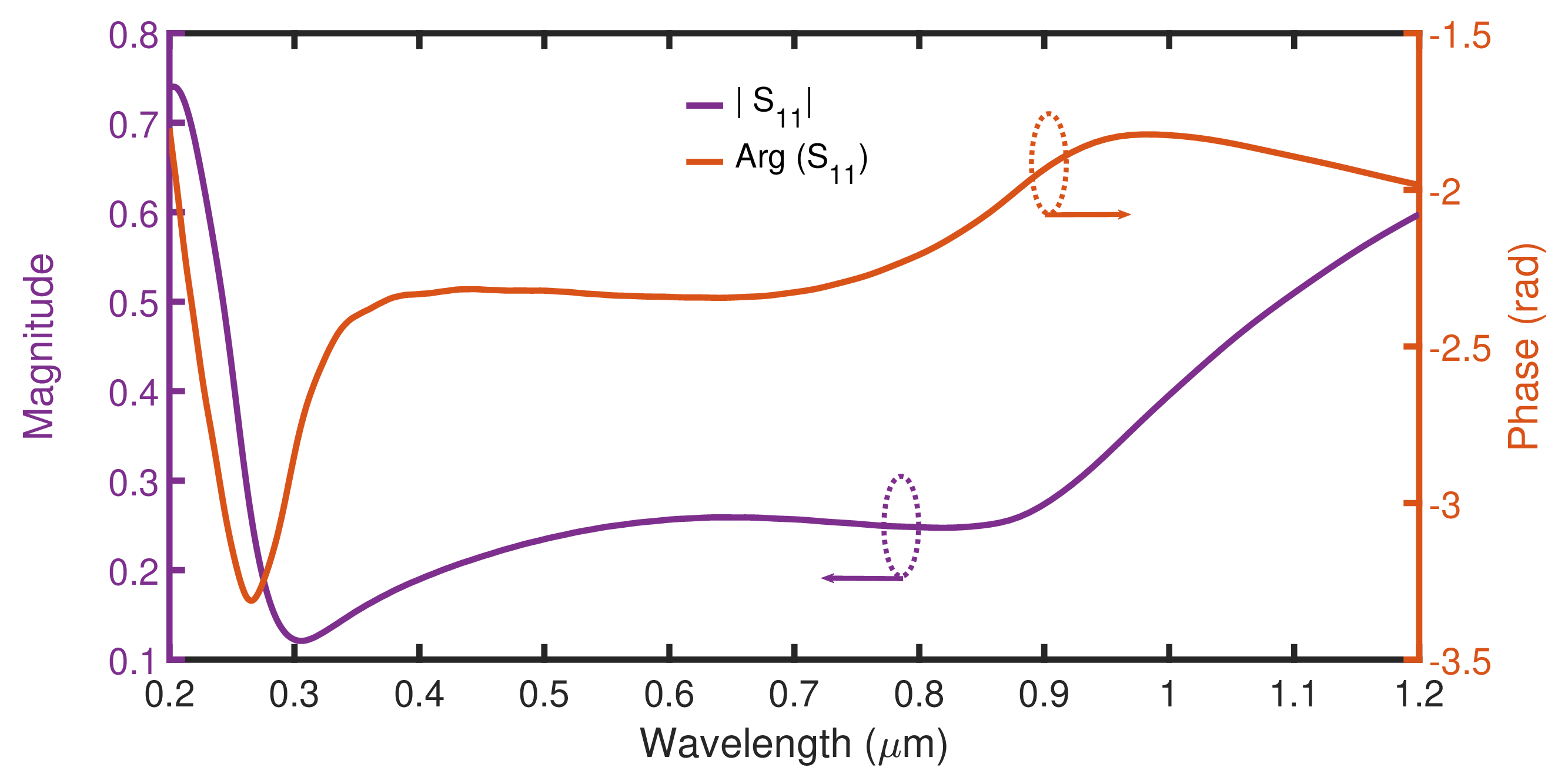}}\hspace{0.5cm}
\subfloat[]{\includegraphics[width=0.475\columnwidth]{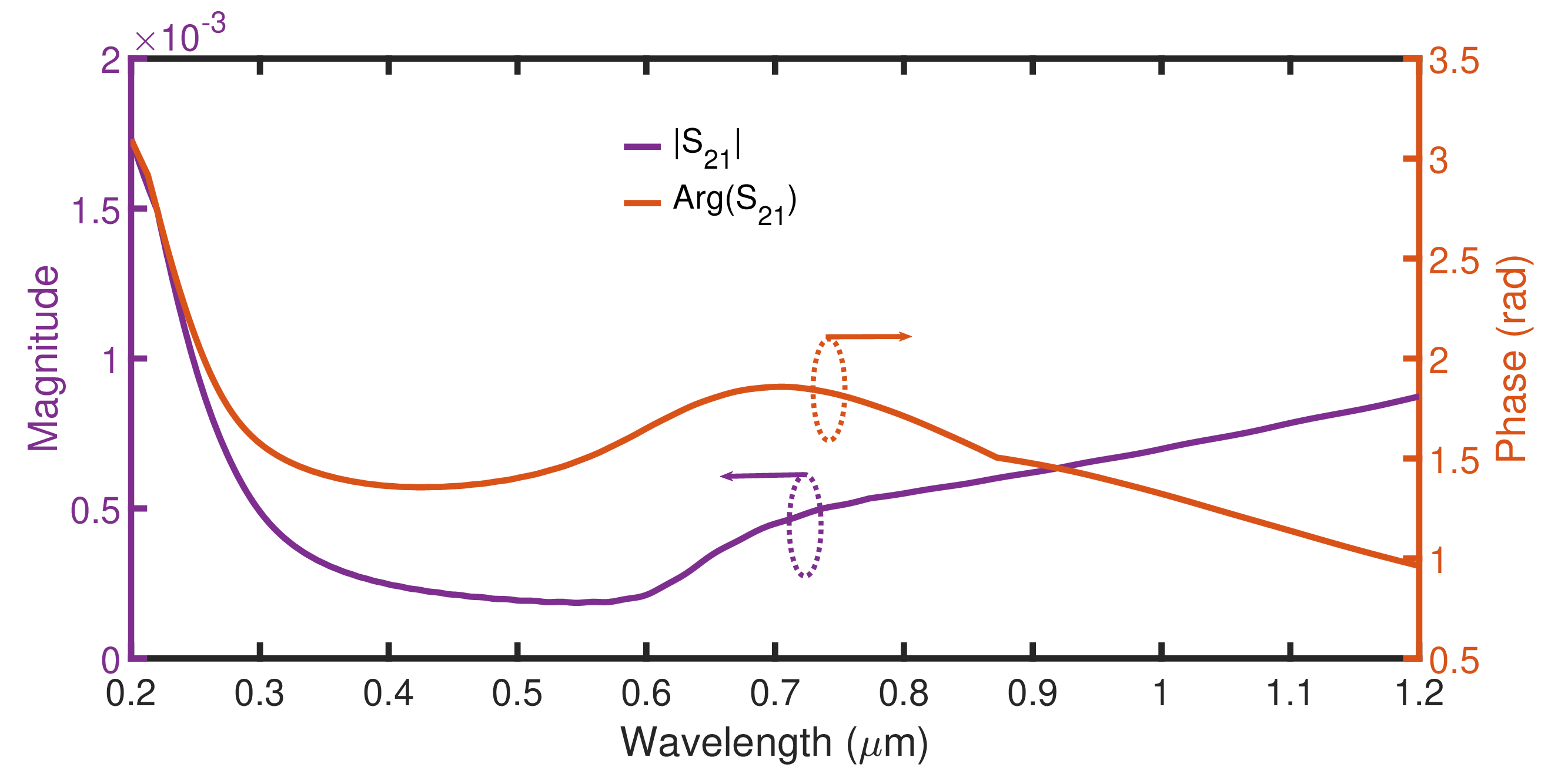}}\\
\subfloat[]{\includegraphics[width=0.475\columnwidth]{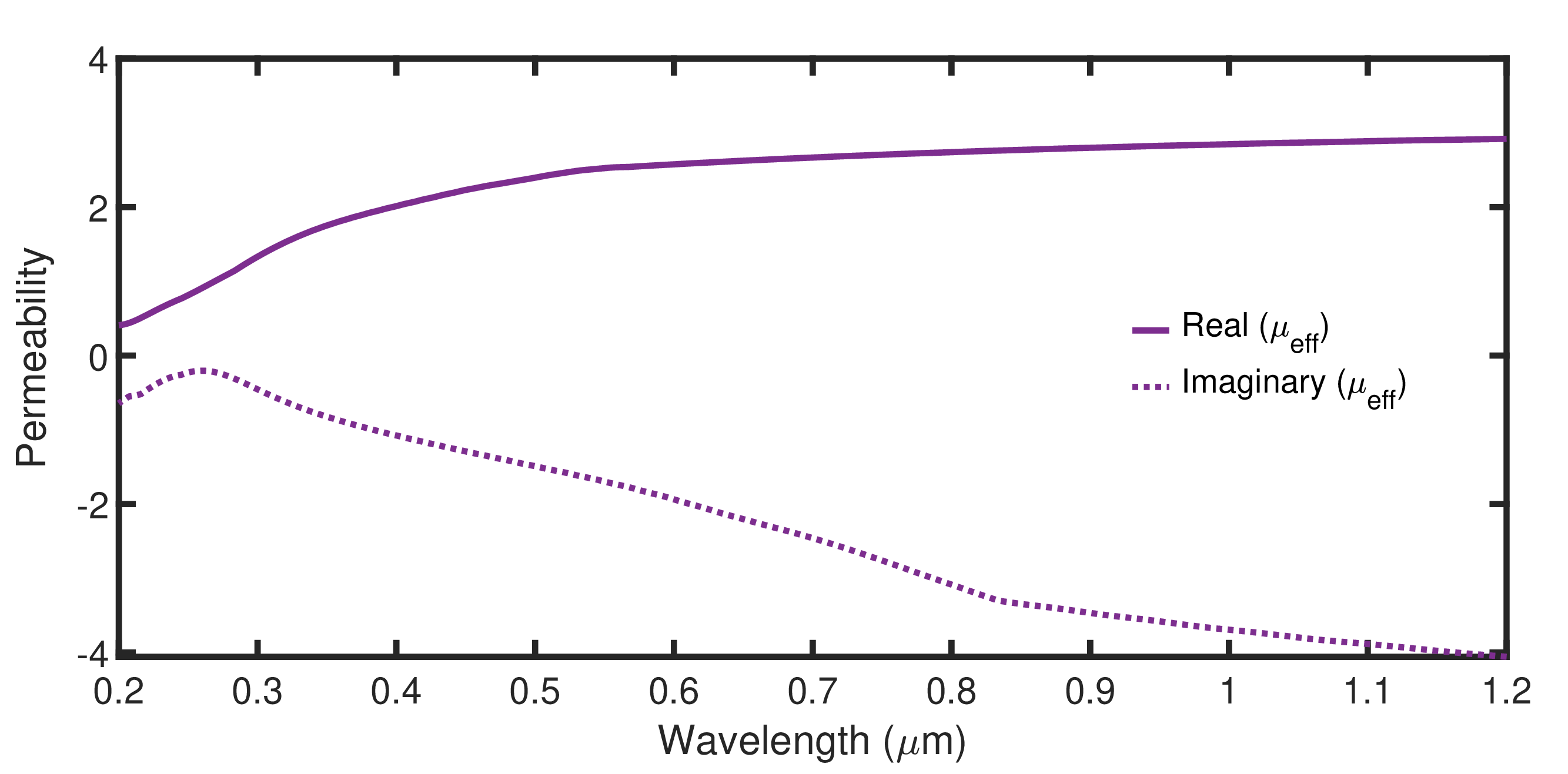}}\hspace{0.5cm}
\subfloat[]{\includegraphics[width=0.475\columnwidth]{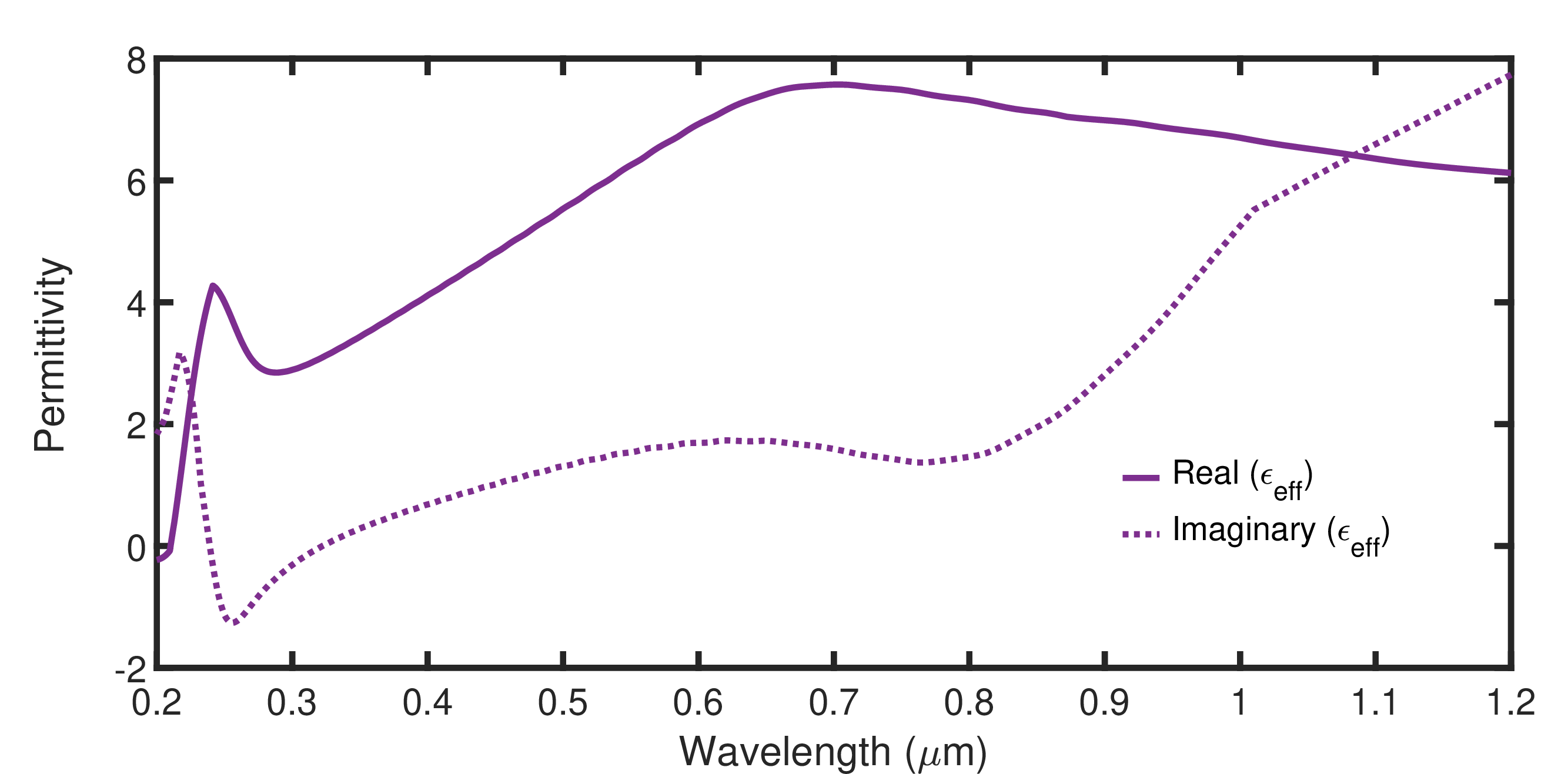}}\\
\subfloat[]{\includegraphics[width=0.475\columnwidth]{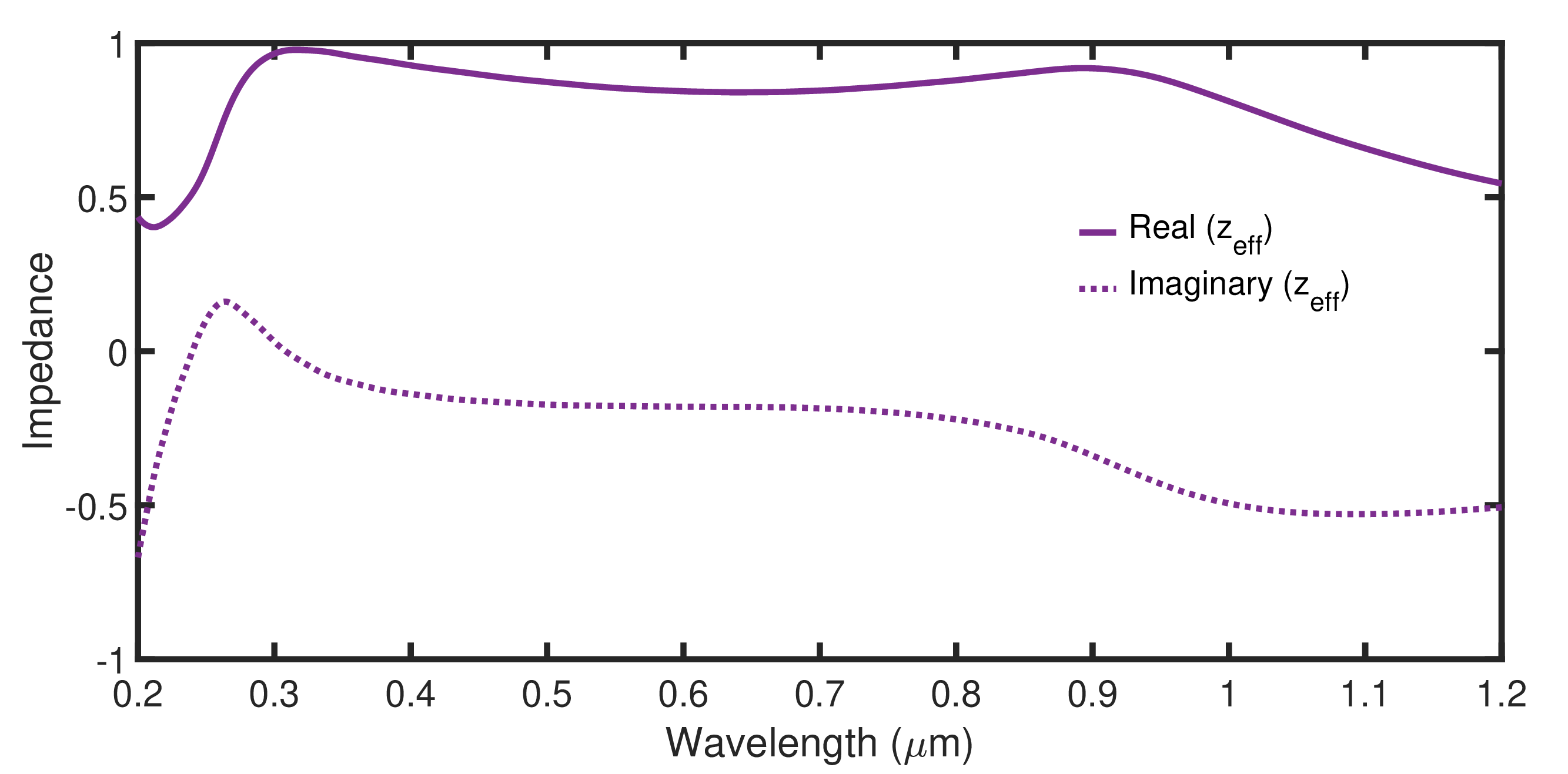}}
  \caption{$S$-parameters and retrieved effective material properties for the optimized design. Magnitude and phase of (a) $S_{11}$ and (b) $S_{21}$. Real and imaginary parts of (c) the relative effective permeability $\mu_{\mathrm{eff}}$, (d) the relative effective permittivity $\varepsilon_{\mathrm{eff}}$, and (e) the relative effective impedance $z_{\mathrm{eff}}$.}
    \label{S_parameter}
\end{figure}

To gain deeper insight into the broadband absorption characteristics, one can examine the effective impedance of the absorber $Z_{\mathrm{eff}}$ relative to that of free-space impedance $Z_0$. Perfect absorption with zero reflection occurs when $Z_{\mathrm{eff}}$ matches $Z_0$~\cite{mehrabi2022ultra}. Let $z_{\mathrm{eff}}=Z_{\mathrm{eff}}/Z_0$ represent the \emph{relative} effective impedance of the absorber. $z_{\mathrm{eff}}$ can be computed from simulated or measured $S$-parameters using~\cite{par_retrieval2005}: 
\begin{equation}
  z_{\mathrm{eff}} = \sqrt{\frac{\mu_{\mathrm{eff}}}{\varepsilon_{\mathrm{eff}}}} = \sqrt{\frac{(1+S_{11})^{2} - S_{21}^{2}}{(1-S_{11})^{2} - S_{21}^{2}}}.
  \label{eqn_effective_Z}
\end{equation}
Here, $S_{11}$ and $S_{22}$ are the reflection coefficients when the incident wave is applied from the top and bottom sides of the device, respectively, while $S_{21}$ and $S_{12}$ are the transmission coefficients corresponding to wave propagation from top to bottom and bottom to top, respectively. The quantities $\mu_{\mathrm{eff}}$ and $\varepsilon_{\mathrm{eff}}$ denote the relative effective permeability and the relative effective permittivity of the absorber, respectively, and are related to the $S$-parameters through the following expressions~\cite{par_retrieval2005}:
\begin{subequations}
  \begin{equation}
  \varepsilon_{\mathrm{eff}} = \frac{2}{\sqrt{-kd}}
  \frac{1-(S_{21}+S_{11})} {1+(S_{21}+S_{11})},
  \label{eqn_epsiln}
  \end{equation}
  \begin{equation}
  \mu_{\mathrm{eff}} = \frac{2}{\sqrt{-kd}}
  \frac{1-(S_{21}-S_{11})} {1+(S_{21}-S_{11})}.
  \label{eqn_mu}
  \end{equation}
\end{subequations}
In the case of perfect absorption for the proposed design, $S_{11}$ and $S_{21}$ approach zero (zero reflection and zero transmission), leading to $z_{\mathrm{eff}}$ = 1 as per Eq.~\ref{eqn_effective_Z}. Thus, maintaining a unity or near-unity value for $z_{\mathrm{eff}}$ over a broadband is essential to achieve wideband absorption. The magnitudes of simulated $S_{11}$ and $S_{21}$ are shown in Figs.~\ref{S_parameter}(a) and (b), respectively. Since the transmission is nearly negligible, the magnitude of $S_{21}$ is effectively zero. Figures~\ref{S_parameter}(c)–(e) present the extracted values of $\mu_{\mathrm{eff}}$, $\varepsilon_{\mathrm{eff}}$, and $z_{\mathrm{eff}}$, respectively, computed from the simulated $S$-parameters using the relations above.

Notably, the magnitude of $z_{\mathrm{eff}}$ remains close to unity across the wavelength range, as shown in Fig.~\ref{S_parameter}(f). For instance, $333\,\mathrm{nm}$, $z_{\mathrm{eff}} = 0.98 + 0.002i$, and at $880.5\,\mathrm{nm}$, $z_{\mathrm{eff}} = 0.94 - 0.2i$, confirming good impedance matching over a broadband spectral range.

\subsection{Effect of surrounding medium on absorption}
The nanostructure is designed under the assumption that it resides in free space, characterized by an impedance of $Z_0$, as described in the previous section. However, depending on the intended application, the surrounding medium may differ from free space. For instance, in the solar cell application discussed in Section~\ref{sec:app}, the nanostructure is embedded within a layered medium comprising indium tin oxide (ITO) and silicon ($\mathrm{Si}$) layers.

\begin{figure}
\centering
\subfloat[]{\includegraphics[width=0.6\columnwidth]{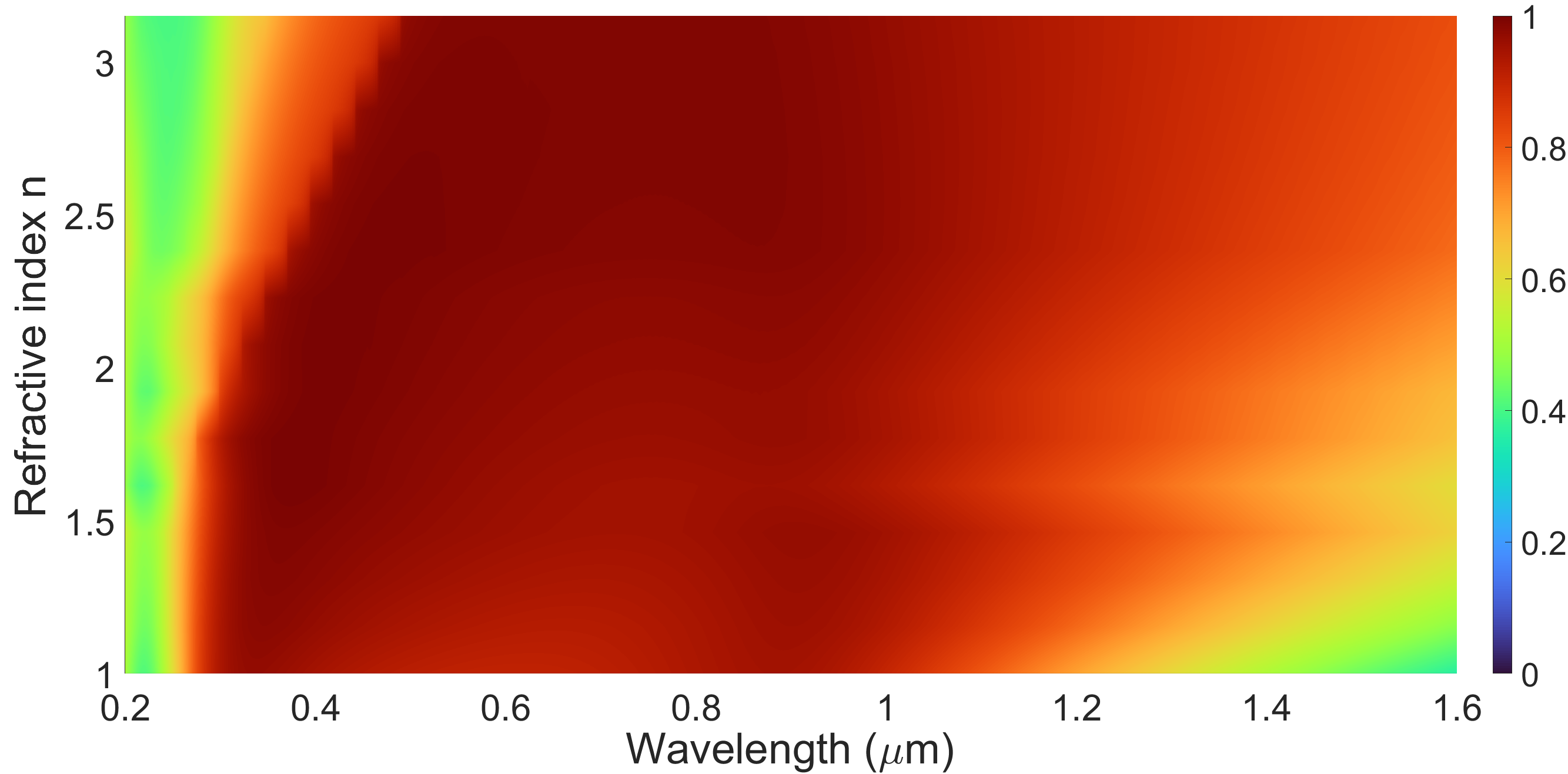}}\\
\subfloat[]{\includegraphics[width=0.6\columnwidth]{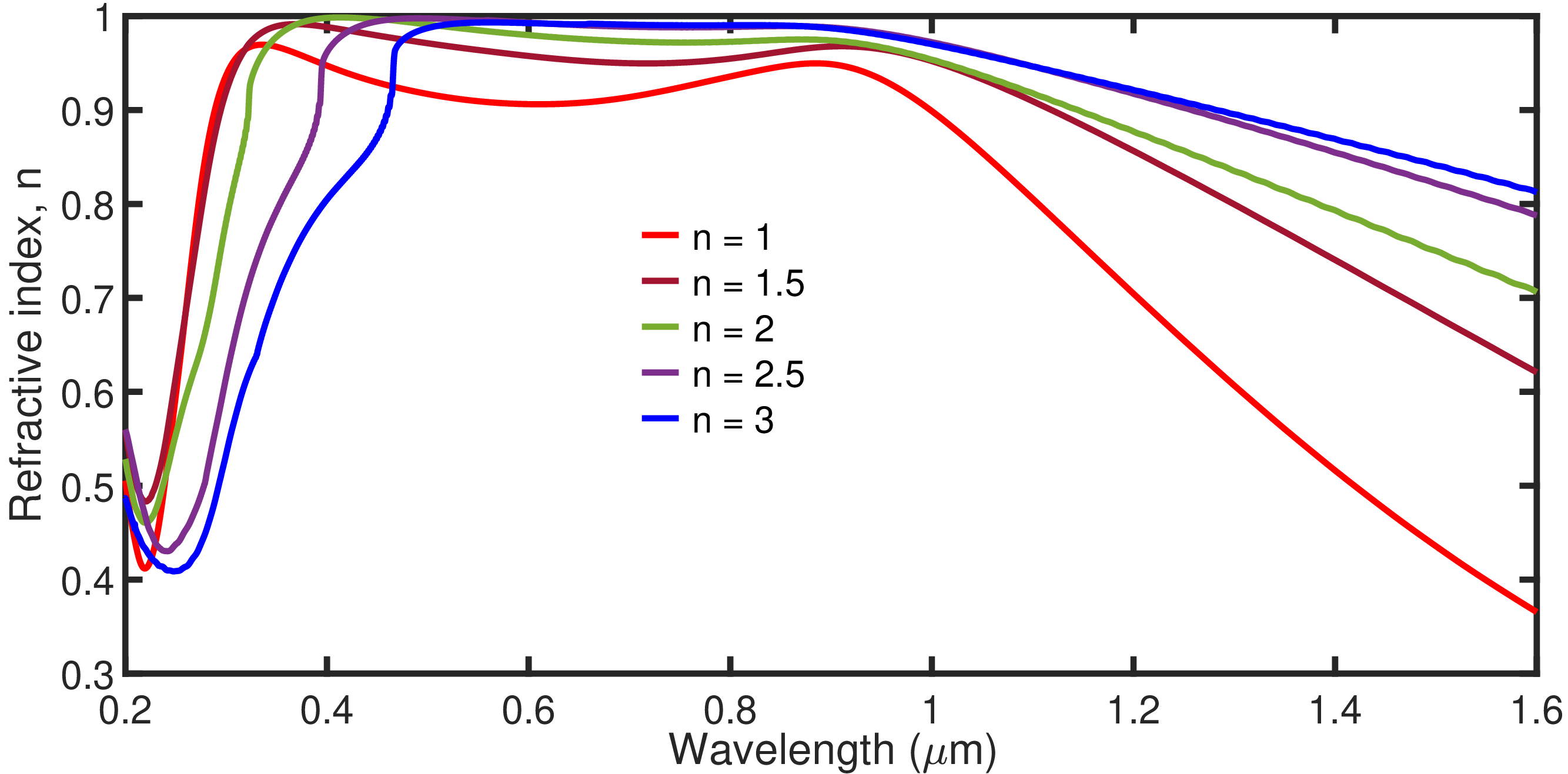}}
\caption{(a) Simulated absorption spectra for the surrounding medium's refractive index varying between $1$ and $3.15$. (b) Cuts of the plot in (a) at refractive index values of $1$, $1.5$, $2$, $2.5$, and $3$.}
\label{RI_variance_sim}
\end{figure}

To assess the impact of the surrounding medium on the absorption performance, simulations are conducted using the optimized geometric parameters of the nanostructure while varying the refractive index of the surrounding medium between $1$ and $3.15$. A normally incident plane wave polarized along the $x$-direction is used for excitation. Figures~\ref{RI_variance_sim}(a) and (b) show the resulting absorptance spectra for different refractive index values. As the refractive index increases, a noticeable redshift in the spectra is observed, accompanied by enhanced absorption in the longer wavelength region. Notably, the nanostructure maintains strong absorption even in high-index environments, highlighting the potential of the proposed nanostructure for integration with photovoltaic platforms.

\section{Fabrication and characterization}
\begin{figure}[t]
  \centering
   \includegraphics[width=0.95\linewidth]{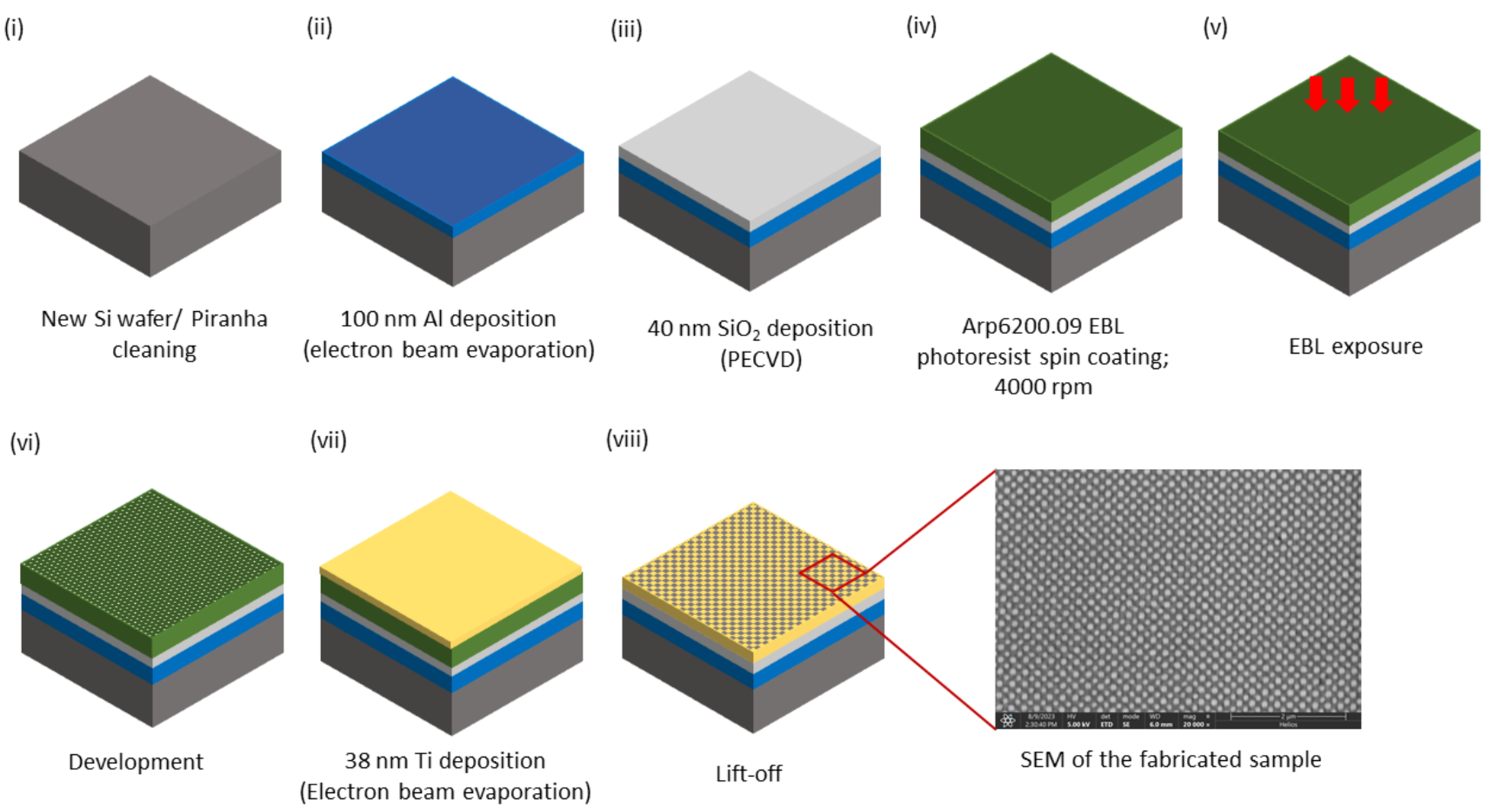}
  \caption{Process flow for the fabrication of the proposed absorber.}
    \label{process_flow}
\end{figure}

The sample is fabricated using a standard electron beam lithography (EBL) technique, followed by a metal lift-off process. The fabrication process flow of the proposed device is illustrated in Fig.~\ref{process_flow}. A silicon ($\mathrm{Si}$) wafer serves as the substrate, onto which a $100\,\mathrm{nm}$ $\mathrm{Al}$ layer is deposited via electron beam evaporation (EBE). Subsequently, a $40\,\mathrm{nm}$ $\mathrm{SiO_2}$ layer is deposited on top of the $\mathrm{Al}$ layer using plasma-enhanced chemical vapor deposition (PECVD). The structure is then patterned using a $\mathrm{JEOL\,JBX-6300FS\,EBL}$ system with $\mathrm{ARP\,6200.09}$ as the photoresist. A $38\,\mathrm{nm}$ $\mathrm{Ti}$ layer is deposited through EBE, followed by a metal lift-off process. The final fabricated sample has a footprint of $500\,\mathrm{\mu m}$ $\times$ $500\,\mathrm{\mu m}$. 

\begin{figure}[t]
  \centering
   \includegraphics[width=0.7\linewidth]{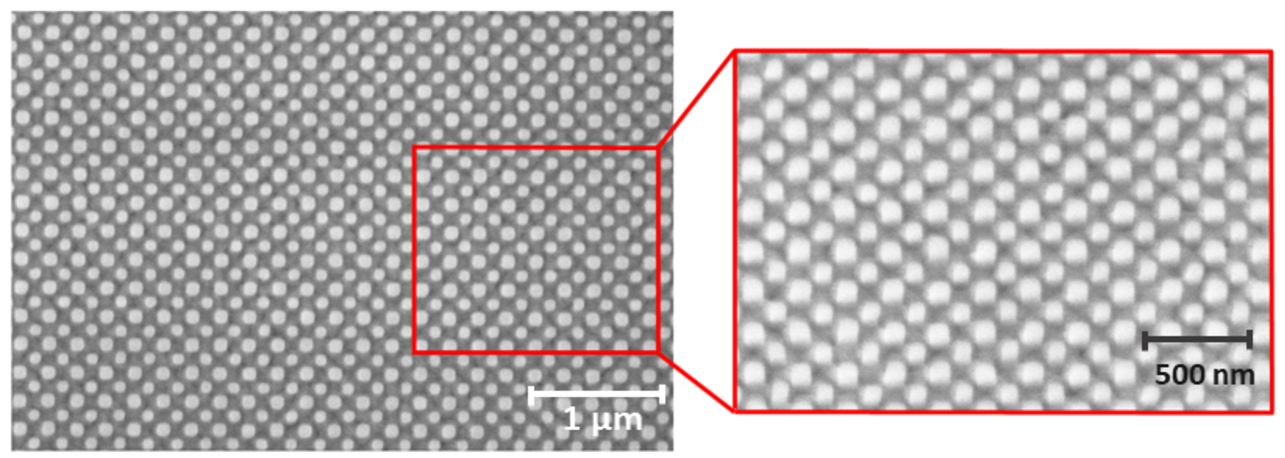}
  \caption{SEM image of the fabricated sample. The inset shows an enlarged view taken at a $30^{\circ}$ tilt angle.}
    \label{SEM_Ti}
\end{figure}

Figure~\ref{SEM_Ti} shows a scanning electron microscope (SEM) image of the fabricated sample, with the inset presenting an enlarged view taken at a $30^{\circ}$ tilt. Optical characterization is performed using a broadband halogen light source ($\mathrm{Thorlabs\,SLS201L}$). The incident light is focused onto the sample using a $50\times$, $0.75$ numerical aperture microscope objective, and the reflected light is collected and analyzed using an optical spectrum analyzer (OSA) ($\mathrm{Yokogawa\,AQ6370B}$). The reflection spectrum is normalized with respect to a gold mirror reference. 

\begin{figure}[t!]
\centering
\subfloat[]{\includegraphics[width=0.55\columnwidth]{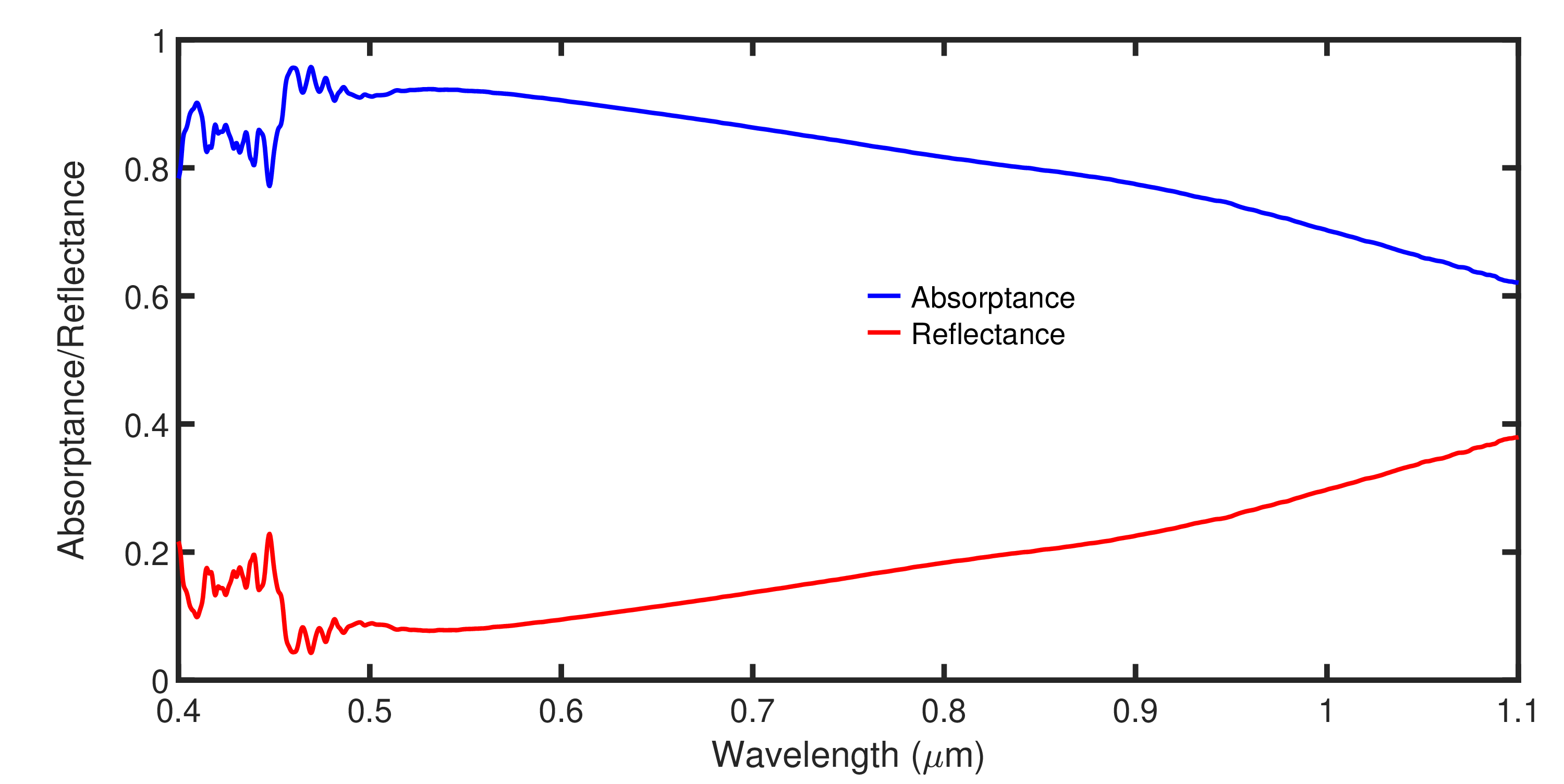}}\\
\subfloat[]{\includegraphics[width=0.55\columnwidth]{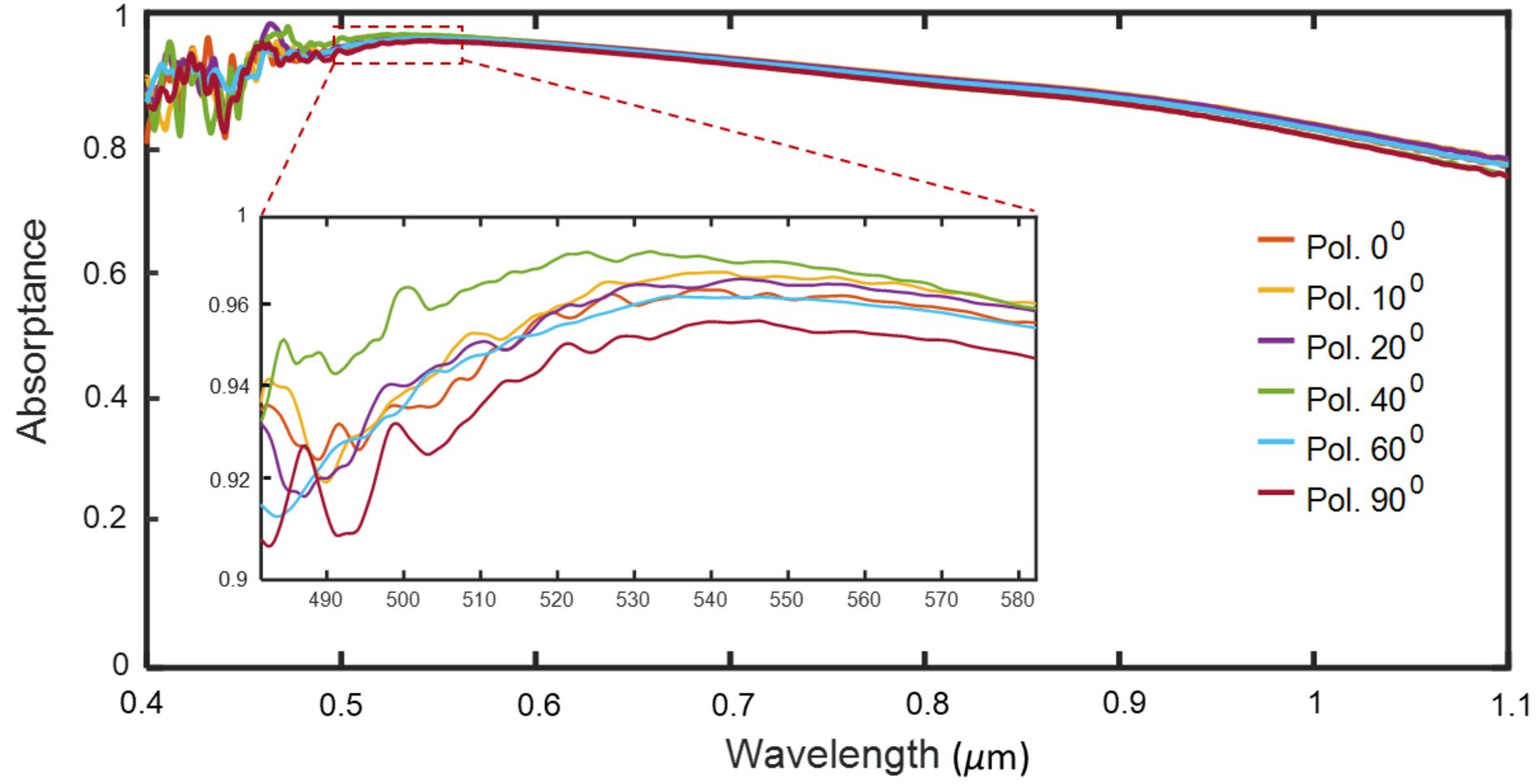}}\\
\subfloat[]{\includegraphics[width=0.55\columnwidth]{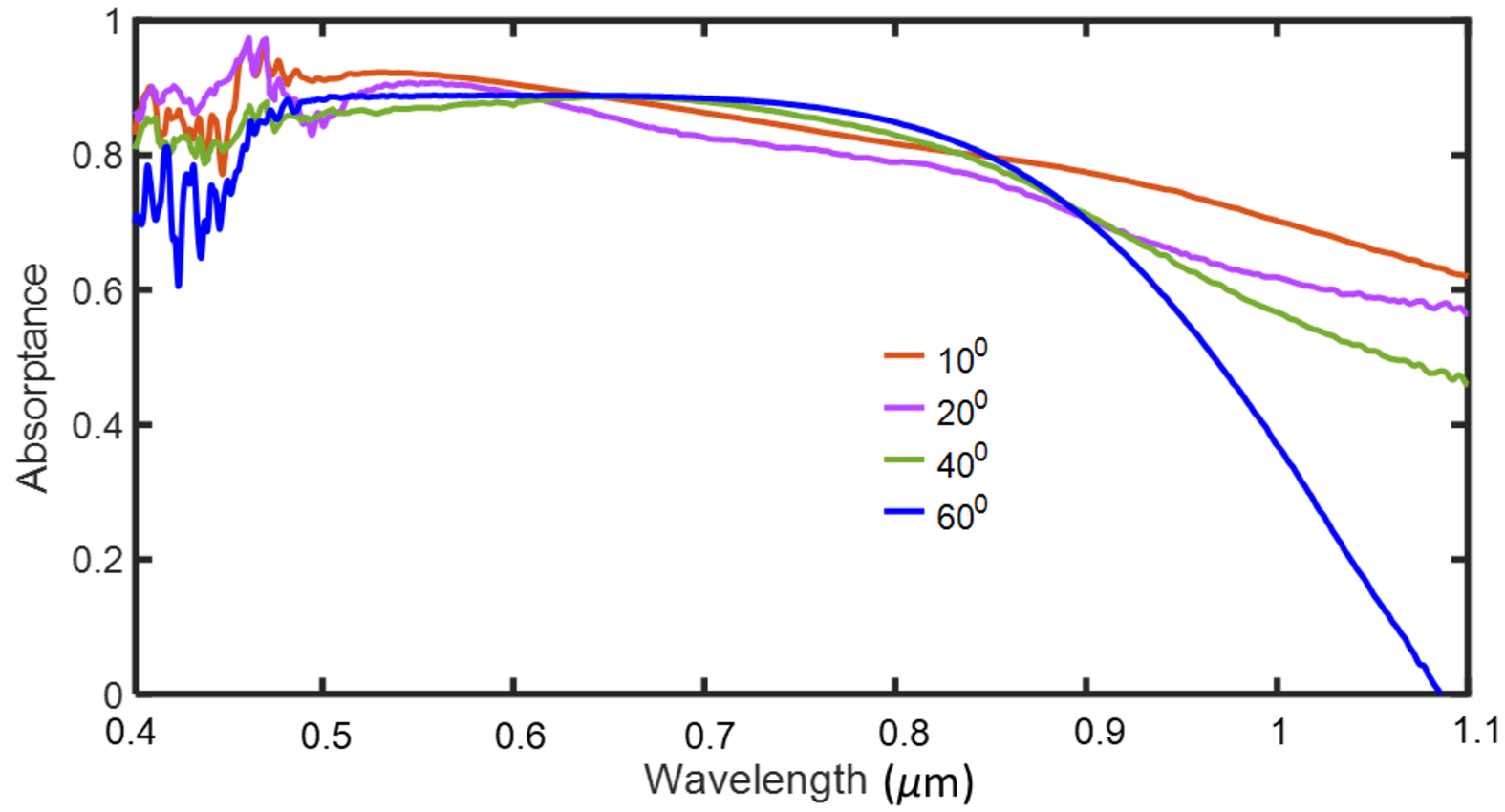}}
\caption{Measurement results of the fabricated sample. (a) Reflectance spectrum under $x$-polarized normally incident light. (b) Absorptance spectra at different polarization angles of the normally incident light. (c) Absorptance spectra under $x$-polarized light for various angles of incidence.}
\label{exp_abs}
\end{figure}

Figure~\ref{exp_abs}(a) shows the reflectance spectrum under $x$-polarized normally incident light, demonstrating efficient broadband absorption over a wide spectral range in good agreement with simulation results. Minor discrepancies between measured and simulated spectra are attributed to fabrication imperfections and measurement uncertainties. 

Polarization dependence is characterized by measuring the absorption at different polarization angles of the normally incident light, as shown in Fig.~\ref{exp_abs}(b). The results confirm the polarization insensitivity, consistent with the simulation results shown in Fig.~\ref{Pol_angle_variance_sim}(a). Figure~\ref{exp_abs}(c) shows the absorptance spectra under $x$-polarized light for various angles of incidence. As the angle increases, a slight reduction in bandwidth is observed, while strong absorption performance is maintained up to $60^{\circ}$.

\section{Fabrication error analysis}
Deviations from the ideal designed nanostructure geometry are inevitable due to the inherent limitations of state-of-the-art nanofabrication techniques. To quantify the impact of these fabrication-induced imperfections, a set of simulations is conducted to evaluate their influence on the absorption performance of the nanostructure with the optimized geometric parameters. A normally incident plane wave polarized along the $x$-direction is used for excitation.

\begin{figure}
  \centering
 \subfloat[] {\includegraphics[width=0.45\columnwidth]{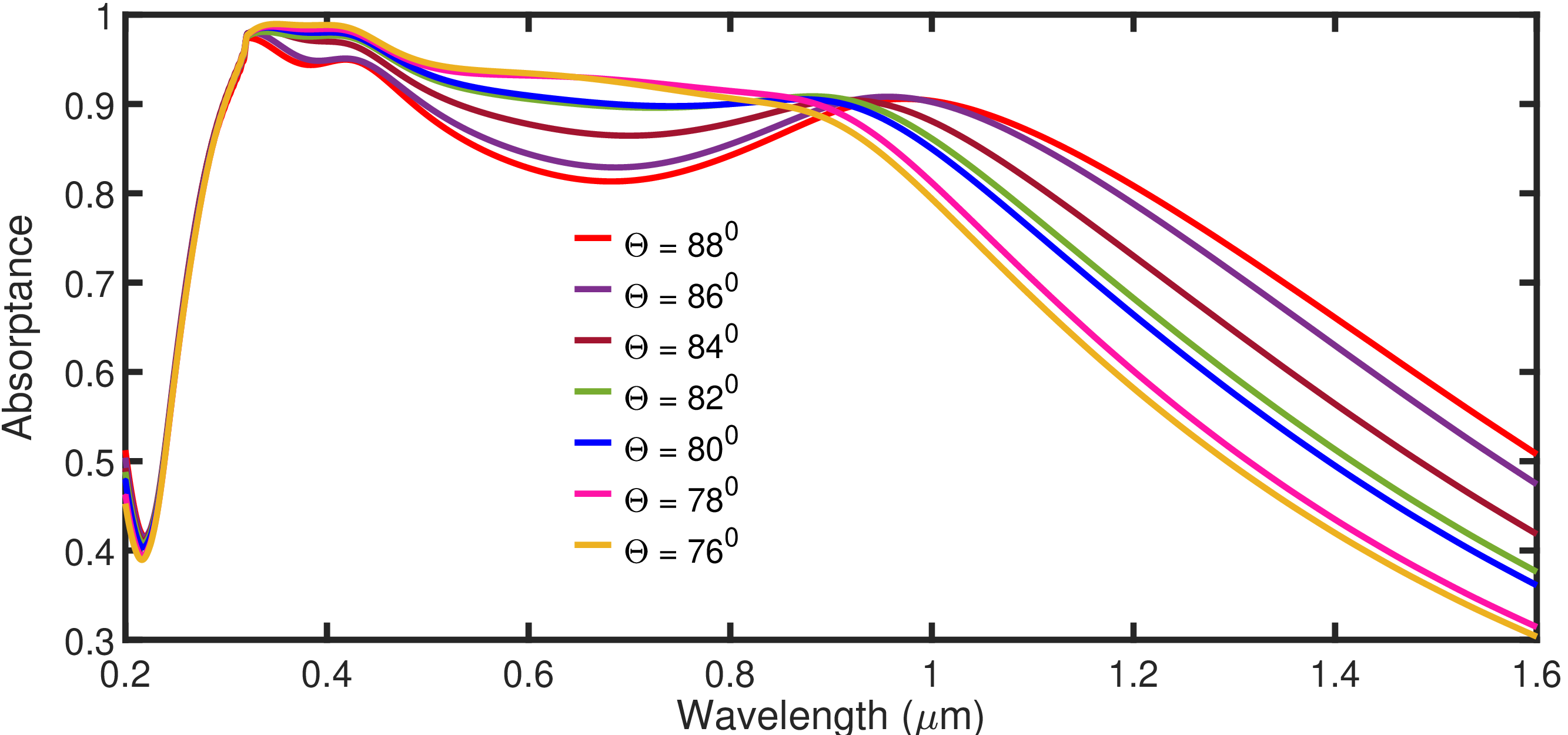}}\hspace{0.5cm}
\subfloat[]{ \includegraphics[width=0.44\columnwidth]{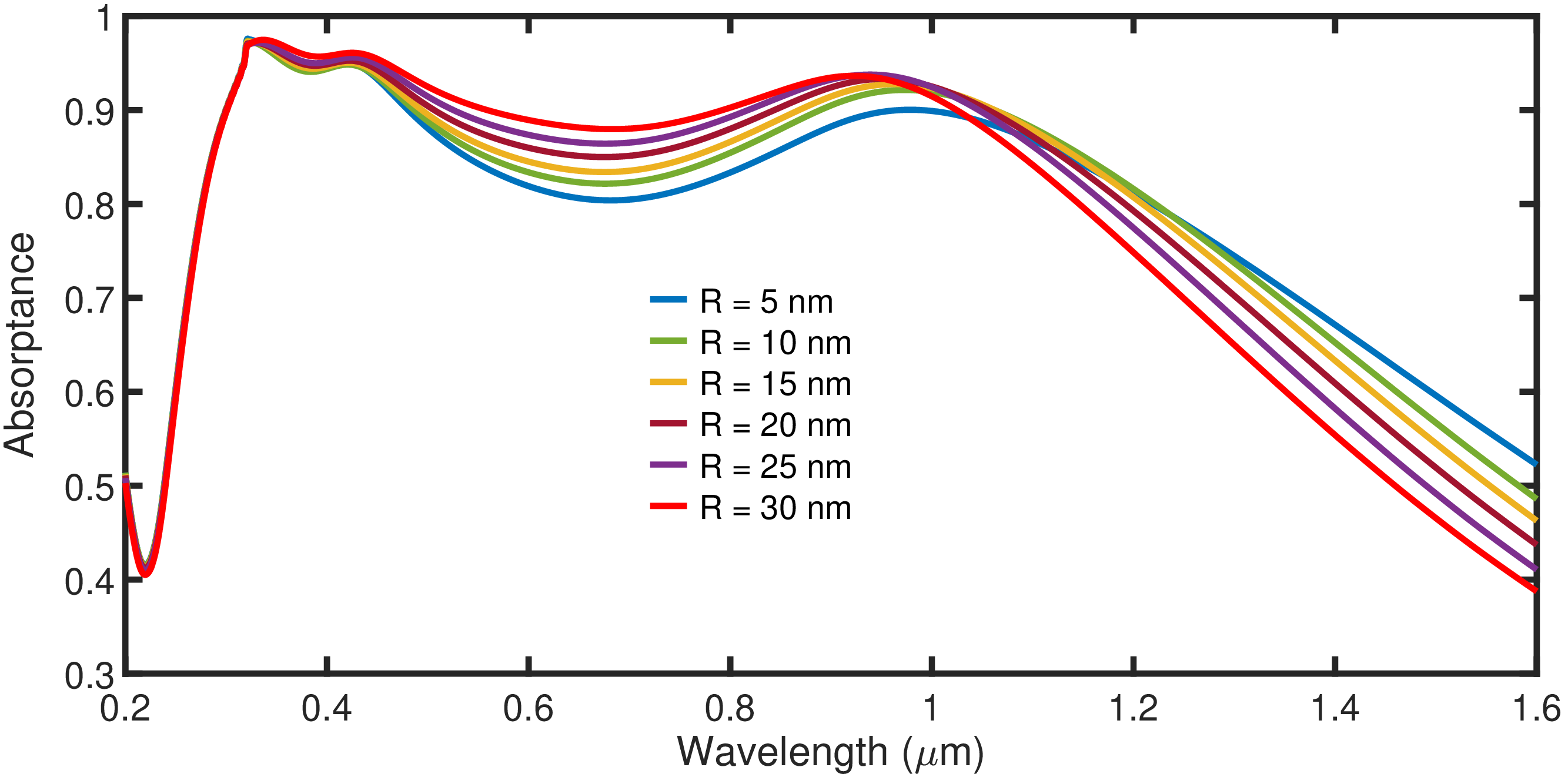}}\\
  \subfloat[]{\includegraphics[width=0.45\columnwidth]{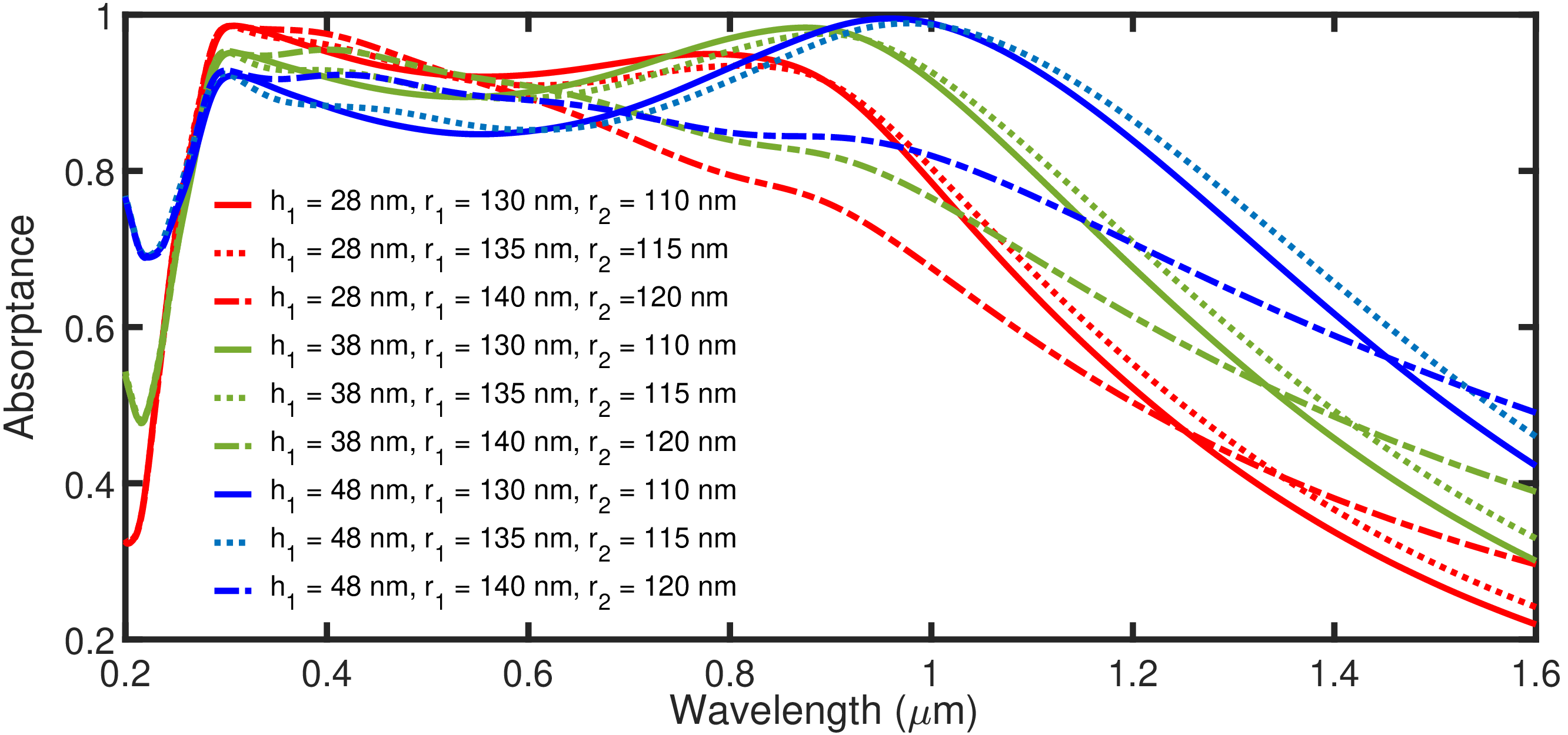}}\hspace{0.5cm}
  \subfloat[]{\includegraphics[width=0.45\columnwidth]{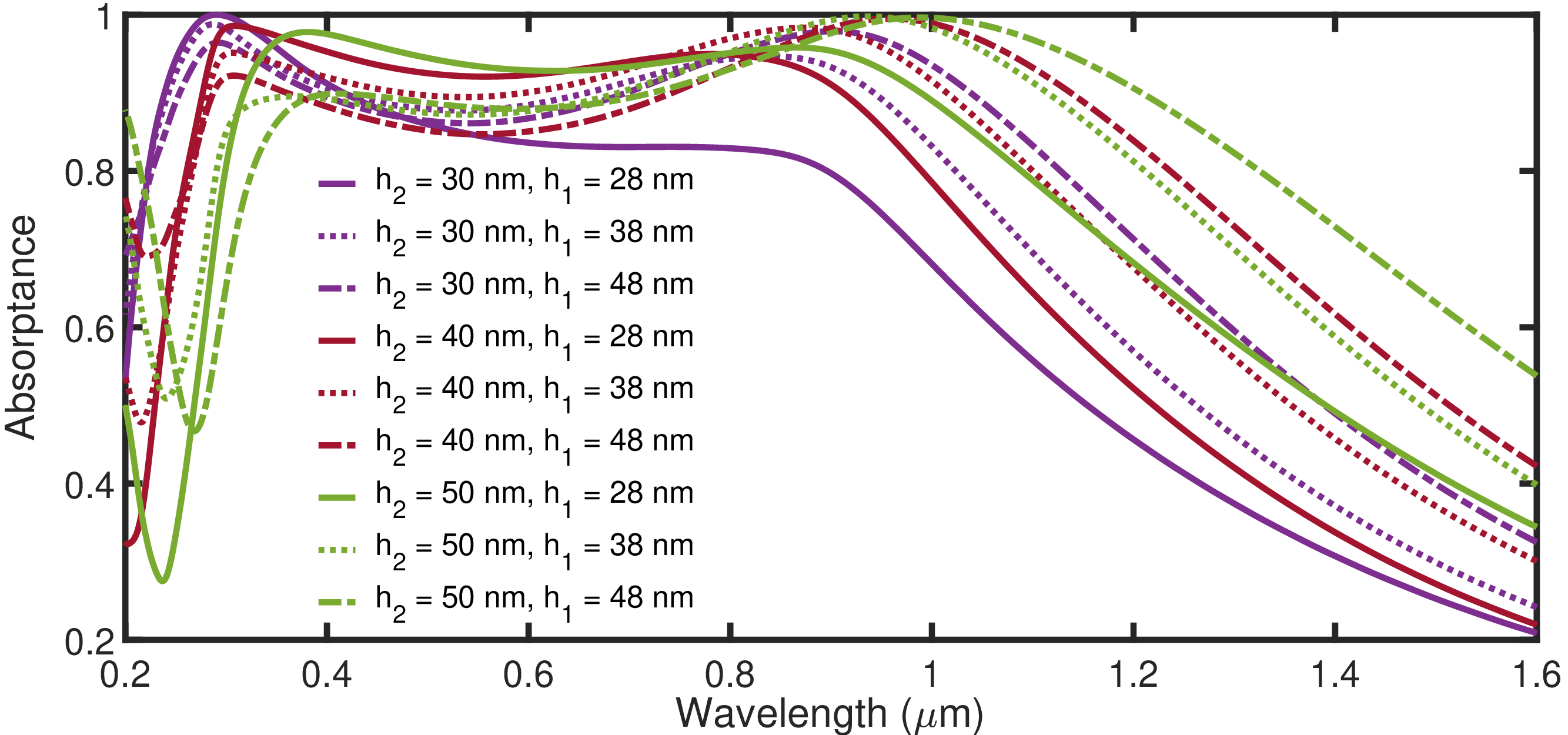}}
  \caption{Fabrication error analysis. Absorptance spectra for (a) varying inclined sidewall angle $\Theta$ (b) varying radius of corner rounding $R$, (c) different combinations of variations in the lateral unit cell dimensions $r_{1}$ and $r_{2}$ and the top layer thickness $h_1$ with spacer layer thickness $h_2$ fixed at its design value of $40\,\mathrm{nm}$, and (d) different combinations of $h_1$ and $h_2$ with $r_1$ and $r_2$ fixed at $130\,\mathrm{nm}$ and $110\,\mathrm{nm}$, respectively.}
  \label{error analysis}
\end{figure}

One notable fabrication imperfection arises from the inclined sidewalls of blocks on the top layer, commonly introduced during reactive ion etching (RIE) or the lift-off process. Ideally, the sidewall angle ($\Theta$) should be $90^\circ$. As illustrated in Fig.~\ref{error analysis}(a), the nanostructure exhibits robust absorption performance across a broad range of sidewall angles. While a slight degradation is observed at larger deviations particularly in the longer wavelength region, the overall spectral characteristics remain largely preserved.

Another common fabrication imperfection is the formation of a tapered, cone-like shape in the blocks, typically resulting from EBL followed by lift-off processes. Instead of exhibiting a perfectly rectangular cross-section, the blocks may acquire a slight conical distortion. Figure~\ref{error analysis}(b) illustrates the effect of this deviation on the absorptance spectrum. The results indicate that the device retains robust absorption performance even in the presence of substantial shape distortion.

In addition to above imperfections, variations in the lateral unit cell dimensions $r_{1}$ and $r_{2}$, the top layer thickness $h_1$, and the spacer layer thickness $h_2$ can also occur due to non-uniform deposition during PECVD and EBE. Figure~\ref{error analysis}(c) shows the impact of variations in $h_1$, $r_1$, and $r_2$ on the absorptance spectrum, with $h_2$ fixed at its design value of $40\,\mathrm{nm}$. The nanostructure maintains high absorption performance for deviations in $r_1$ and $r_2$ of up to $15\,\mathrm{nm}$ from their respective design values ($r_1 = 120\,\mathrm{nm}$, $r_2 = 100\,\mathrm{nm}$), beyond which the absorption efficiency begins to decline. However, the device exhibits strong tolerance to variations in $h_{1}$. Figure~\ref{error analysis}(d) shows the absorptance under simultaneous variations in $h_{1}$ and $h_{2}$, with $r_1$ and $r_2$ fixed at $130\,\mathrm{nm}$ and $110\,\mathrm{nm}$, respectively. These results confirm that the device exhibits strong tolerance to thickness deviations in both the metal and dielectric layers.

This analysis demonstrates that the proposed design exhibits robust absorption performance in the presence of common fabrication imperfections, including sidewall inclination, conical distortions, and dimensional deviations in the block geometry.

\section{Application to Si solar cells}\label{sec:app}


Nanoscale light trapping and localization offer significant advantages for photovoltaic applications~\cite{wei2025metasurface,guo2014}. Enhanced optical confinement in the active layer increases light absorption, enabling the design of efficient, ultrathin photovoltaic devices. Furthermore, strong electromagnetic localization improves carrier collection due to the reduced transport distances in thin films~\cite{olaimat2021using, shameli2018absorption,shameli2018light, mellor2017}. 

\begin{figure}
  \centering
   \includegraphics[width=0.5\linewidth]{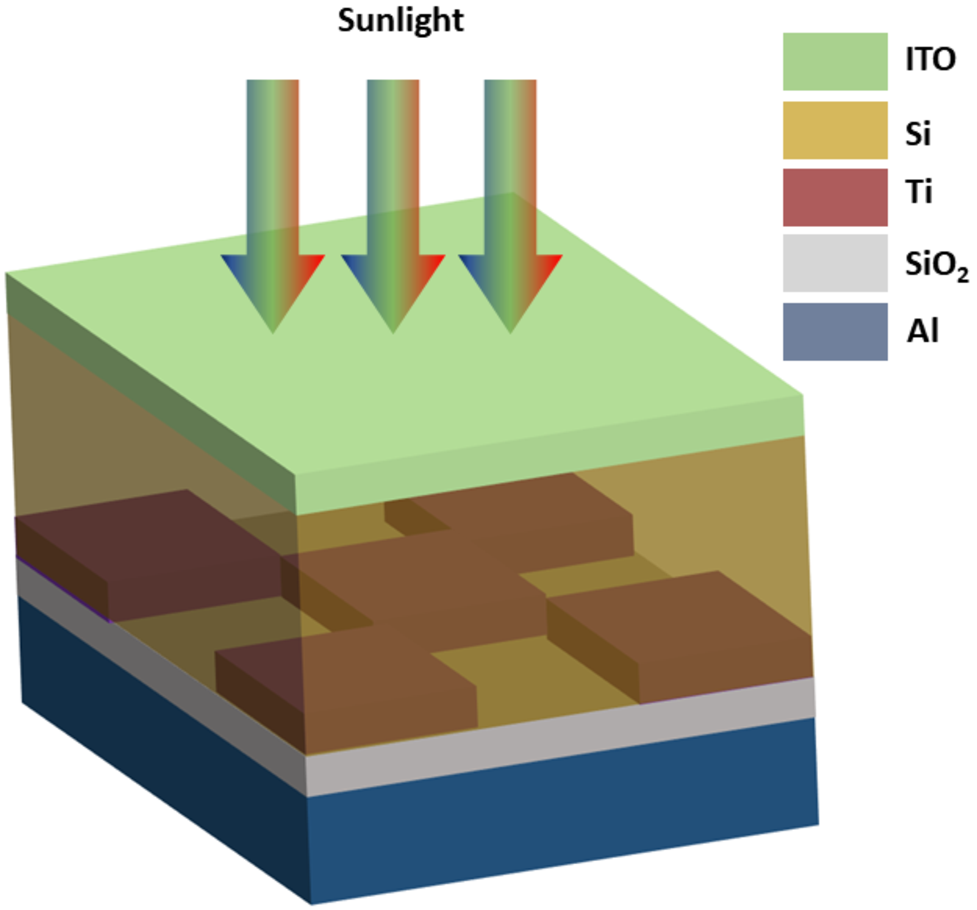}
  \caption{Schematic of the thin-film $\mathrm{Si}$ solar cell integrating the proposed nanostructured absorber beneath the active layer to enhance broadband absorption.}
    \label{solar_cell_f}
\end{figure}

This section investigates the integration of the proposed absorber into a thin-film solar cell architecture. The resulting device structure is shown in Fig.~\ref{solar_cell_f}. A $75\,\mathrm{nm}$-thick ITO layer is deposited on the top surface to minimize reflection losses while serving as a transparent conducting electrode. The active layer made of $\mathrm{Si}$ is deposited above the absorber designed to enhance light absorption. The absorber consists of a $100\,\mathrm{nm}$ $\mathrm{Al}$ back reflector, a $40\,\mathrm{nm}$ $\mathrm{SiO_2}$ spacer, and a $38\,\mathrm{nm}$ nanostructured $\mathrm{Ti}$ layer, following the previously optimized configuration.

\begin{figure}
\centering
\subfloat[]{\includegraphics[width=0.48\columnwidth]{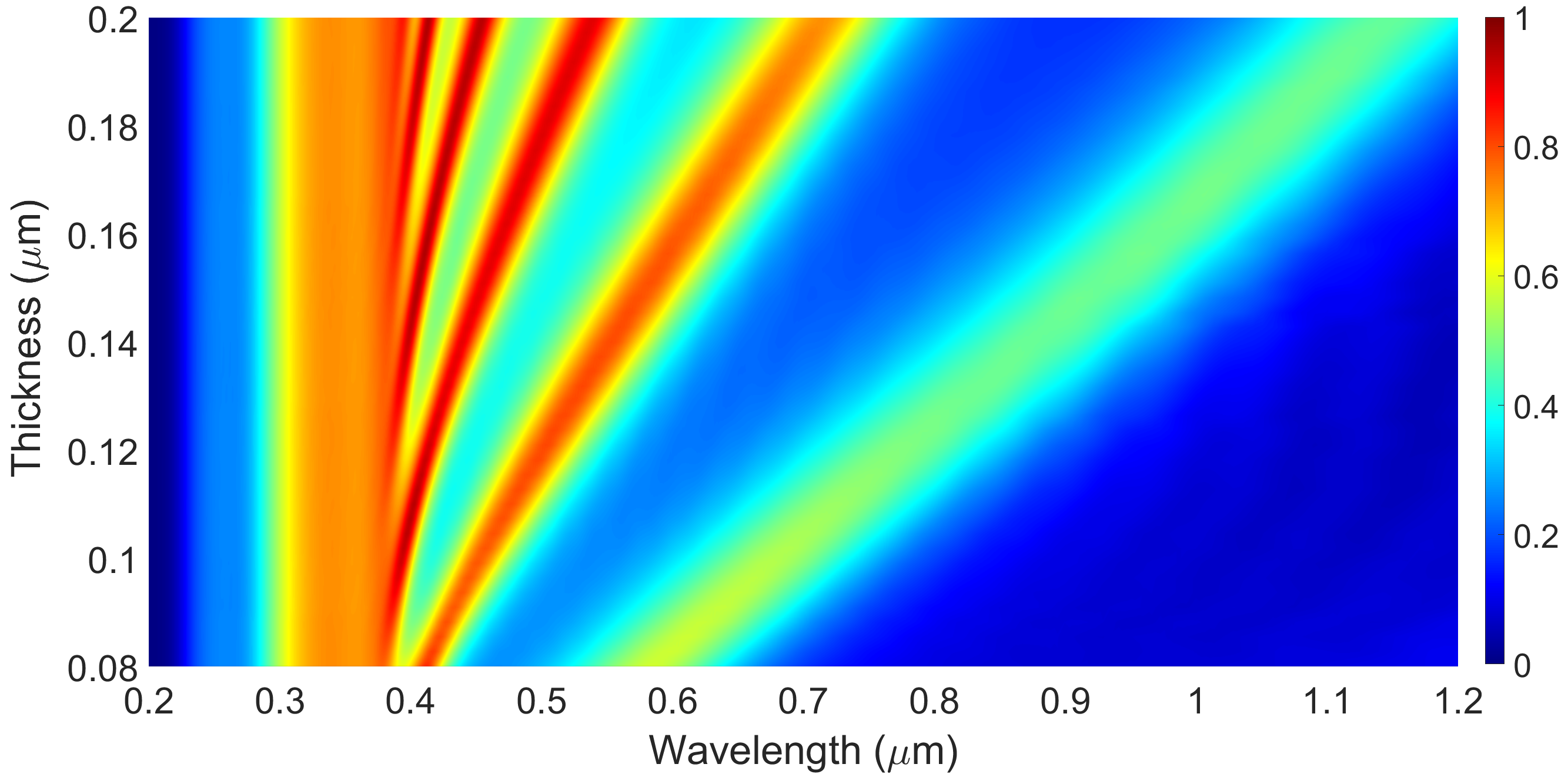}}\hspace{0.25cm}
\subfloat[]{\includegraphics[width=0.48\columnwidth]{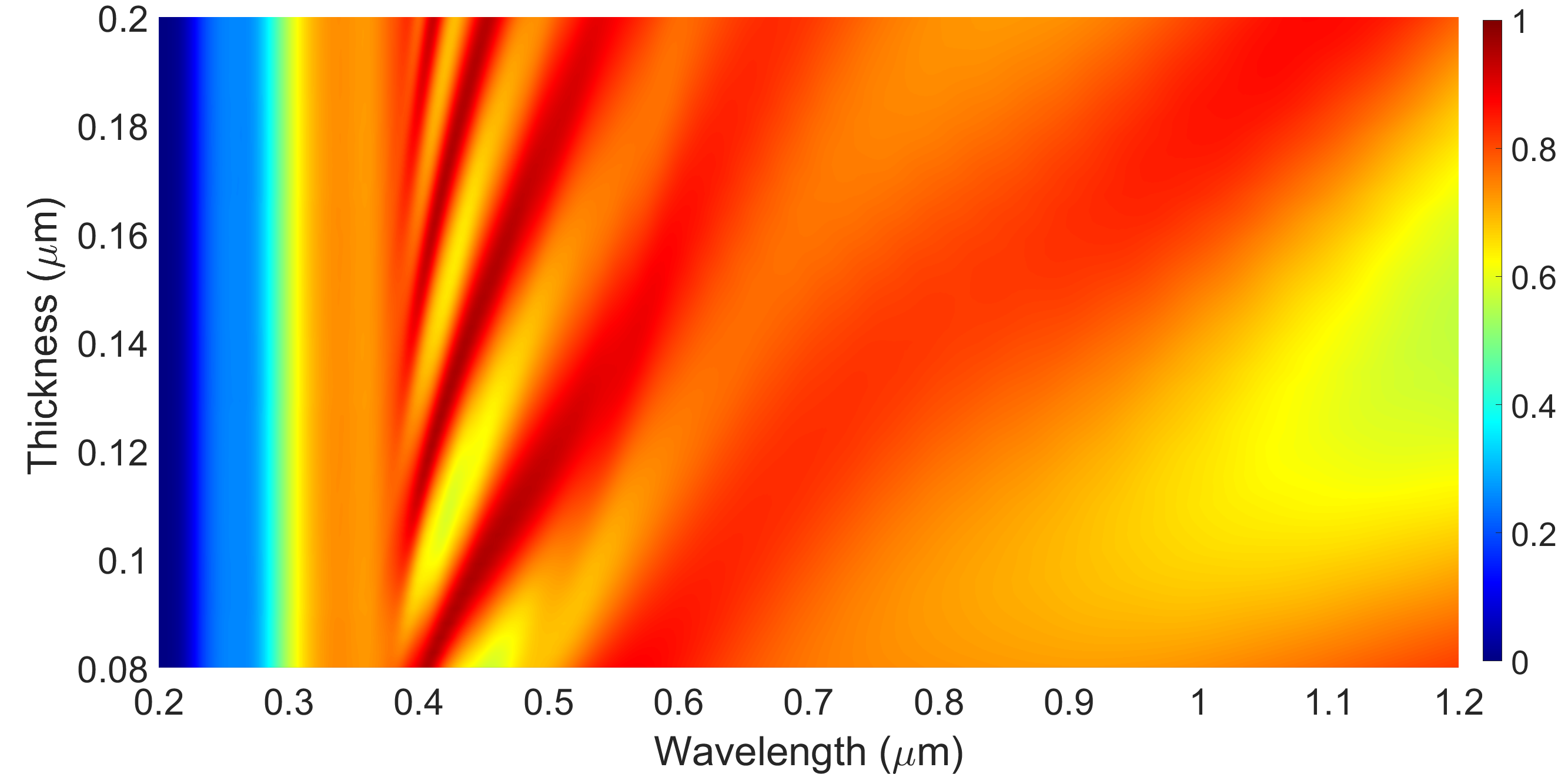}}\\
\subfloat[]{\includegraphics[width=0.48\columnwidth]{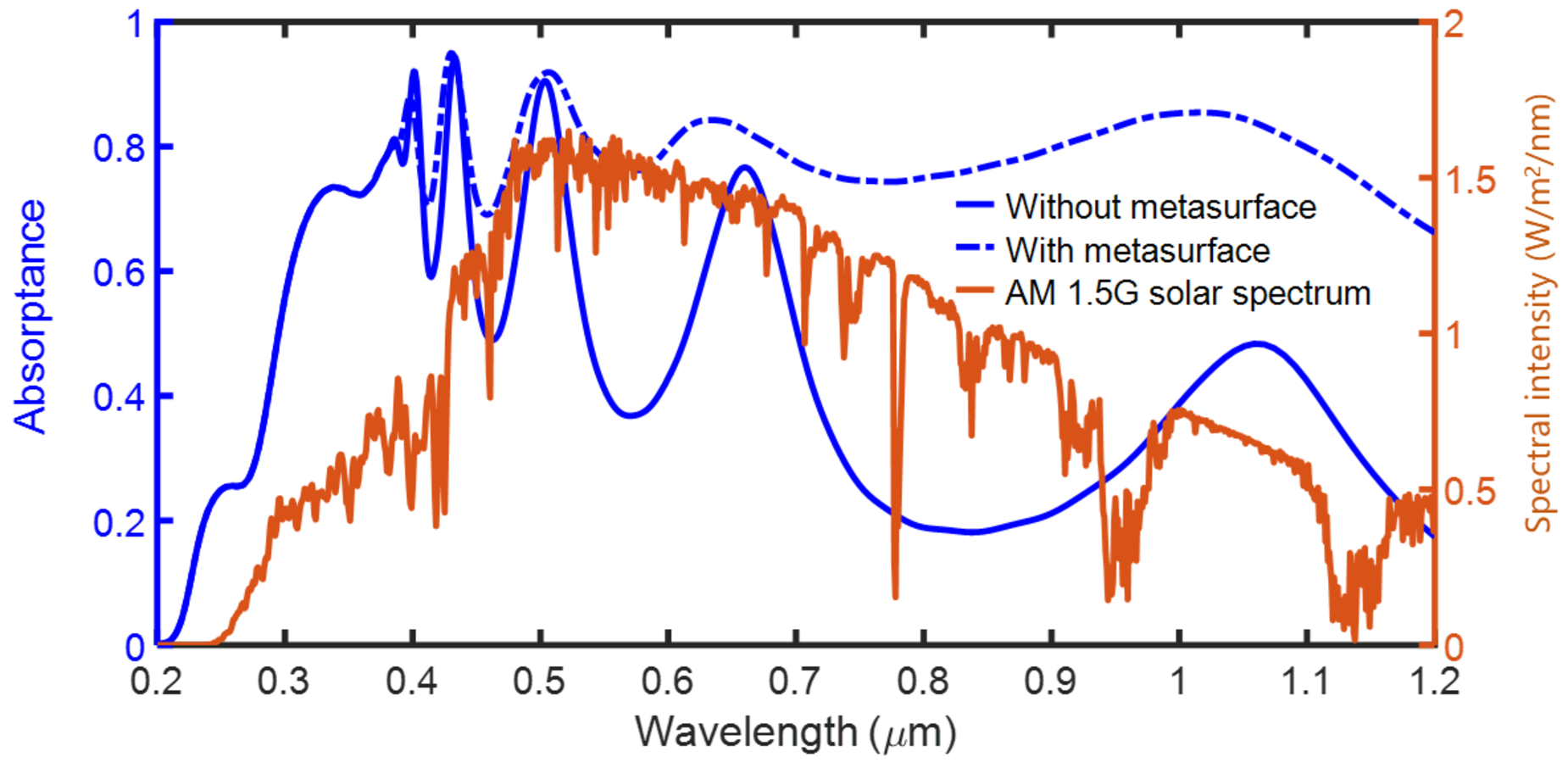}}\hspace{0.25cm}
\subfloat[]{\includegraphics[width=0.48\columnwidth]{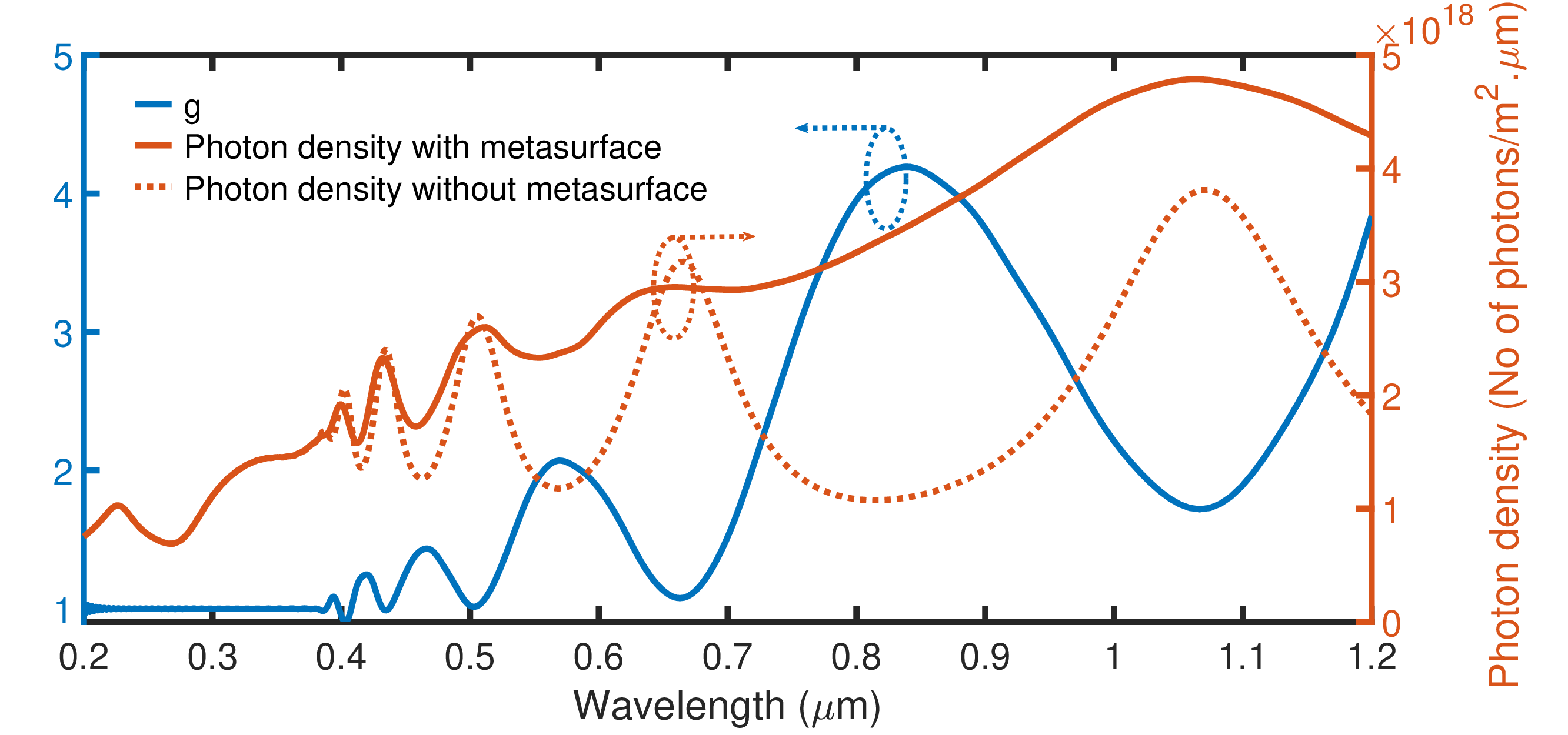}}
 \caption{Absorptance spectra of the solar cell (a) without and (b) with the proposed absorber for varying $\mathrm{Si}$ layer thickness. (c) Absorptance spectra of the solar cell with a $180,\mathrm{nm}$-thick $\mathrm{Si}$ layer, comparing configurations with and without the absorber. (d) The left axis represents the absorption enhancement factor $g$ as a function of wavelength, while the right axis displays the photon density with and without the absorber.}\label{abs_solarcell_combined}
\end{figure}

The time-averaged power absorbed in the active layer is computed using~\cite{kumawat2020plasmonic}
\begin{equation}
  P_{\mathrm{abs}} =\frac{1}{2} \omega \mathfrak{Im}\{\varepsilon\} \int \mathbf{E} \cdot \mathbf{E}\,dv
  \label{eqn_abs_solarcell}
\end{equation}
where $\omega$ is the angular frequency, $\mathfrak{Im}\{\varepsilon\}$ is the imaginary part of the permittivity (assumed homogeneous inside the active region), and $\mathbf{E}$ is the electric field, obtained by converting the FDTD results into the frequency domain~\cite{refFDTD}. The absorptance in the active medium is obtained by normalizing the time-averaged absorbed power to the incident power:
\begin{equation}
  A =\frac{P_{\mathrm{abs}}}{P_{\mathrm{in}}}
  \label{eqn_Alamda_solarcel}
\end{equation}
where $P_{\mathrm{in}}$ is the incident power. Figures~\ref{abs_solarcell_combined}(a) and (b) show the absorptance spectra in the $\mathrm{Si}$ layer with varying thickness for the solar cell without and with the proposed absorber, respectively. For the solar cell without the absorber, absorption peaks redshift and increase modestly with increasing thickness; however, the overall absorption remains low, particularly at longer wavelengths. In contrast, for the solar cell with the absorber, broadband absorption is significantly enhanced. 

The solar cell with the $\mathrm{Si}$ layer of thickness $180\,\mathrm{nm}$ is selected for further analysis. Figure~\ref{abs_solarcell_combined}(c) compares the absorptance in the $\mathrm{Si}$ layer of the solar cell with and without the absorber, demonstrating a marked improvement with the integrated absorber. Let $g$ denote the absorption enhancement factor, defined as the ratio of the absorption in the solar cell with the absorber to that without. In Fig.~\ref{abs_solarcell_combined}(d), the left axis displays the absorption enhancement factor $g$ as a function of wavelength, while the right axis shows the photon density with and without the absorber. The variation in the enhancement factor arises from the strong wavelength-dependent absorption characteristics of the planar structure. These results clearly demonstrate that the inclusion of the absorber significantly increases the photon density, highlighting enhanced light confinement and improved potential for electricity generation.

\begin{figure}[t!]
\centering
\subfloat[]{\includegraphics[width=0.45\columnwidth]{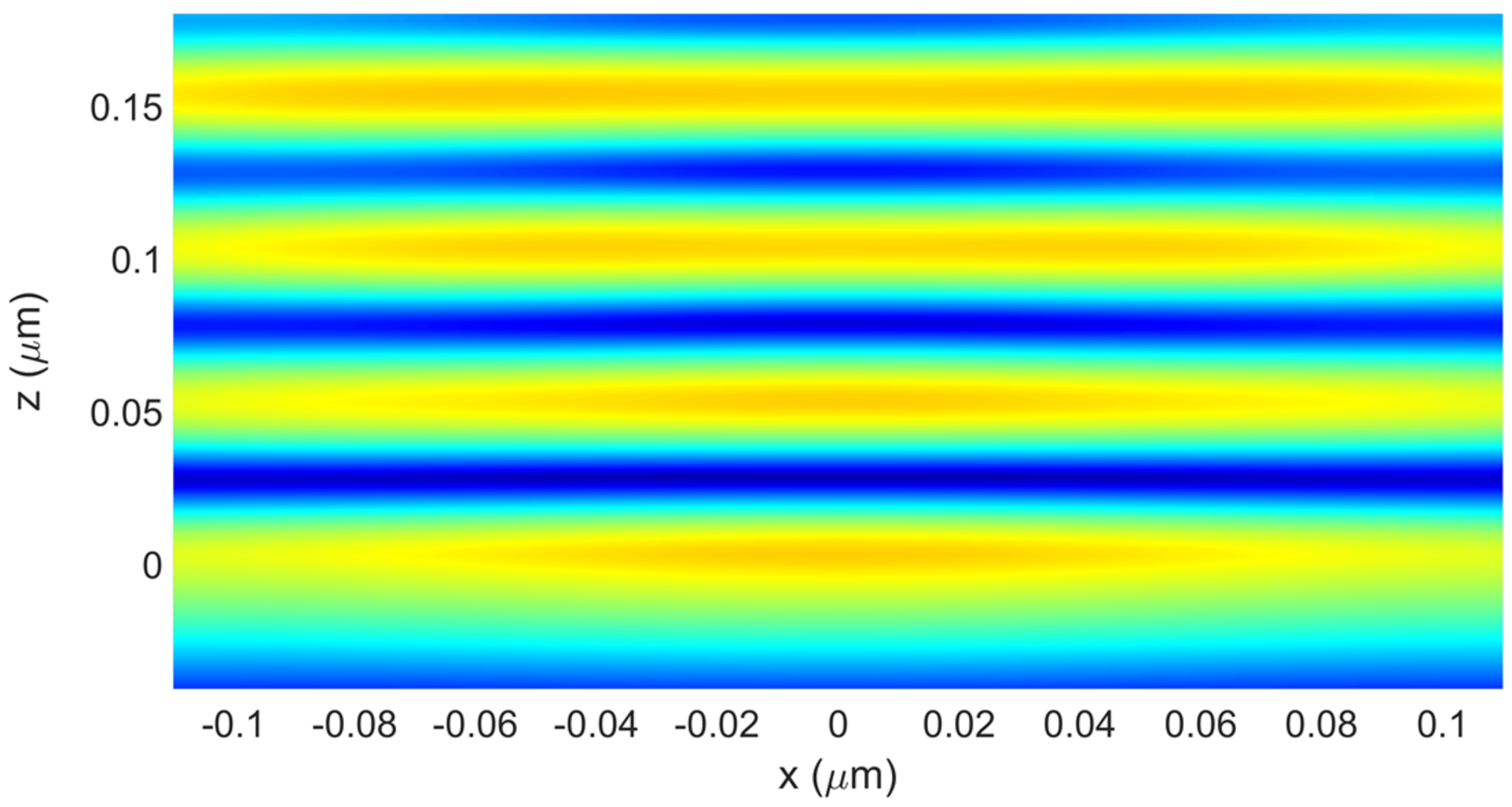}}\hspace{0.5cm}
\subfloat[]{\includegraphics[width=0.45\columnwidth]{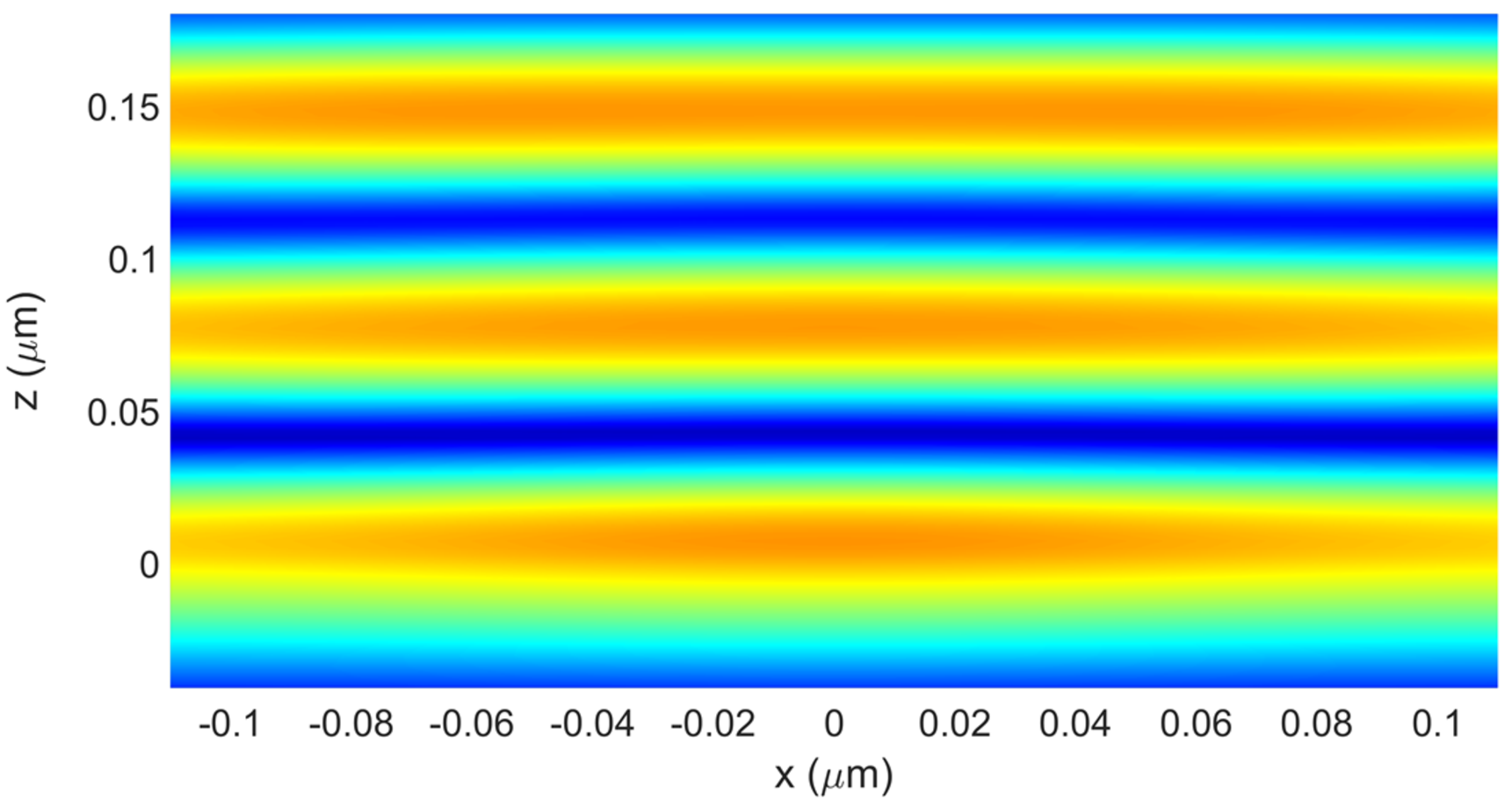}}\\
\subfloat[]{\includegraphics[width=0.45\columnwidth]{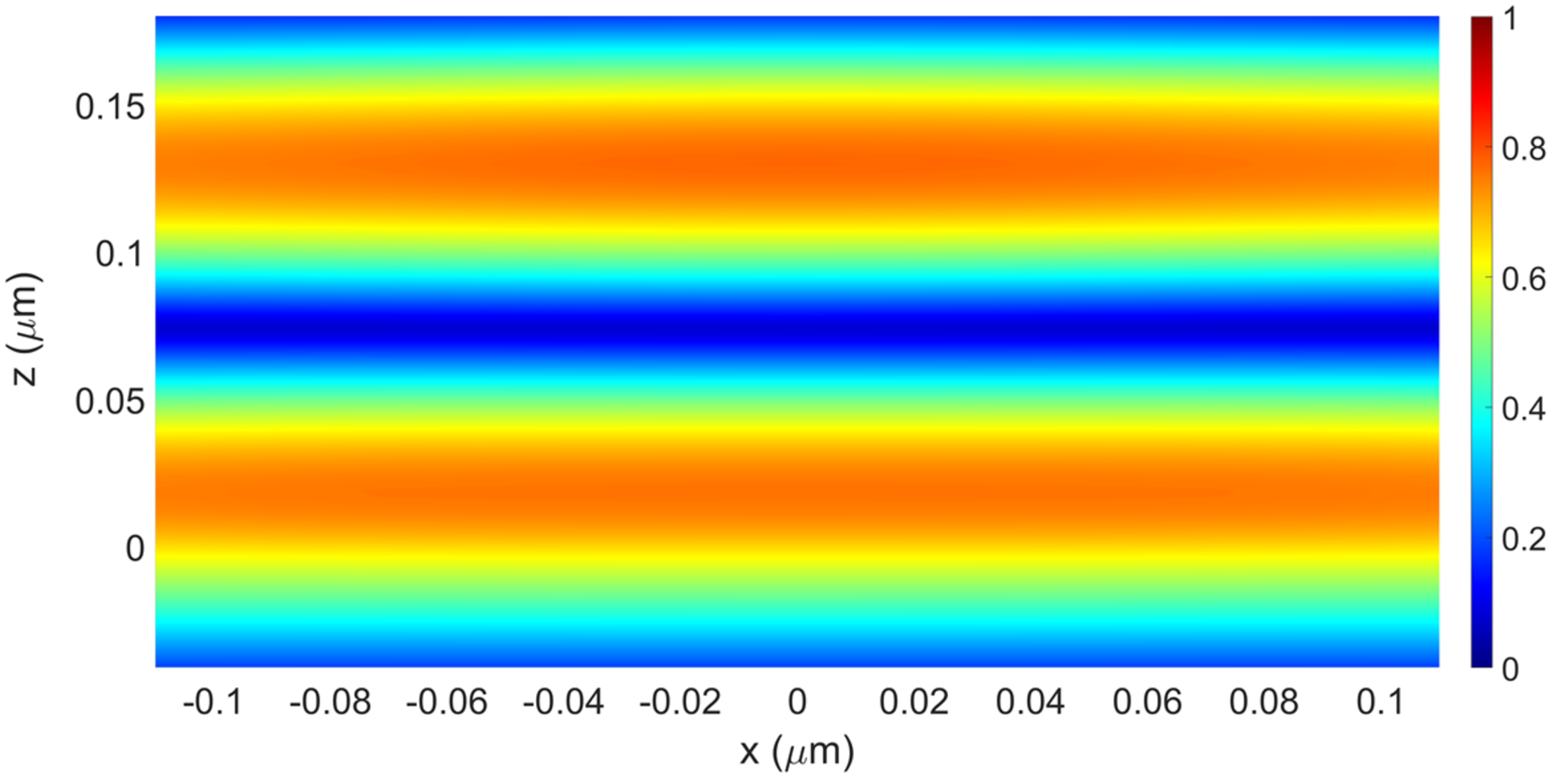}}\hspace{0.5cm}
\subfloat[]{\includegraphics[width=0.45\columnwidth]{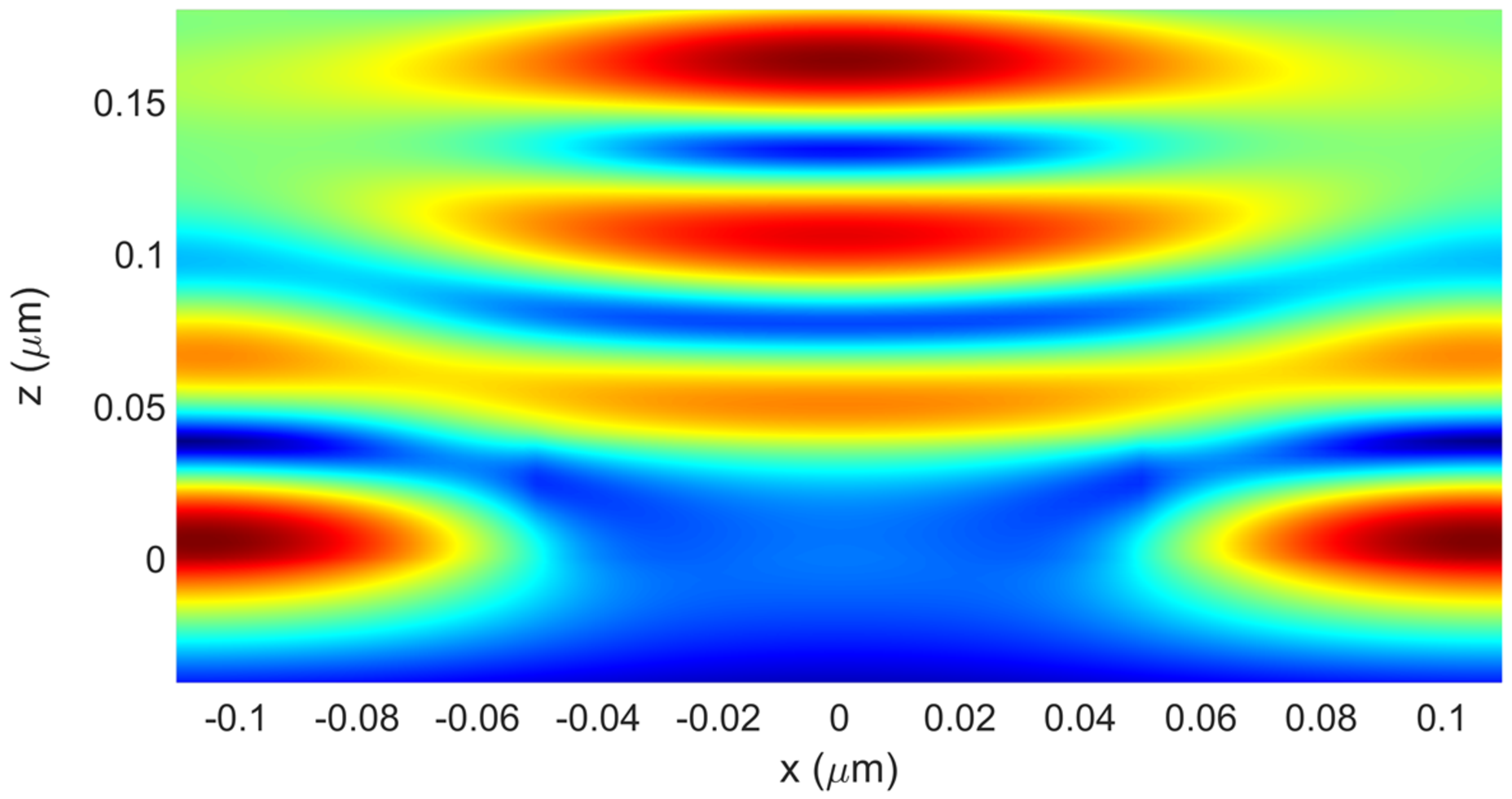}}\\
\subfloat[]{\includegraphics[width=0.45\columnwidth]{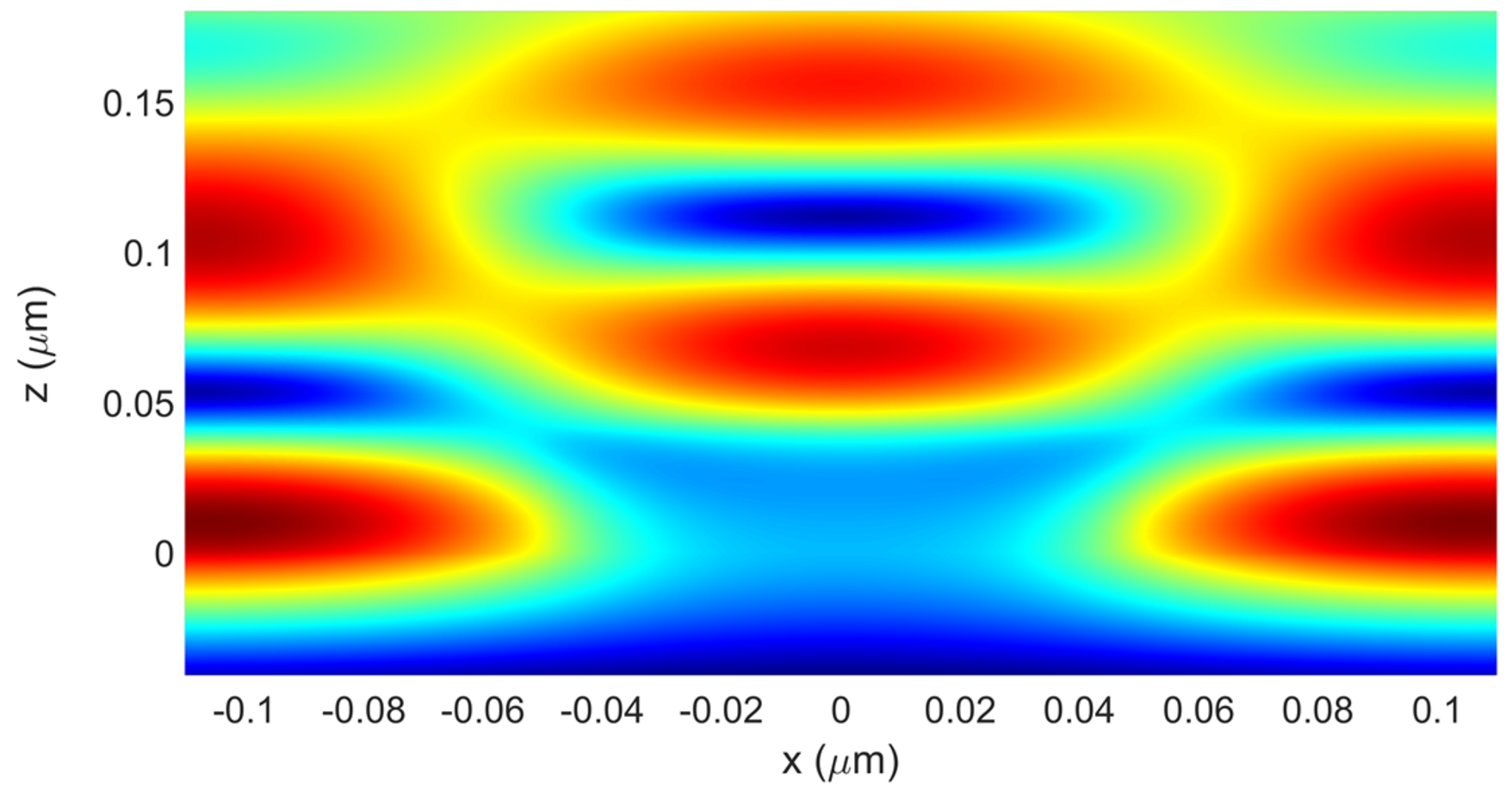}}\hspace{0.5cm}
\subfloat[]{\includegraphics[width=0.45\columnwidth]{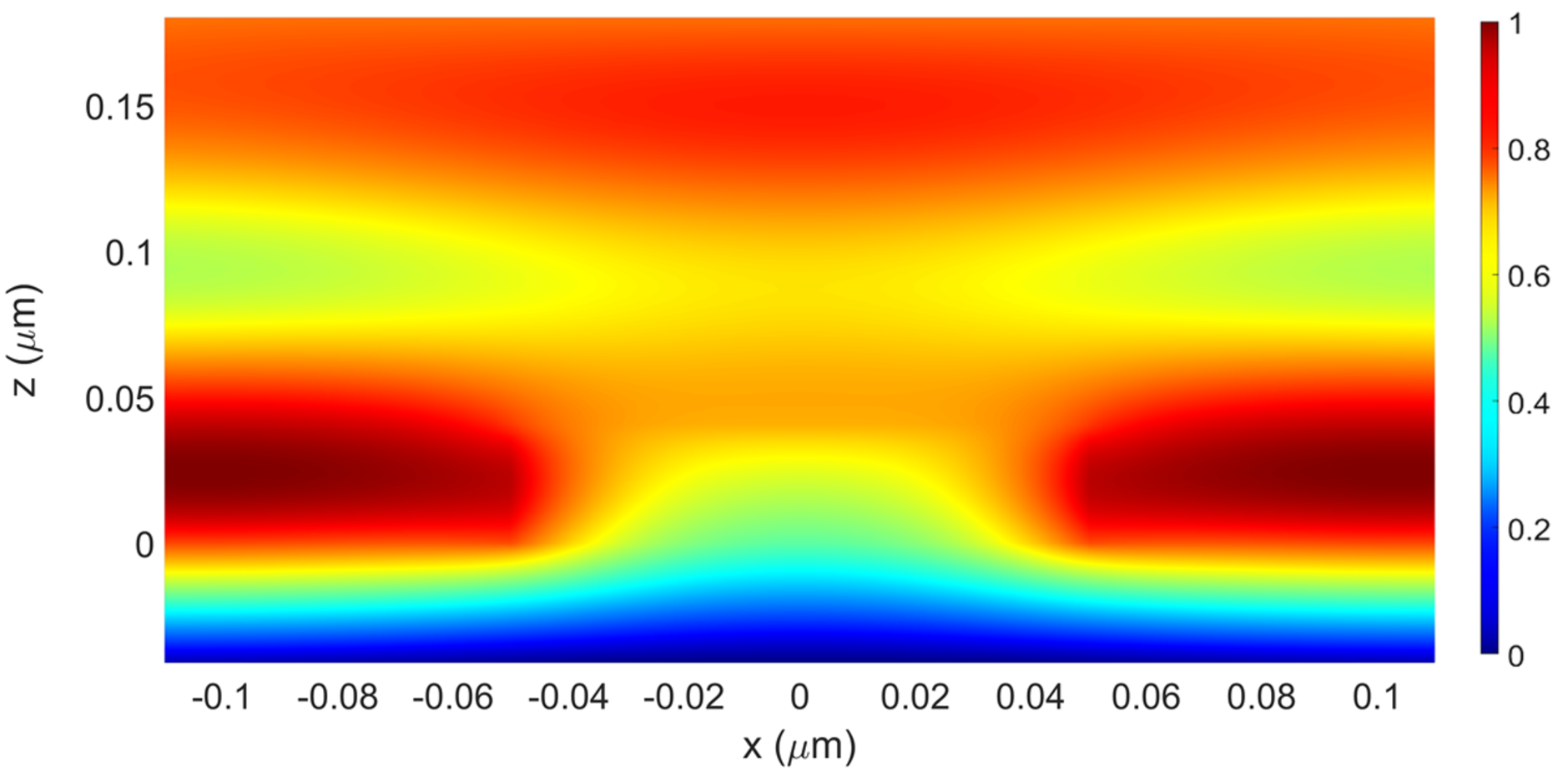}}

\caption{Normalized electric field intensity in the $xz$-plane for the solar cell without the absorber at (a) $466\,\mathrm{nm}$, (b) $568\,\mathrm{nm}$, and (c) $836\, \mathrm{nm}$, and with the absorber at (d) $466\,\mathrm{nm}$, (e) $568\,\mathrm{nm}$, and (f) $836\, \mathrm{nm}$.}
    \label{combined_solarcell_abs_plot}
\end{figure}

Next, the electric field distributions in the $\mathrm{Si}$ layer of the solar cell with and without the absorber are investigated. Figures~\ref{combined_solarcell_abs_plot}(a)-(c) and (d)-(f) show the normalized magnitude of the electric field in the $xz$-plane at $466\,\mathrm{nm}$, $568\,\mathrm{nm}$, and $836\,\mathrm{nm}$ for the solar cell without and with the absorber, respectively. The figures clearly show that the inclusion of the absorber leads to a significant enhancement of the electric field intensity within the active region, highlighting the role of nanostructure-based absorbers in boosting light confinement and absorption in thin-film solar cells. 
Absorbed solar energy can be harvested through systems such as thermophotovoltaic (TPV)~\cite{ni2019highly,isobe2020spectral,shin2018thermoplasmonic} converters. In these converters, the plasmonic nanostructure, upon heating under solar irradiation, serves as a selective thermal emitter. By engineering its emissivity spectrum to align with the bandgap of a nearby TPV cell, the emitted infrared radiation can be efficiently converted into electrical power.
 
\section{Conclusion}
A nanostructured absorber based on an MDM configuration is designed and fabricated. The structure consists of a periodic $\mathrm{Ti}$ array as the top layer, a thin $\mathrm{SiO_{2}}$ spacer, and a continuous $\mathrm{Al}$ back reflector. Broadband absorption covering the UV to near-IR range of the spectrum is achieved through the excitation of LSPs and cavity modes, as confirmed by FDTD simulations. The influence of key structural parameters, polarization, and angle of incidence on the absorption performance is systematically analyzed.

The optimized absorber is fabricated using electron beam lithography and characterized using a broadband light source. 
Measured results show good agreement with numerical simulations, validating the proposed design. Numerical simulations further demonstrate that integrating the absorber into a thin-film $\mathrm{Si}$ solar cell significantly enhances the active layer's absorption across a broad wavelength range. The proposed absorber design holds promise for solar energy harvesting, thermal emission, and next-generation photovoltaic applications.

\section*{Acknowledgments}
The authors gratefully acknowledge the KAUST Nanofabrication Core Lab for providing the fabrication facilities. They also extend their sincere thanks to Prof. Boon S. Ooi for granting access to the experimental setup in his Photonics Laboratory.

\bibliographystyle{IEEEtran}

\newpage\clearpage

\section*{Figures}

\newpage\clearpage

\newpage\clearpage

\newpage\clearpage

\newpage\clearpage

\newpage\clearpage
\newpage\clearpage

\newpage\clearpage

\newpage\clearpage

\newpage\clearpage

\newpage\clearpage

\newpage\clearpage

\begin{figure*}
\centering
 \subfloat{\includegraphics[width=0.3\columnwidth]{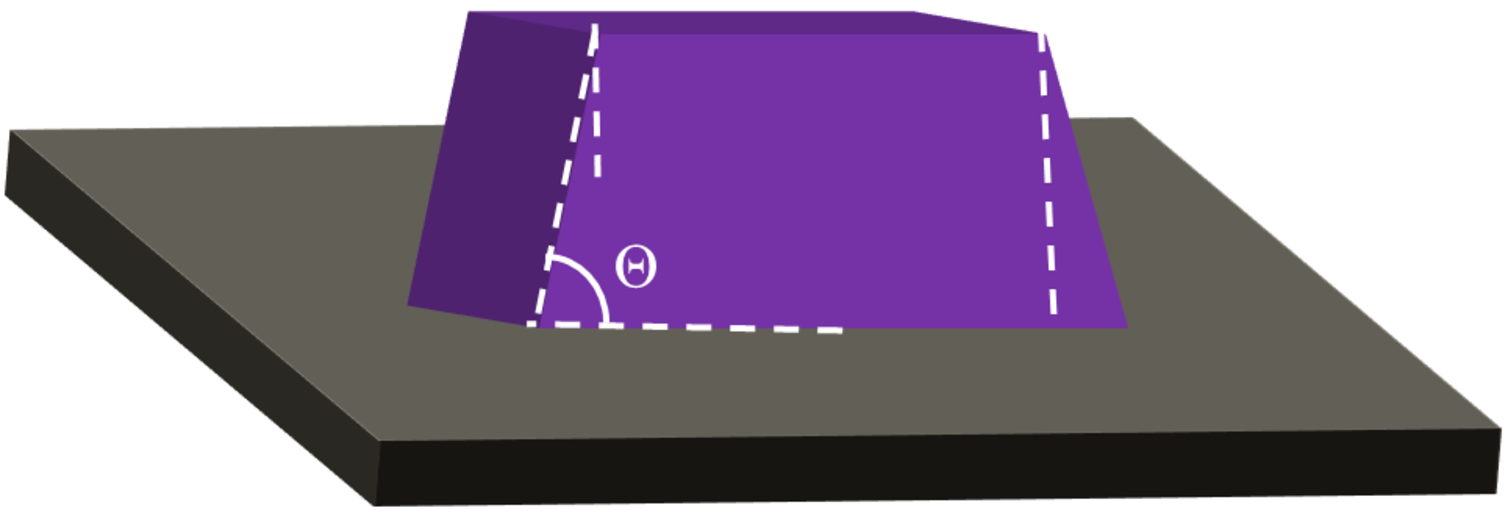}}\hspace{2cm}
\subfloat{ \includegraphics[width=0.3\columnwidth]{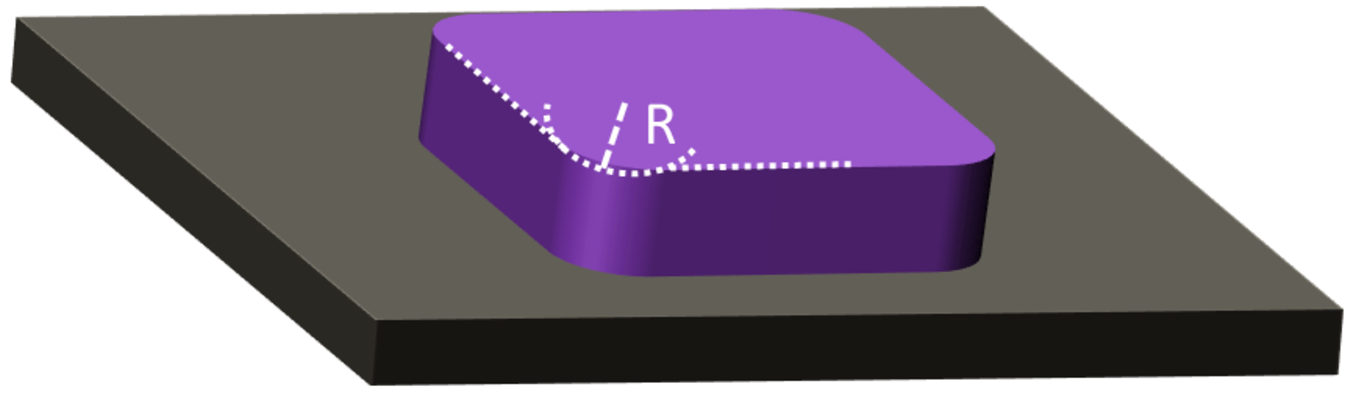}}
\vspace{-2cm}
\end{figure*}

\newpage\clearpage

\newpage\clearpage

\newpage\clearpage

\end{document}